\documentclass[iop,tighten]{emulateapj}
\usepackage{natbib}
\usepackage{graphicx}
\usepackage{amssymb}
\usepackage{amsmath}
\usepackage{enumerate}
\usepackage{xcolor}
\usepackage{hyperref}
\hypersetup{
	colorlinks = true,
	linkcolor = black,
	citecolor = blue,
	filecolor= black,
	runcolor= black,
	urlcolor= blue
}
\usepackage{url}

\defcitealias{steidel2016}{S16}
\defcitealias{steidel2014}{S14}
\defcitealias{pilyugin2012}{Pil12}
\defcitealias{pettini2004}{PP04}
\defcitealias{perez-montero2009}{PMC09}

\shorttitle{Ionization, excitation, and N/O in $z\simeq2-3$ galaxies}
\shortauthors{Strom et al.}

\begin{document}

\title{Nebular emission line ratios in $z\simeq2-3$ star-forming galaxies with KBSS-MOSFIRE\altaffilmark{1}: \\ exploring the impact of ionization, excitation, and nitrogen-to-oxygen ratio}
\altaffiltext{1}{The data presented in this paper were obtained at the W.M. Keck Observatory, which is operated as a scientific partnership among the California Institute of Technology, the University of California, and the National Aeronautics and Space Administration. The Observatory was made possible by the generous financial support of the W.M. Keck Foundation.}
\author{Allison L. Strom\altaffilmark{2}}
\email{astrom@astro.caltech.edu}
\author{Charles C. Steidel\altaffilmark{2}}
\altaffiltext{2}{Cahill Center for Astronomy and Astrophysics, California Institute of Technology, MS 249-17, Pasadena, CA 91125, USA}
\author{Gwen C. Rudie\altaffilmark{3}}
\altaffiltext{3}{Carnegie Observatories, 813 Santa Barbara Street, Pasadena, CA 91101, USA}
\author{Ryan F. Trainor\altaffilmark{4}}
\altaffiltext{4}{Department of Astronomy, University of California, Berkeley, New Campbell Hall, Berkeley, CA 94720, USA}
\author{Max Pettini\altaffilmark{5}}
\altaffiltext{5}{Institute of Astronomy, Madingley Road, Cambridge CB3 OHA, UK}
\author{Naveen A. Reddy\altaffilmark{6}}
\altaffiltext{6}{Department of Physics and Astronomy, University of California, Riverside, 900 University Avenue, Riverside, CA 92521, USA}
\begin{abstract}
We present a detailed study of the rest-optical ($3600-7000$\AA) nebular spectra of $\sim380$ star-forming galaxies at $z\simeq2-3$ obtained with Keck/MOSFIRE as part of the Keck Baryonic Structure Survey (KBSS). The KBSS-MOSFIRE sample is representative of star-forming galaxies at these redshifts, with stellar masses M$_{\ast}=10^9-10^{11.5}$M$_{\odot}$ and star formation rates SFR$=3-1000$~M$_{\odot}$~yr$^{-1}$. We focus on robust measurements of many strong diagnostic emission lines for individual galaxies: [\ion{O}{2}]$\lambda\lambda 3727$,3729, [\ion{Ne}{3}]$\lambda3869$, H$\beta$, [\ion{O}{3}]$\lambda\lambda$4960,5008, [\ion{N}{2}]$\lambda\lambda 6549,6585$, H$\alpha$, and [\ion{S}{2}]$\lambda\lambda 6718$,6732. Comparisons with observations of typical local galaxies from the Sloan Digital Sky Survey (SDSS) and between subsamples of KBSS-MOSFIRE show that high-redshift galaxies exhibit a number of significant differences in addition to the well-known offset in log([\ion{O}{3}]$\lambda5008$/H$\beta$) and log([\ion{N}{2}]$\lambda6585$/H$\alpha$). We argue that the primary difference between \ion{H}{2} regions in $z\sim2.3$ galaxies and those at $z\sim0$ is an enhancement in the degree of nebular excitation, as measured by [\ion{O}{3}]/H$\beta$ and ${\rm R23}\equiv\log$[([\ion{O}{3}]$\lambda\lambda4960,5008$+[\ion{O}{2}]$\lambda\lambda3727,3729$)/H$\beta$].  At the same time, KBSS-MOSFIRE galaxies are $\sim10$ times more massive than $z\sim0$ galaxies with similar ionizing spectra and have higher N/O (likely accompanied by higher O/H) at fixed excitation. These results indicate the presence of harder ionizing radiation fields at fixed N/O and O/H relative to typical $z\sim0$ galaxies, consistent with Fe-poor stellar population models that include massive binaries, and highlight a population of massive, high-specific star formation rate galaxies at high-redshift with systematically different star formation histories than galaxies of similar stellar mass today.
\end{abstract}
\maketitle

\section{Introduction}

The strong emission lines of elements such as hydrogen, oxygen, and nitrogen observed in the rest-optical ($3600-7000$\AA) spectra of galaxies come primarily from the \ion{H}{2} regions surrounding massive O and B stars. Thus, attempts to characterize this emission ultimately reveal information regarding the properties of both the massive stars \textit{and} the ionized gas immediately surrounding them. Furthermore, we know from studies of individual \ion{H}{2} regions in the Milky Way and nearby galaxies that changes in the nebular spectrum are tightly correlated with one another, manifesting the underlying correlations between the physical properties of the ionized regions---including electron density, electron temperature, gas chemistry, and ionization parameter---that determine the strength of emission lines.

Studies conducted using integrated-light spectra of large samples of nearby galaxies, such as the Sloan Digital Sky Survey \citep[SDSS,][]{york2000}, reveal similar behavior in galaxies spanning a wide range in stellar mass and star formation rate, with star-forming galaxies occupying a tight locus in many parameter spaces. For example, low-$z$ star-forming galaxies form a relatively narrow sequence in the $\log$([\ion{O}{3}]$\lambda5008$/H$\beta$) vs. $\log$([\ion{N}{2}]$\lambda6585$/H$\alpha$) plane (hereafter the N2-BPT diagram), extending from high [\ion{O}{3}]/H$\beta$ and low [\ion{N}{2}]/H$\alpha$ to low [\ion{O}{3}]/H$\beta$ and high [\ion{N}{2}]/H$\alpha$ as gas-phase metallicity increases and the level of ionization decreases. The position of objects in the N2-BPT plane can therefore be used as a tool for distinguishing galaxies powered by young stars from galaxies with significantly harder ionizing radiation fields (such as those produced by active galactic nuclei, or AGN). The very hard ionizing spectrum of AGN produces enhanced collisional line emission in an extended partially-ionized region, resulting in a diffuse ``plume" that extends to both high [\ion{O}{3}]/H$\beta$ and high [\ion{N}{2}]/H$\alpha$ in the N2-BPT plane.

The use of line ratios to distinguish between sources of ionizing radiation was first proposed by \cite{baldwin1981}, but \cite{veilleux1987} introduced the version of the N2-BPT diagram commonly used today (along with two other diagnostic diagrams using [\ion{S}{2}]$\lambda\lambda$6718,6733/H$\alpha$ and [\ion{O}{1}]$\lambda6300$/H$\alpha$ instead of [\ion{N}{2}]$\lambda6585$/H$\alpha$). \citet{kewley2001} and \citet{kauffmann2003} separately provided classification curves that are now frequently used to separate $z\sim0$ star-forming galaxies from AGN in the N2-BPT plane; the former relied on predictions from photoionization models, but the latter specifically considered the importance of correlations between the physical conditions in star-forming galaxies, particularly between metallicity and ionization parameter.

Advances in our knowledge of the conditions (and the important correlations between them) in galaxies at earlier epochs---specifically $z\simeq2-3$, when both cosmic star formation \citep[e.g.,][]{madau1996,hopkins2006,reddy2008,madau2014} and supermassive black hole accretion reached their peak values \citep{richards2006}---have been slower to arrive, due in large part to limitations in the sensitivity and multiplexing capabilities of near-infrared spectrographs on $8-10$~m class telescopes. Still, early observations of small samples of high-redshift galaxies \citep[e.g.,][]{shapley2005,erb2006metal,liu2008} suggested that the situation at high-$z$ might be radically different, with galaxies exhibiting lower gas-phase oxygen abundance at fixed stellar mass and nebular line ratios inconsistent with observations of the majority of local galaxies. Since the commissioning of efficient multi-object near-infrared spectrographs like the K-band Multi-Object Spectrograph on the VLT \citep[KMOS,][]{sharples2013} and the Multi-Object Spectrometer for InfraRed Exploration on the Keck I telescope \citep[MOSFIRE, ][]{mclean2012}, the number of $z\simeq2-3$ galaxies with high-quality rest-optical spectra has increased dramatically \citep{steidel2014,kriek2015,wisnioski2015}. The results from galaxy surveys using these instruments have confirmed that, while high-redshift star-forming galaxies also occupy a relatively tight locus in the N2-BPT diagnostic diagram, they show a clear offset toward higher [\ion{O}{3}]$\lambda5008$/H$\beta$ at a given [\ion{N}{2}]$\lambda6585$/H$\alpha$ compared to their low-redshift counterparts \citep{masters2014,steidel2014,shapley2015}.

Systematic differences between typical star-forming galaxies in the local universe and those present during the peak epoch of galaxy growth are not surprising---on average $z\sim2$ galaxies have higher star formation rates at fixed mass by at least a factor of 10 and cold gas masses higher by at least a factor of $\sim5$, all in smaller volumes compared to local galaxies \citep[e.g.,][]{erb2006mass,law2012,tacconi2013}. However, understanding the physical cause of the differences in their nebular spectra has proven both challenging and controversial \citep[c.f.][]{masters2014,steidel2014,sanders2016,masters2016}, with much of the difficulty stemming from the diversity (and degeneracy) of possible explanations. The differences between nebular diagnostics observed at $z\sim2$ relative to $z\sim0$ can be attributed to differences in the underlying stellar populations (e.g., EUV ionizing spectrum, main sequence lifetime, metallicity, binarity, rotation) and/or differences in the conditions of the interstellar medium (ISM; e.g., density, temperature, metallicity).

From our analysis of the initial KBSS-MOSFIRE sample of $\sim200$ $z \simeq 2.3$ galaxies in \citet[][hereafter S14]{steidel2014}, we concluded that the high-redshift locus of galaxies in the N2-BPT plane is most easily explained by a harder stellar ionizing radiation field than applies to galaxies occupying the low-redshift sequence, accompanied by slightly elevated ranges in ionization parameter $U$($\equiv n_\gamma/n_H$) and electron density ($n_e$). In \citetalias{steidel2014}, we noted that the systematic offset of high-$z$ galaxies relative to star-forming galaxies in the low-redshift universe in the N2-BPT plane cautions against using the common ``strong-line" metallicity relations (generally calibrated using galaxy and \ion{H}{2} region samples at $z\sim0$) for high-redshift galaxies, since the calibrations are designed to reproduce the local N2-BPT sequence. Additionally, an important consequence of the high excitation of the $z\sim2.3$ KBSS-MOSFIRE sample is that the strong-line ratios become relatively insensitive to the ionized gas-phase oxygen abundance (often what one intends to measure) and more dependent on the spectral shape of the integrated ionizing radiation field produced by massive stars. 

In \citetalias{steidel2014}, we also investigated the effect of differences in nitrogen-to-oxygen abundance ratio on the rest-optical nebular spectra of high-$z$ galaxies, but ultimately found that systematically different values of N/O between $z\sim2.3$ and $z\sim0$ galaxies were not required to reproduce the observed N2-BPT offset using the photoionization models presented in that paper. Other recent work based on independent (but smaller) galaxy samples at similar redshifts \citep{masters2014,shapley2015,cowie2016,sanders2016,masters2016} has argued that high-redshift objects have higher N/O at a given O/H---i.e., that the N2-BPT offset is primarily a shift toward higher [\ion{N}{2}]/H$\alpha$ and is confined to lower-mass galaxies in those samples. The primary basis for this assertion was that an offset of the BPT locus is observed in [\ion{N}{2}]/H$\alpha$ but not in [\ion{S}{2}]/H$\alpha$. However, the behavior of N/O versus O/H for many individual galaxies at high-redshift has yet to be established.

Although much can be learned from rest-optical nebular spectra regarding the properties of the ISM and massive stars, it is not clear that the physical factors most important for determining the location of typical local star-forming galaxies in the nebular diagnostic diagrams are the same as those at $z\sim2$. The goal of this paper is to leverage measurements from the rest-optical spectra of an expanded sample of $\langle z \rangle = 2.3$ KBSS-MOSFIRE galaxies to characterize the properties of high-$z$ galaxies in terms of their stellar masses, star formation rates, ionization and excitation conditions, electron densities, and the variation in N/O as a function of O/H. We will show that the seemingly discrepant behavior in the N2-BPT and S2-BPT diagrams (and other diagnostic line ratios) is expected in the context of physically-motivated photoionization models anchored by observations of the rest-UV spectra of the same galaxies \citep[][hereafter S16]{steidel2016}.

The remainder of this paper is structured as follows: \S\ref{kbss} describes the Keck Baryonic Structure Survey, the near-infrared spectroscopic observations, and the detailed spectral fitting of the data. \S\ref{ratios} reports the measurements of strong-line ratios from the nebular spectra of KBSS-MOSFIRE galaxies; \S\ref{offset_text} examines the properties of $z\sim2.3$ galaxies as a function of their offset from the $z\sim0$ relation in the N2-BPT diagram, \S\ref{nitrogen_ratios} presents the N/O measurements and their correlation with galaxy properties, and \S\ref{models} describes the inferences that can be made by comparing a subset of the photoionization models from \citetalias{steidel2016} with the observed line ratios and inferred abundance ratios. Finally, \S\ref{whyoffset} revisits the nature of the BPT ``offset" in the context of the KBSS-MOSFIRE results. The conclusions are summarized in \S\ref{summary}.

Throughout the paper, we assume a $\Lambda$CDM cosmology when necessary, with $H_0=70$\,km s$^{-1}$ Mpc$^{-1}$, $\Omega_{\Lambda}=0.7$, and $\Omega_{\rm m} = 0.3$. Stellar masses and star formation rates are reported assuming a \citet{chabrier2003} stellar initial mass function (IMF). When a solar metallicity scale is needed for comparison, we adopt 12+log(O/H)$_{\odot}=8.69$, log(N/O)$_{\odot} = -0.86$, and $Z_{\odot} = 0.0142$ \citep{asplund2009}. Specific spectral features are referred to using their vacuum wavelengths.

\section{The Keck Baryonic Structure Survey}
\label{kbss}

The Keck Baryonic Structure Survey \citep[KBSS,][]{rudie2012,steidel2014} is a large, targeted spectroscopic survey designed to jointly probe galaxies and their gaseous environments at the peak of galaxy assembly, $z \simeq 2-3$. The survey comprises 15 independent fields, with a total survey area of 0.24~deg$^2$. The galaxies themselves are selected from deep optical and near-infrared (NIR) imaging and subsequently followed up with spectroscopic observations in the rest-UV with the Low Resolution Imaging Spectrometer \citep[LRIS,][]{oke1995,steidel2004} and, since 2012 April, in the rest-optical with the Multi-Object Spectrometer For InfraRed Exploration \citep[MOSFIRE,][]{mclean2012,steidel2014}.

In addition to the physical motivation for studying galaxies at $z\simeq2-3$, this redshift range also offers important practical advantages: many of the strong rest-optical ($3600-7000$\AA) nebular emission lines originating in the galaxies' \ion{H}{2} regions are well-positioned with respect to the $J$, $H$, and $K$ band atmospheric windows. Additionally, the rest-frame far-ultraviolet ($1000-2000$\AA) spectra of the same galaxies are accessible to ground-based telescopes, which is critically important for complementary studies of their massive stellar populations, which is one of the main goals of our recent work in \citetalias{steidel2016} and is discussed briefly in \S\ref{models}.

\subsection{Photometry and Sample Selection}
\label{sample_text}

The photometric data available in the KBSS fields have been described in detail elsewhere \citep[e.g., by][]{steidel2004,reddy2012,steidel2014}. To summarize, all of the survey regions have optical imaging in the $U_n$, $G$, and $\mathcal{R}$ bands, as well as broad-band NIR imaging in $J$ and $K_s$. For 14 fields, imaging was also obtained using \textit{Spitzer}/IRAC (typically including coverage in two channels per field); 10 fields have at least one deep pointing obtained using \textit{Hubble}/WFC3-IR F160W; and intermediate-band NIR imaging in $J1$, $J2$, $J3$, $H1$, and $H2$ was collected using Magellan-FourStar \citep{persson2013} for 8 fields accessible from the southern hemisphere.

\begin{figure}
\centering
\includegraphics{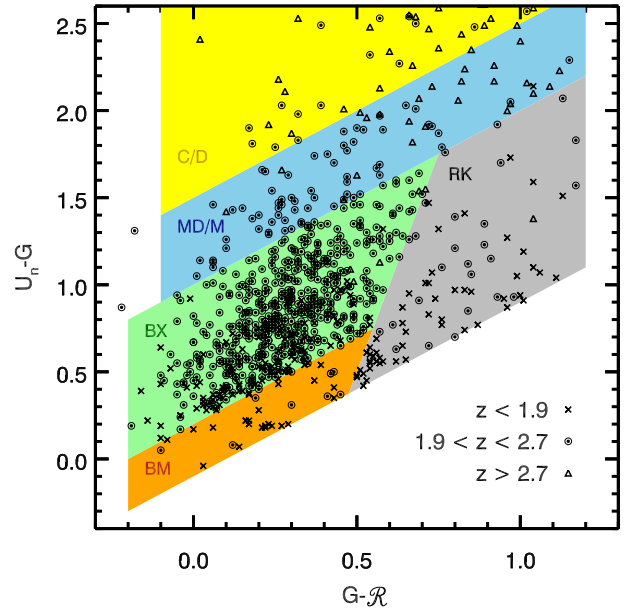}
\caption{The optical color selection windows from \citet{steidel2004} used to identify candidate star-forming galaxies with $z\simeq2.0-2.7$ (``BX" in green) and $z \simeq2.7-3.4$ (``MD/M" and ``C/D" in blue and yellow, respectively); the grey region indicates the new ``RK" color selection at redder $G-\mathcal{R}$ colors. Black symbols represent spectroscopically-confirmed galaxies in KBSS-MOSFIRE sample, with the symbol shape indicating the nebular redshift.}
\label{rk_select}
\end{figure}

The majority of KBSS galaxies are selected by their rest-UV colors, based on a $U_nG\mathcal{R}$ color selection designed to identify Lyman Break Galaxy analogues at $z \simeq 2-2.7$ \citep{adelberger2004,steidel2004}. However, this rest-UV color selection may be biased against massive galaxies and galaxies whose UV continua are heavily reddened due to extinction by dust $(E(B-V)_{\rm cont}>0.3$). Fortunately, these biases can be mitigated by using information about the combined rest-UV and rest-optical shape of the galaxies' spectra. The 4000~\AA\ and Balmer breaks, which probe both age and stellar mass, lie within the $J$ band at $z\sim2.3$ but are still well-traced by $\mathcal{R}-K_s$ color. Therefore, combining a $U_nG\mathcal{R}$ color selection with a $(\mathcal{R}-K_s)_{\rm AB} > 2$ color cut preferentially selects galaxies at the redshifts of interest that occupy the high end of the stellar mass distribution. We also extend the optical color selection to include sources with redder $U_nG\mathcal{R}$ colors and simultaneously impose a $\mathcal{R}-K_s$ cut, in order to identify galaxies which may be at similar redshifts but are scattered out of the \citet{steidel2004} selection windows due to the effects of substantial reddening by dust.

We defer a more detailed description of this sample of galaxies---referred to as ``RK" objects---and their effect on the stellar mass-completeness of the KBSS-MOSFIRE sample to a forthcoming paper (A. Strom et al., in prep.), but Figure~\ref{rk_select} offers a visual illustration of the color selection windows described here. With the inclusion of RK objects, the KBSS targets sources throughout the full rest-UV color space occupied by high-$z$ galaxies. Although the sample only includes galaxies with $\mathcal{R}<25.5$ ($\mathcal{R}$ samples the rest-frame at $\sim2100$\AA), the color selection allows for galaxies with significantly fainter rest-frame FUV emission to be selected, down to an AB magnitude of $26.7$ in $G$ band ($\lambda_e\sim1500$\AA\, at $z\sim2.3$).

\begin{figure}
\centering
\includegraphics{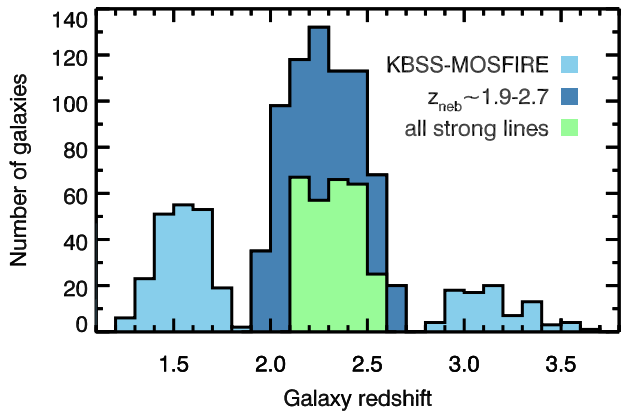}
\caption{The redshift distribution of spectroscopically-confirmed KBSS-MOSFIRE galaxies, as of 2016 May. The blue histograms represent galaxies with confirmed nebular redshifts from MOSFIRE observations (1060 total). Of these, 733 are known to be at $1.9\leq z\leq2.7$ and have MOSFIRE observations covering at least one atmospheric band (dark blue). The subsample with good signal-to-noise detections of strong rest-optical emission lines in $J$, $H$, and $K$ band is shown in green (279 galaxies).}
\label{zhist}
\end{figure}

\subsection{NIR Spectroscopic Observations}
\label{observations}

\citetalias{steidel2014} introduced the NIR spectroscopic observations of KBSS galaxies with MOSFIRE, and readers are referred to Section 2 of that paper for details related to data acquisition and reduction. One particularly salient aspect of the KBSS-MOSFIRE observing strategy is that, due to the dense spatial sampling of the KBSS fields, multiple slitmask configurations in each band are usually required to obtain observations of all high-priority targets in a given field. Because galaxies remain on masks until the strongest lines (H$\alpha$, [\ion{N}{2}]$\lambda$6585, [\ion{O}{3}]$\lambda$5008 H$\beta$, and [\ion{O}{2}]$\lambda\lambda$3727,3729) are measured with ${\rm S/N}>5$, an individual galaxy is often observed at least 2 times in a single band. \S\ref{slitcors} describes the method used to correct observations of KBSS-MOSFIRE galaxies for slit losses, which takes advantage of the multiple observations of individual galaxies that result from this survey strategy.

As of 2016 May, 1060 KBSS galaxies have been confirmed spectroscopically with observations from MOSFIRE (Figure~\ref{zhist}); of these, 733 are in the primary targeted redshift range $1.9\lesssim z\lesssim 2.7$, with the remainder split roughly 2:1 between $1.3\lesssim z\lesssim 1.7$ and $z\gtrsim2.9$. The majority of the $z\sim2.3$ galaxies have been observed in at least two NIR bands, and 279 have high-quality observations in $J$, $H$, and $K$ band, providing good S/N measurements of most of the strong rest-optical emission lines: [\ion{O}{2}]$\lambda\lambda 3727$,3729, [\ion{Ne}{3}]$\lambda3869$ in $J$ band; H$\beta$, [\ion{O}{3}]$\lambda\lambda$4960,5008 in $H$ band; [\ion{N}{2}]$\lambda\lambda 6549,6585$, H$\alpha$, [\ion{S}{2}]$\lambda\lambda 6718$,6732 in $K$ band.

\subsection{Interactive 1D Spectral Extraction \\and Emission-line Fitting with MOSPEC}

The two-dimensional (2D) spectrograms produced by the MOSFIRE data reduction pipeline\footnote{\url{http://www2.keck.hawaii.edu/inst/mosfire/drp.html}} are flux-calibrated and corrected for telluric absorption using wide- and narrow-slit observations of A0V stars and then shifted to account for the heliocentric velocity at the start of each exposure sequence. Separate MOSFIRE observations of the same object are combined using inverse-variance weighting to produce the final 2D spectrograms from which 1D spectra are extracted.

One-dimensional (1D) spectra are extracted from the 2D spectrograms using MOSPEC, an interactive analysis tool developed in IDL specifically for MOSFIRE spectroscopy of faint emission-line galaxies. By default, MOSPEC uses a boxcar extraction aperture (determined by the user, with a median value of 10~pixels or $1\farcs80$, corresponding to 14.8 kpc at $z=2.3$). Other extraction algorithms, including optimal extraction, were tested and found not to significantly impact the measured line fluxes or their significance.

As part of the extraction process, the 1D spectrum is simultaneously fit for the redshift, line width, and fluxes of a user-specified list of emission lines. The continuum level is estimated using a reddened high-resolution stellar population synthesis model \citep[from ][]{bruzual2003} that best matches the full spectral energy distribution (SED) of the galaxy and is scaled to match the median level of the observed spectrum excluding regions $<5$\AA\, from strong emission lines. This method self-consistently accounts for the effects of age and dust extinction on the underlying stellar continuum. If this method is not possible (most frequently due to very faint continua), MOSPEC uses a linear fit to the continuum, again excluding regions around strong emission lines. These two estimates agree well for most galaxies, although the model-continuum method has the advantage of including the stellar Balmer absorption features. 

The correction for absorption in H$\alpha$ and H$\beta$ for individual objects is determined by fitting the galaxy spectrum with a modified SED model continuum that linearly interpolates between the continuum on either side of regions around Balmer features and comparing the result with measurements from the full fit that includes stellar absorption features. For the $\sim13\%$ of the sample where a stellar continuum model could not be used to automatically correct the measured line fluxes, we assume a multiplicative correction of $C_{{\rm H}\beta} = 1.06$, the median of the measured corrections. No correction is applied to H$\alpha$ (although, formally, the median measured correction is $C_{{\rm H}\alpha} = 1.01$).

The emission lines are fit using Gaussian profiles with a single redshift ($z$) and observed velocity width ($\sigma$, in km\,s$^{-1}$) in a given band (i.e., all lines in the $K$ band spectrum are fit with a single $z$ and $\sigma$, but these may differ from the parameters measured in $J$ or $H$ band). For objects with observations covering $H$ and $K$ band, $\langle |\Delta z_{H-K}|\rangle=2\times10^{-4}$ ($\Delta v \approx18$~km/s) and $\langle \frac{|\Delta \sigma_{H-K}|}{\sigma_H} \rangle=0.22$. For 19\% of galaxies with observations covering $J$ band, the redshift was fixed to match the redshift observed in $H$ or $K$ band to ensure an accurate measurement of the [\ion{O}{2}] doublet ratio in cases where the lines are only partially resolved; for all other objects, $\langle |\Delta z_{K-J}| \rangle=4\times10^{-4}$ and $\langle | \Delta z_{H-J} |\rangle=4\times10^{-4}$. Except for emission lines with particularly high signal-to-noise (S/N $\sim 50$), the Gaussian approximation agrees with the directly integrated flux (summed over a range $\pm3\sigma$ as defined by the fitted velocity width) at the few percent level. The ratios of the nebular [\ion{O}{3}]$\lambda\lambda$4959,5008 and [\ion{N}{2}]$\lambda\lambda$6549,6585 doublets are fixed at 1:3 and the ratios of the density-sensitive [\ion{O}{2}]$\lambda\lambda$3727,3729 and [\ion{S}{2}]$\lambda\lambda$6718,6732 doublets are restricted to be within 20\% of the range of physically allowed values, which is $0.43-1.5$ for [\ion{S}{2}]$\lambda$6718/[\ion{S}{2}]$\lambda$6732 \citep{tayal2010,mendoza2014} and $0.35-1.5$ for [\ion{O}{2}]$\lambda$3729/[\ion{O}{2}]$\lambda$3727 \citep{kisielius2009}. The 1$\sigma$ errors on all measurements account for uncertainties in the fit parameters as well as covariance between parameters.
\\

\subsection{Slit-loss Corrections}
\label{slitcors}

\begin{figure}
\centering
\includegraphics{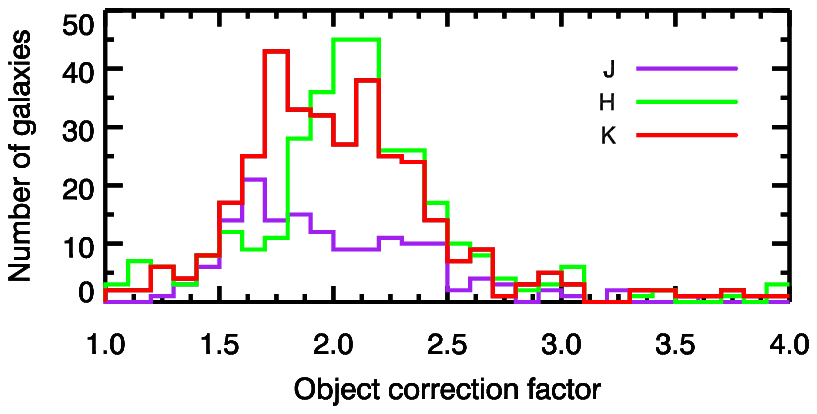}
\includegraphics{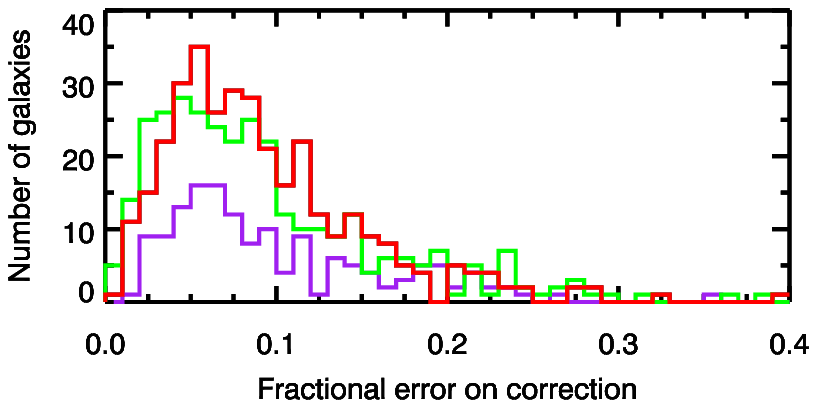}
\caption{The distributions of individual object slit correction (top panel) and fractional error on the correction (bottom panel) broken down by NIR atmospheric band for 826 measurements of $z\sim2.3$ KBSS-MOSFIRE galaxies. Overall, the median slit correction is 2.03, with a median uncertainty of 8\%.}
\label{slitcor_hist}
\end{figure}

Because some rest-frame optical diagnostics rely on ratios of emission lines widely separated in wavelength, it is important to ensure that the relative flux-calibration between the NIR atmospheric bands is correct for individual objects. For example, without such cross-calibration, extinction corrections using the Balmer decrement are impossible, as H$\beta$ falls in $H$ band and H$\alpha$ falls in $K$ band for galaxies at $z\simeq2-2.7$. 

The overall atmospheric transparency and image quality for each MOSFIRE mask are monitored using concurrent observations of a bright star placed in a dedicated slit (with all slit widths $=0\farcs7$), and an initial guess at the slit-loss correction factor for observations from that mask is determined by comparing the measured flux of the star with its broadband photometry. At the typical seeing (between $0\farcs5-0\farcs7$), most science targets are unresolved or only marginally resolved. However, there are still several factors which can affect throughput of sources differently depending on their physical size and location on the mask, including minor glitches in tracking the rotation of high-elevation pointings and ``slit drift" due to the combined effects of differential atmospheric refraction and differential flexure of the guider field relative to the science field. Consequently, it is important to optimize the mask correction factors for all objects on a given mask in a given atmospheric band, rather than rely exclusively on the correction determined using observations of the bright comparison star.

Optimal mask correction factors are calculated for all individual masks (213 total) using a Markov Chain Monte Carlo (MCMC) routine. The MCMC routine searches parameter space for the combination of individual mask correction factors that minimizes the scatter between independent measurements of strong lines (typically H$\alpha$, [\ion{O}{3}]$\lambda5008$, or the sum of the [\ion{O}{2}]$\lambda\lambda3727,3729$ doublet) in galaxies' spectra, as many galaxies are observed more than once in a given band. The routine uses the comparison star measurements to construct priors on the mask corrections, which are allowed to sample a range from 1 to 3.5, chosen empirically to coincide with the range of corrections observed for the comparison stars. In cases where an individual mask includes observations of more than one bright star, the preliminary correction factors are compared to ensure good agreement. Outlier rejection is performed on a mask-by-mask basis prior to initializing the chain by censoring individual galaxies that require an anomalously high or low correction factor relative to other galaxies on the same mask. The principal advantage of using this method to solve for the mask correction factors in parallel is that it provides a quantitative measure of how well or how poorly a single mask correction factor performs for all objects observed on a mask and provides the means to flag masks with unreliable correction factors based on the shape of the posterior distribution.

On average, the posteriors for the mask corrections are positively skewed, and $\sim60\%$ of masks exhibit a significant degree of skewness (positive or negative). As a result, the characteristic value for a given mask correction is taken to be the center of the 68\% highest posterior density interval rather than the mean value, with the width of the interval representing the uncertainty in the mask correction determination.

For galaxies observed on a single mask, the optimal mask correction is used to correct the measured line fluxes. When a galaxy has been observed more than once, line fluxes from the 2D weighted average spectrum are corrected to match the weighted average of the corrected line flux measurements from the constituent 2D spectral observations; the scatter around the average is the reported uncertainty on the individual object slit correction. The statistics of the 826 individual measurements of galaxies and their uncertainties are presented in Figure~\ref{slitcor_hist}; the median slit-loss correction factor is 2.03, with a median uncertainty of 8\%.

\subsection{Stellar Masses and Star-formation Rates}
\label{sfrs+masses}

\begin{figure}
\centering
\includegraphics{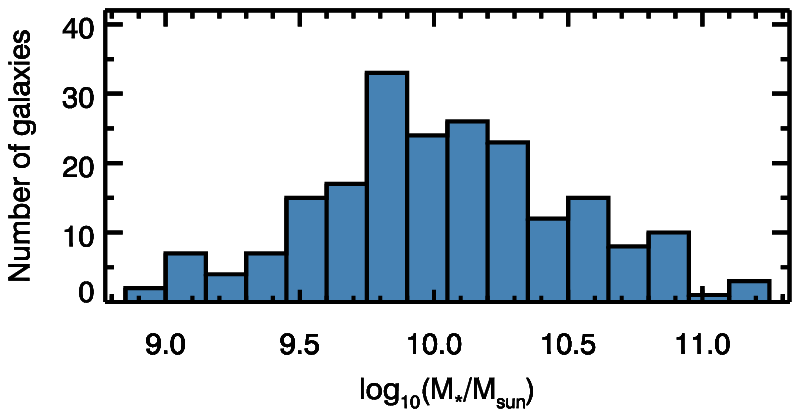}
\includegraphics{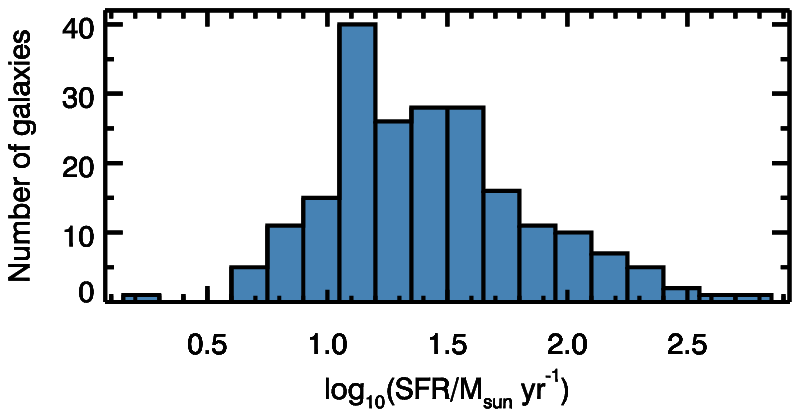}
\includegraphics{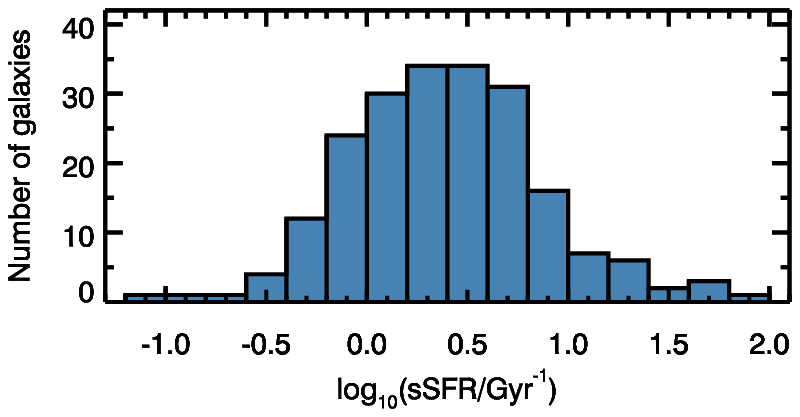}
\caption{Stellar mass, star formation rate, and specific star formation rate distributions for 207 $z\sim2.3$ KBSS-MOSFIRE galaxies with S/N~$>5$ Balmer decrement measurements. Stellar masses are determined from SED fits, and SFRs are calculated from dust-corrected H$\alpha$ emission line measurements. The median stellar mass is $1.0\times10^{10}$~M$_{\odot}$, and the median SFR is $24$~M$_{\odot}$~yr$^{-1}$, with a median sSFR of 2.4~Gyr$^{-1}$.}
\label{mstar+sfr}
\end{figure}
\label{sfrs_and_masses}

\begin{deluxetable*}{rccccll}
\tablecaption{KBSS-MOSFIRE samples used for line-ratio analysis}
\tablehead{
\colhead{subsample} &\colhead{$N$} & \colhead{$\langle z \rangle$} &\colhead{$M_{\ast}$}\footnotemark[1] &\colhead{SFR}\footnotemark[1] &\colhead{} &\colhead{} \\
\colhead{ } &\colhead{ } & \colhead{ } &\colhead{(M$_{\odot}$)} &\colhead{(M$_{\odot}$~yr$^{-1}$)} &\colhead{Emission Line Criteria} &\colhead{Notes}}
\startdata
N2-BPT & 377 & 2.31 & $1.09\times10^{10}$ & 23.6 & H$\alpha \geq 5\sigma$, H$\beta$, [\ion{O}{3}]$\lambda5008 \geq 3\sigma$, [\ion{N}{2}]$\lambda6585$ observed & \footnotemark[2]\footnotemark[3] \\
S2-BPT & 370 & 2.30 &  $1.13\times10^{10}$ & 23.5 & H$\alpha \geq 5\sigma$, H$\beta$, [\ion{O}{3}]$\lambda5008 \geq 3\sigma$, [\ion{S}{2}]$\lambda\lambda6718,6733$ observed & \footnotemark[3] \\
O32-R23 & 171 & 2.35 &  $9.49\times10^{9}$ & 22.9 & H$\alpha \geq 5\sigma$, H$\beta$, [\ion{O}{3}]$\lambda5008$, [\ion{O}{2}]$\lambda\lambda3727,3729 \geq 3\sigma$ & \footnotemark[3]\footnotemark[4] \\
N/O & 151 & 2.34 &  $9.94\times10^{9}$ & 23.5 & H$\alpha \geq 5\sigma$, H$\beta$, [\ion{O}{2}]$\lambda\lambda3727,3729 \geq 3\sigma$, [\ion{N}{2}]$\lambda6585$ observed & \footnotemark[3]\footnotemark[4] \\
Ne3O2 & 69 & 2.33 &  $7.14\times10^{9}$ & 22.6 & H$\alpha \geq 5\sigma$, H$\beta$, [\ion{O}{3}]$\lambda5008$, [\ion{O}{2}]$\lambda\lambda3727,3729$, [\ion{Ne}{3}]$\lambda3869 \geq 3\sigma$,  & \footnotemark[3]\footnotemark[4]
\enddata
\footnotetext[1]{Median values}
\footnotetext[2]{\citetalias{steidel2014} used H$\alpha \geq 5\sigma$, H$\beta\geq3\sigma$, [\ion{O}{3}]$\lambda5008 \geq 5\sigma$, [\ion{N}{2}]$\lambda6585$ observed, taken as an upper limit when the ${\rm S/N}<2$}
\footnotetext[3]{We adopt $2\sigma$ upper limits for emission lines without a listed S/N threshold and $<2\sigma$ significance}
\footnotetext[4]{Emission line errors incorporate uncertainties in the slit-loss corrections, and the Balmer decrement must have ${\rm S/N}>5$}
\label{samples}
\end{deluxetable*}

Stellar mass (M$_{\ast}$) estimates for galaxies in the KBSS-MOSFIRE sample are inferred from reddened stellar population synthesis models fit to broadband photometry, as described by \citet{reddy2012} and \citetalias{steidel2014}. The SED fitting employs solar metallicity models from \citet{bruzual2003}, a \citet{chabrier2003} IMF, the \citet{calzetti2000} attenuation curve, and constant star formation histories for all galaxies, with a minimum allowed age of 50~Myr. This age is a dynamically-reasonable minimum imposed to prevent best-fit solutions with unrealistically young ages, as discussed by \citet{reddy2012}. Typical uncertainties in log($M_{\ast}$/$M_{\odot}$) are estimated to be $\pm0.1-0.2$ dex, with a median uncertainty of 0.16 dex \citep{shapley2005,erb2006mass}.

Because star formation rate (SFR) estimates from SED-fitting are naturally correlated with other fit parameters such as age and M$_{\ast}$, the SFRs used for analysis of the KBSS-MOSFIRE sample in this paper are obtained from dust-corrected measurements of the H$\alpha$ recombination line using the relation from \citet{kennicutt1998} and assuming a \citet{chabrier2003} IMF. \citet{shivaei2016} reported that for $z\sim2$ galaxies from the MOSFIRE Deep Evolution Field survey \citep[MOSDEF,][]{kriek2015}, H$\alpha$-based estimates of SFR are consistent with those estimated from the full panchromatic SED (including UV and far-IR photometry). Dust extinction toward star-forming regions is estimated by comparing the Balmer decrement (H$\alpha/H\beta$) with the expected Case B value of 2.86 \citep{osterbrock1989}. Galaxies with Balmer decrements $<2.86$ are assumed to have zero extinction, with $\sim46$\% of such galaxies exhibiting Balmer decrements consistent with the nominal Case B value within 3$\sigma$.  Adopting the Galactic extinction curve from \citet{cardelli1989} for galaxies with $H\alpha$/H$\beta\geq2.86$, the interquartile range for extinction along the line-of-sight to \ion{H}{2} regions in the full sample of $z\sim2.3$ KBSS-MOSFIRE galaxies is $E(B-V)_{\rm neb} =  0.06-0.47$, with a median value of 0.25 and a corresponding median $A_{\rm H\alpha} = 0.63$~mag. 

The distributions of M$_{\ast}$, H$\alpha$-based SFR, and specific star formation rate (sSFR$\equiv$SFR/M$_{\ast}$) for 207 $z\sim2.3$ KBSS-MOSFIRE galaxies with S/N~$>5$ Balmer decrement measurements are presented in Figure~\ref{mstar+sfr}. Table~\ref{samples} provides an overview of the sample statistics for the subsamples of KBSS-MOSFIRE galaxies discussed in the following sections and notes the corresponding emission line selection criteria. In cases where extinction corrections are necessary, we require the measurement of H$\alpha$/H$\beta$ to have ${\rm S/N}>5$, including the uncertainty from the relative slit corrections. This S/N cut is well-motivated by the fact that extinction corrections for individual emission lines scale non-linearly with the measured value of the Balmer decrement, going approximately as (H$\alpha$/H$\beta$)$^{k_{\lambda}}$, where $k_{\lambda}$ is the value of the reddening curve at a specific wavelength ($k_{{\rm H}\alpha} = 2.52$ for the \citealt{cardelli1989} extinction curve). Thus, even relatively small uncertainties in the Balmer decrement translate to much larger uncertainties in the corrected line ratio.

\section{Nebular Emission Line Ratios}
\label{ratios}

\begin{figure*}
\centering
\includegraphics{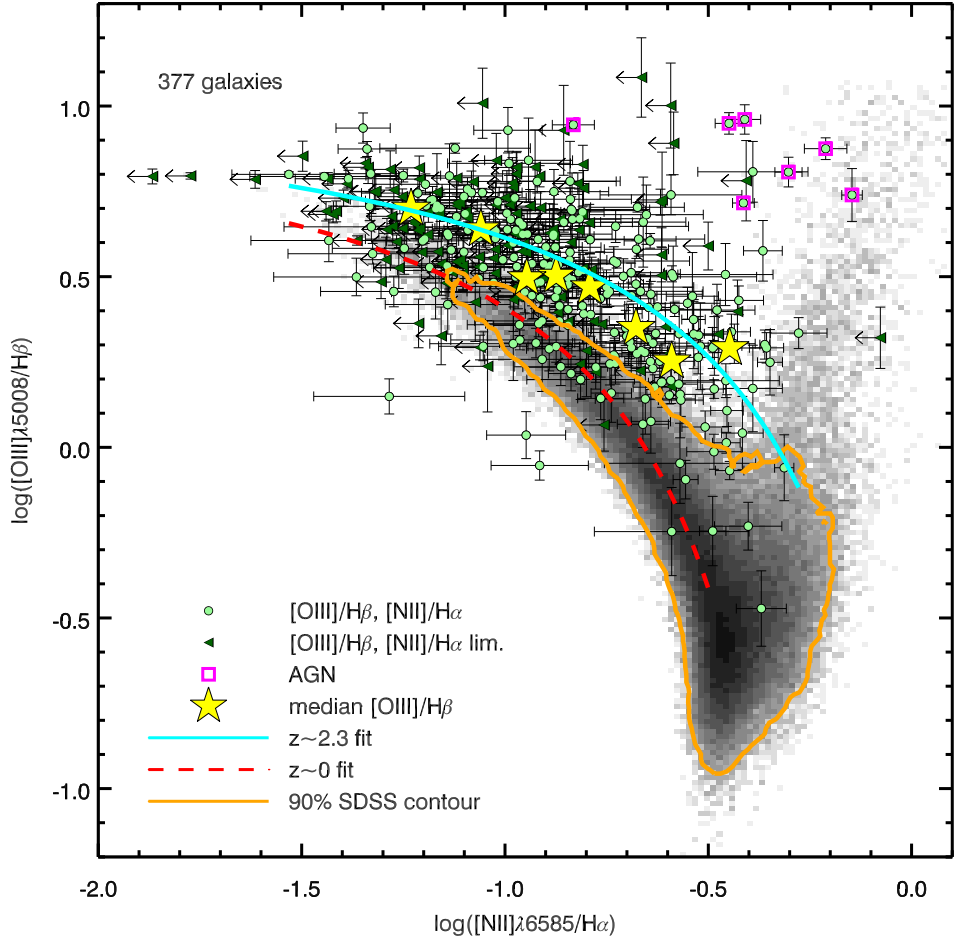}
\caption{KBSS-MOSFIRE galaxies in the N2-BPT plane, compared with local galaxies from SDSS (in greyscale, with the orange contour enclosing 90\% of the total sample). KBSS-MOSFIRE galaxies with $>2\sigma$ detections of [\ion{N}{2}]$\lambda6585$ are plotted in light green, with $2\sigma$ upper limits shown instead as dark green triangles. Magenta squares denote objects identified as AGN/QSOs. The ridge-line of the high-$z$ locus occurs far outside the 90\% SDSS contour, demonstrated not only by the location of the median log([\ion{O}{3}]$\lambda5008$/H$\beta$) values in equal-number bins of log([\ion{N}{2}]$\lambda6585$/H$\alpha$) (yellow stars, see also Table~\ref{medianratios}) but also by the formal fit to the distribution of KBSS-MOSFIRE galaxies (cyan curve). The $z\sim0$ locus is represented by the red dashed curve.}
\label{n2_bpt}
\end{figure*}

\subsection{The N2- and S2-BPT Diagrams}
\label{bpt_text}

\begin{figure*}
\centering
\includegraphics{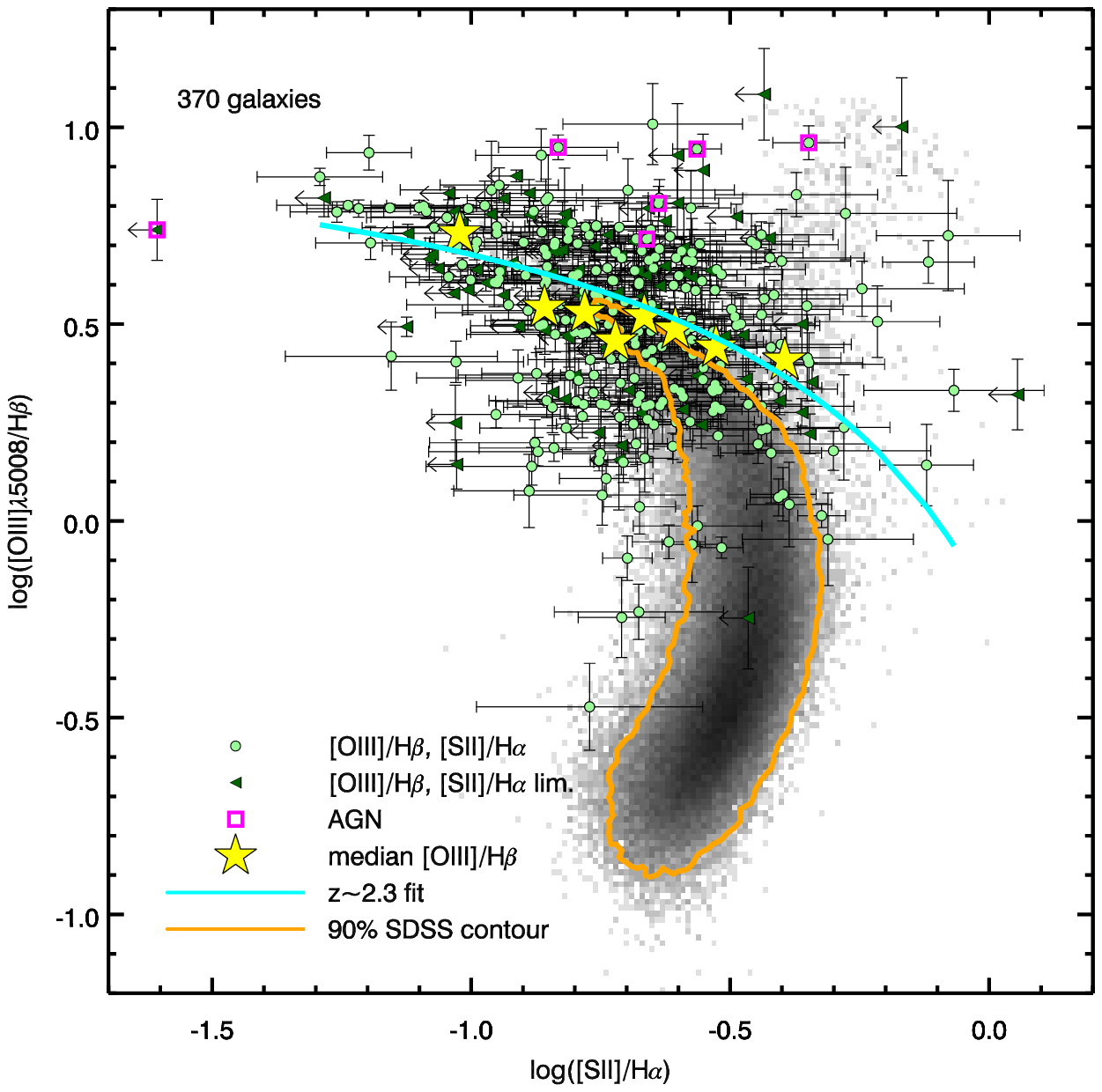}
\caption{The S2-BPT diagram displayed in the same manner as Figure~\ref{n2_bpt}. Note that while $z\sim2.3$ galaxies are noticeably offset relative to the $z\sim0$ sample in the N2-BPT diagram, a similar offset is not observed in the S2-BPT plane. Instead, the log([\ion{O}{3}]/H$\beta$) medians (yellow stars, Table~\ref{medianratios}) and the fit to the KBSS-MOSFIRE locus (cyan curve) coincide with the upper envelope of the orange contour, which encloses 90\% of SDSS galaxies.}
\label{s2_bpt}
\end{figure*}

Figure~\ref{n2_bpt} shows the location of 377 KBSS-MOSFIRE galaxies with $1.9<z<2.7$ ($\langle z\rangle=2.3$) in the N2-BPT plane, including only those galaxies where H$\alpha$ is detected at $>5\sigma$ and H$\beta$ and [\ion{O}{3}]$\lambda$5008 are detected at $>3\sigma$ (light green points). Galaxies with [\ion{N}{2}]$\lambda6585$ measurements with S/N~$<2$ are assigned 2$\sigma$ upper limits (dark green triangles), accounting for $\sim36\%$ of the sample. As we originally reported in \citetalias{steidel2014} for a smaller sample, KBSS-MOSFIRE galaxies occupy a region of the N2-BPT plane almost entirely distinct from the majority of local galaxies, represented by SDSS-DR8 \citep[in greyscale,][]{aihara2011}.

The SDSS comparison sample employed in this paper takes emission line measurements and ancillary physical parameters from the MPA-JHU catalogs\footnote{\url{https://www.sdss3.org/dr10/spectro/galaxy_mpajhu.php}} \citep[c.f. Sections~4.3.1 and 4.3.2 of][]{aihara2011} and has been selected to be similar to KBSS-MOSFIRE in terms of detection properties. The SDSS comparison galaxies have $0.04\leq z\leq 0.1$ (to avoid severe aperture effects), $> 50\sigma$ measurements of H$\alpha$, and a ``reliable" flag signaling good results from the MPA-JHU pipeline. Stellar masses, SFRs, and sSFRs are reported using the median estimates of the PDFs for the ``total" values.

Although the existence of an offset in the N2-BPT plane has been widely reported for other samples of high-redshift galaxies \citep{masters2014,shapley2015,sanders2016}, the degree to which the KBSS-MOSFIRE sample differs from typical local galaxies is quite remarkable. Nearly all KBSS-MOSFIRE galaxies have larger values of log([\ion{O}{3}]/H$\beta$) at fixed log([\ion{N}{2}]/H$\alpha$) than typical SDSS galaxies in the N2-BPT diagram (represented by the 90\% of $z\sim0$ galaxies enclosed by the orange contour in Figure~\ref{n2_bpt}). Furthermore, the ridge-line of the $z\sim2.3$ sample with $>2\sigma$ detections of [\ion{N}{2}]$\lambda6585$, traced by median values of log([\ion{O}{3}]/H$\beta$) in equal-number bins of log([\ion{N}{2}]/H$\alpha$) (yellow stars, Table~\ref{medianratios}), falls well outside the same contour. 

The N2-BPT locus can be described analytically using the following functional form:
\begin{equation}
\log (\textrm{[\ion{O}{3}]/H$\beta$}) = \frac{p_0}{\log (\textrm{[\ion{N}{2}]/H$\alpha$})+p_1}+p_2.
\end{equation}
However, because the fit parameters are degenerate, interpreting them is difficult. To be consistent with the literature, we fit the KBSS-MOSFIRE locus with $p_0$ fixed to the value reported by \citet{kewley2001} for the N2-BPT extreme starburst classification line ($p_0=0.61$). We determine the best-fit curve describing the KBSS-MOSFIRE N2-BPT locus (cyan curve in Figure~\ref{n2_bpt}) using the IDL routine MPFIT \citep{markwardt2009}, with the following result:
\begin{equation}
\log (\textrm{[\ion{O}{3}]/H$\beta$}) = \frac{0.61}{\log (\textrm{[\ion{N}{2}]/H$\alpha$})-0.22}+1.12.
\label{kbss_n2_bpt_eq}
\end{equation}
The intrinsic scatter relative to the best-fit curve is 0.18~dex, estimated from the amount of additional uncertainty required to achieve $\chi^2/$DOF$=1$ after accounting for individual measurement errors. Galaxies with upper limits on [\ion{N}{2}] and objects identified as AGN (magenta squares; see also \S\ref{agn_text}) are excluded from the fit.

We determine the $z\sim0$ locus for the SDSS comparison sample using the same form (red dashed line), including only star-forming galaxies identified using the \citet{kauffmann2003} selection. When $p_1$ and $p_2$ are fixed separately, we find that the observed offset between the $z\sim0$ N2-BPT locus and the KBSS-MOSFIRE ridge-line can be described by a shift of either 
\begin{eqnarray}
\Delta \log (\text{[\ion{O}{3}]/H}\beta)&=&0.26{\rm ~dex~or} \nonumber \\
\Delta \log (\text{[\ion{N}{2}]/H}\alpha)&=&0.37{\rm ~dex.} \nonumber
\end{eqnarray}
Even when $p_0$ is treated as a free parameter, the amplitudes of these shifts are similar, so long as the same $p_0$ is adopted for both the low-$z$ SDSS and high-$z$ KBSS-MOSFIRE samples.

\begin{deluxetable}{cccc}
\tablecolumns{4}
\tablecaption{Median line ratios}
\tablehead{
\multicolumn{2}{c}{N2-BPT} & \multicolumn{2}{c}{S2-BPT}\\
\colhead{$\log$[\ion{N}{2}]/H$\alpha$} & \colhead{$\log$[\ion{O}{3}]/H$\beta$} & \colhead{$\log$[\ion{S}{2}]/H$\alpha$} & \colhead{$\log$[\ion{O}{3}]/H$\beta$}}
\startdata
-1.23 & 0.70 & -1.02 & 0.73 \\
-1.06 & 0.64 & -0.86 & 0.54 \\
-0.95 & 0.50 & -0.78 & 0.53 \\
-0.87 & 0.50 & -0.72 & 0.46 \\
-0.79 & 0.47 & -0.67 & 0.52 \\
-0.68 & 0.35 & -0.61 & 0.49 \\
-0.59 & 0.25 & -0.53 & 0.44 \\
-0.45 & 0.29 & -0.39 & 0.41
\enddata
\tablecomments{The above values of $\log$[\ion{O}{3}]/H$\beta$ are the medians in equal-number bins of $\log$[\ion{N}{2}]/H$\alpha$ and $\log$[\ion{S}{2}]/H$\alpha$, where only galaxies with $>2\sigma$ detections of [\ion{N}{2}]$\lambda6585$ or [\ion{S}{2}]$\lambda\lambda6718,6732$ have been included. The reported values of $\log$[\ion{N}{2}]/H$\alpha$ and $\log$[\ion{S}{2}]/H$\alpha$ are the median values of those quantities in each bin.}
\label{medianratios}
\end{deluxetable}

The difference between $z\sim0$ and $z\sim2.3$ galaxies observed in the S2-BPT diagram (Figure~\ref{s2_bpt}) is less notable than in the N2-BPT plane, a result that was previously shown by \citet{masters2014} for stacked spectra and by \citet{shapley2015} and \citet{sanders2016} for individual high-$z$ galaxies from the MOSDEF survey. The KBSS-MOSFIRE galaxies appear to trace the high [\ion{O}{3}]/H$\beta$ tail of SDSS, although with larger scatter due to measurement errors. The best fit to the S2-BPT locus (cyan curve in Figure~\ref{s2_bpt}) with $p_0= 0.72$ \citep{kewley2001} grazes the upper edge of the distribution of local galaxies, taking the form
\begin{equation}
\log (\textrm{[\ion{O}{3}]/H$\beta$}) = \frac{0.72}{\log (\textrm{[\ion{S}{2}]/H$\alpha$})-0.53}+1.15,
\end{equation}
with an intrinsic scatter of 0.22~dex. Note that the median values of log([\ion{O}{3}]/H$\beta$) in equal-number bins of log([\ion{S}{2}]/H$\alpha$) (yellow stars) also lie close to the upper envelope of the orange contour enclosing 90\% of SDSS galaxies. 

As we will discuss in \S\ref{twin_text}, the same behavior persists when one compares the $z\sim2.3$ KBSS-MOSFIRE sample with SDSS galaxies that have similar ionization and excitation properties: the KBSS-MOSFIRE sample exhibits a substantial offset relative to SDSS in the N2-BPT plane, but virtually no separation is observed in the S2-BPT diagram. In \S\ref{models}, we show that the same self-consistent photoionization models can explain the observations in both cases.

\subsection{AGN Contamination}
\label{agn_text}

Given the differences observed in the N2-BPT diagram between $z\sim2.3$ galaxies and local star-forming galaxies, it is important to understand the risk of AGN contamination, which could lead to misinterpretation of the results described above. Although the BPT diagrams are used to separate star-forming galaxies and AGN at low-$z$, a similar application at high-redshift is less straightforward. \cite{kewley2013} used photoionization models to explore differences in the BPT plane at higher redshifts and noted that the distinct ``branches" seen in the N2-BPT diagram become less discernible with changes in ISM conditions and AGN properties. From observations alone, the problem is also clear: the majority of $z\sim2.3$ KBSS galaxies fall in the region of the N2-BPT diagram (Figure~\ref{n2_bpt}) generally reserved for ``composite" objects---those that show signs of AGN activity as well as ongoing star formation---at $z\sim0$.

We considered this issue in some detail in \citetalias{steidel2014}, but summarize it again here. KBSS galaxies are flagged as AGN or QSOs if (1) their rest-UV (LRIS) spectrum shows significant emission in high ionization lines (such as \ion{C}{4}$\lambda1549$, \ion{C}{3}]$\lambda1908$, \ion{N}{5}$\lambda1240$; \citealt{hainline2011}) or (2) a combination of their rest-optical emission-line ratios and line widths suggest the presence of an AGN. These objects (7 in total) generally fall near the upper envelope of log([\ion{O}{3}]/H$\beta$) observed for KBSS galaxies and in the upper half of the stellar mass range sampled by the survey; they are excluded from fits to the galaxy sample (including those reported in \S\ref{bpt_text}) and from analyses that require extinction corrections.

\subsection{The Stellar Mass-Excitation Relation}

\begin{figure}
\centering
\includegraphics{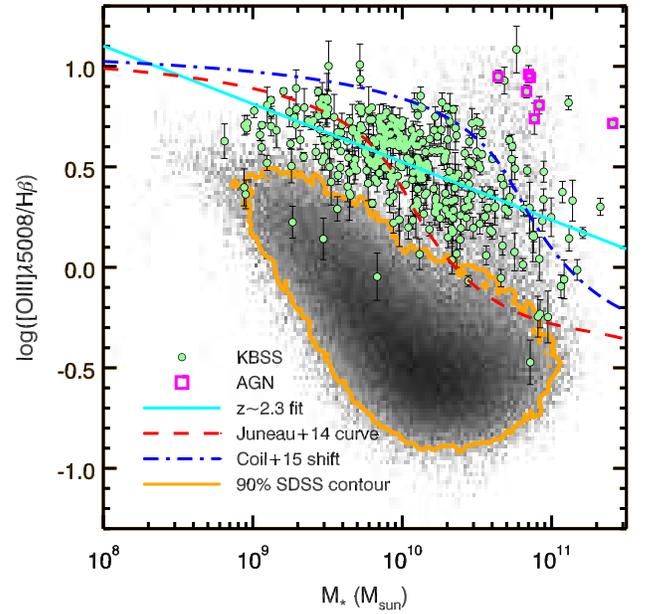}
\caption{The mass-excitation relation (MEx) for 365 KBSS-MOSFIRE galaxies with M$_{\ast}$ estimates (green) and SDSS galaxies (greyscale), with the division between star-forming/composite galaxies and AGN in the local universe from \citet{juneau2014} illustrated by the red dashed curve. The dot-dashed blue curve represents the shift proposed by \citet{coil2015} to separate $z\sim2$ star-forming galaxies from AGN. KBSS-MOSFIRE galaxies identified as AGN are denoted by magenta squares. The correlation between [\ion{O}{3}]$\lambda5008$/H$\beta$ and M$_{\ast}$ is one of the strongest observed for the KBSS-MOSFIRE sample (with the best-fit linear relation shown by the cyan line) and also highlights one of the largest differences between $z\sim2.3$ galaxies and those at $z\sim0$.}
\label{mex_plots}
\end{figure}

Similar to the offset observed in the N2-BPT plane, the behavior of [\ion{O}{3}]/H$\beta$ with respect to M$_{\ast}$ \citep[referred to as the MEx,][]{juneau2011} highlights dramatic differences between $z\sim2.3$ KBSS-MOSFIRE galaxies and local galaxies from SDSS. Figure~\ref{mex_plots} shows the M$_{\ast}$-excitation relation for both samples. 

As for local galaxies (greyscale in Figure~\ref{mex_plots}), $z\sim2.3$ KBSS-MOSFIRE galaxies (green points) exhibit a significant inverse correlation between [\ion{O}{3}]/H$\beta$ and M$_{\ast}$, with a Spearman correlation coefficient of $\rho=-0.47$ and $p=1.4\times10^{-21}$. However, the distribution of the $z\sim2.3$ sample is almost entirely disjoint with respect to the majority of SDSS galaxies (enclosed by the 90\% contour in orange), with 82\% of KBSS-MOSFIRE objects falling above the updated curve from \citet{juneau2014} used to divide star-forming/composite galaxies from AGN in the local universe (red dashed curve in Figure~\ref{mex_plots}). This empirical division was calibrated using $z\sim0.1$ SDSS galaxies and designed to function as an alternative method of distinguishing between AGN and star-forming (or composite) galaxies when not all of the BPT lines are available. Figure~\ref{mex_plots} also shows that AGN in the KBSS-MOSFIRE sample are offset to even higher [\ion{O}{3}]/H$\beta$ than the majority of the $z\sim2.3$ galaxies at fixed M$_{\ast}$, although there are $z\sim0$ AGN from SDSS that occupy the same region of parameter space. These results are largely consistent with those reported by \citet{coil2015}, who explicitly tested the success of the MEx in identifying $z\sim2$ AGN using observations from the MOSDEF survey. In order to construct a diagnostic that more accurately distinguishes between high-$z$ star-forming galaxies and AGN, \citet{coil2015} proposed shifting the MEx division by $\Delta\log($M$_{\ast}/$M$_{\odot})=0.75$, as illustrated by the dot-dashed blue curve in Figure~\ref{mex_plots}; to cleanly separate AGN from star-forming galaxies in the current KBSS-MOSFIRE sample, an even larger shift ($\gtrsim1.0$~dex) would be required.

\begin{figure*}
\centering
\includegraphics{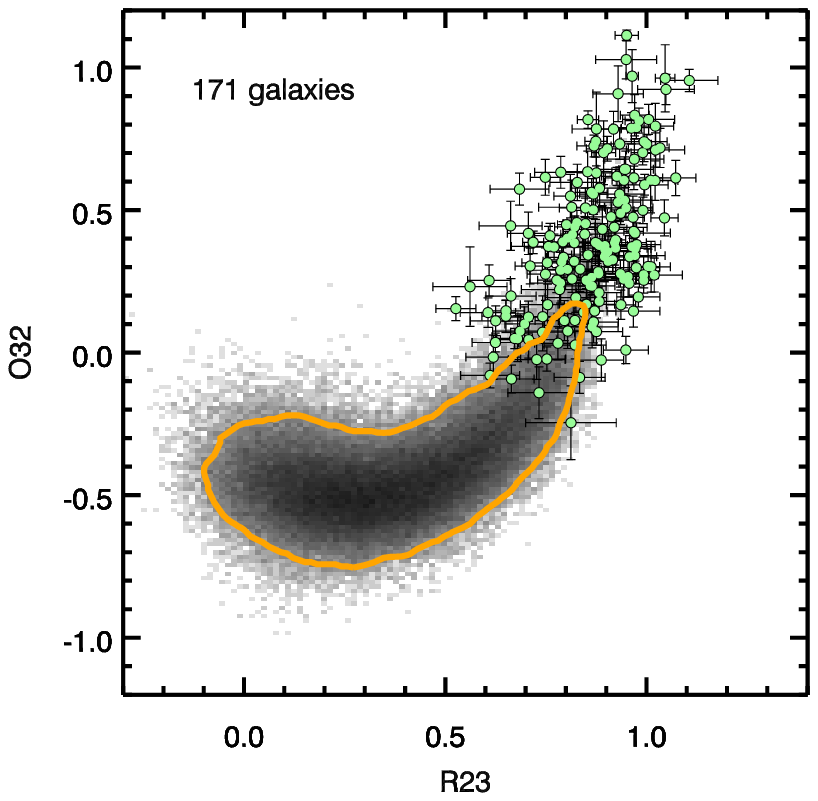}
\includegraphics{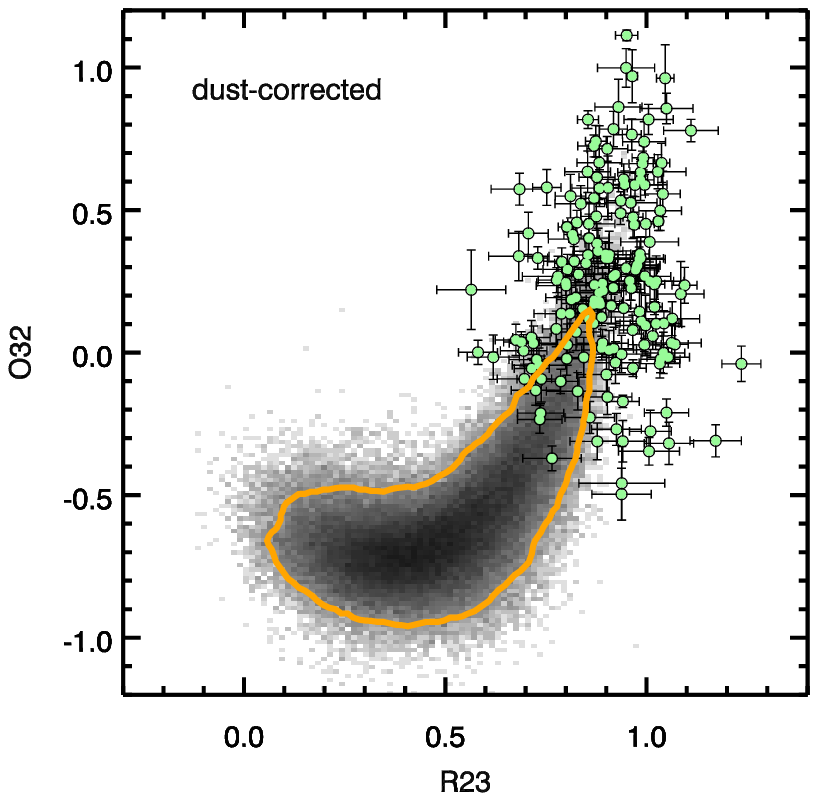}
\caption{\textit{Left:} The distribution of the O32 and R23 indices for KBSS-MOSFIRE galaxies (excluding AGN), using line fluxes that have not been corrected for dust extinction, compared with the locus of $z\sim0$ SDSS galaxies in greyscale (the orange contour encloses 90\% of the sample). Here, in contrast to earlier figures, the SDSS sample contains only star-forming galaxies, separated from AGN by the curve from \citet{kauffmann2003}. \textit{Right:} The same parameter space as the lefthand panel, but with line ratios that have been corrected for dust extinction. As in the S2-BPT diagram, KBSS-MOSFIRE galaxies largely follow the trend established by typical local galaxies, despite overlapping with only the most extreme tail of SDSS.}
\label{o32r23_plots}
\end{figure*}

We calculate a linear fit to the locus of $z\sim2.3$ KBSS-MOSFIRE star-forming galaxies in Figure~\ref{mex_plots} using the MPFITEXY IDL routine\footnote{\url{http://purl.org/mike/mpfitexy}} \citep{williams2010}, and it has the form:
\begin{equation}
\log (\textrm{[\ion{O}{3}]/H$\beta$}) = 0.52-0.29\times[\log(M_{\ast}/M_{\odot})-10],
\end{equation}
with $\sigma_{\rm RMS} = 0.19$~dex scatter about the best-fit relation and an implied intrinsic scatter of $\sigma_{\rm int}=0.17$~dex. The correlation between excitation, as probed by [\ion{O}{3}]/H$\beta$, and M$_{\ast}$ is therefore one of the tightest correlations between the nebular spectrum and a global galaxy property.

Compared to SDSS, KBSS-MOSFIRE galaxies exhibit an offset of up to 0.8~dex toward higher log([\ion{O}{3}]/H$\beta$) at fixed M$_{\ast}$. The largest separation occurs at the high mass end of the M$_{\ast}$ distribution, where there are few local analogs to KBSS-MOSFIRE galaxies. Although some of the offset likely results from decreased gas-phase oxygen abundances in high-$z$ \ion{H}{2} regions \citep[e.g.,][]{pettini2004,maiolino2008}, we show in the following sections that the higher values of [\ion{O}{3}]/H$\beta$ observed in $z\sim2.3$ galaxies reflect differences in the typical shape of the ionizing radiation field in galaxies at fixed M$_{\ast}$ as a function of redshift.

\subsection{Ionization and Excitation: O32 and R23}
\label{o32_r23_text1}

\begin{deluxetable}{cl}
\tablecaption{Strong-line indices}
\tablehead{
\colhead{index} & \colhead{definition}}
\startdata
N2 & $\log$([\ion{N}{2}]$\lambda6585$/H$\alpha$) \\
O3N2 & $\log$([\ion{O}{3}]$\lambda5008$/H$\beta$)$-\log$([\ion{N}{2}]$\lambda6585$/H$\alpha$) \\
R23 & $\log$[([\ion{O}{3}]$\lambda\lambda4960,5008$+[\ion{O}{2}]$\lambda\lambda3727,3729$)/H$\beta$] \\
O32 & $\log$([\ion{O}{3}]$\lambda\lambda4960,5008$/[\ion{O}{2}]$\lambda\lambda3727,3729$) \\
N2O2 & $\log$([\ion{N}{2}]$\lambda6585$/[\ion{O}{2}]$\lambda\lambda3727,3729$) \\
N2S2 & $\log$([\ion{N}{2}]$\lambda6585$/[\ion{S}{2}]$\lambda\lambda6718,6732$) \\
Ne3O2 & $\log$([\ion{Ne}{3}]$\lambda3869$/[\ion{O}{2}]$\lambda\lambda3727,3729$)
\enddata
\tablecomments{The $\lambda\lambda$ notation refers to the sum of both lines.}
\label{strongline}
\end{deluxetable}

We also consider the behavior of O32 and R23; the definitions for both indices as used in this paper can be found in Table~\ref{strongline}, together with other strong-line indices. O32 is often employed as a proxy for ionization \citep{penston1990} and can be used to calculate the ionization parameter ($U\equiv n_{\gamma}/n_H$, the dimensionless ratio of the number density of incident H-ionizing photons to the number density of hydrogen atoms in the gas) given an assumed ionizing spectrum. R23 is used to estimate oxygen abundance \citep[e.g.,][]{pagel1979,mcgaugh1991,kewley2002}, although because of its double-valued nature, an additional parameter (like N2 or O3N2) must be used to break the degeneracy. R23 may also be used as probe of the overall degree of excitation in \ion{H}{2} regions with moderate gas-phase metallicity, as it compares the emission from two collisionally-excited metallic ions (O$^+$ and O$^{++}$) with the emission from a recombination line of hydrogen, thus providing a rough constraint on the amount of kinetic energy in the gas relative to the number of ionizing photons in a similar manner to the line ratios used in the BPT diagrams.

The clear advantage of indices like O32 and R23 over the BPT ratios is that they include only one element (oxygen) in addition to hydrogen and should be insensitive to differences in the abundance ratio patterns of $z\sim0$ and $z\sim2.3$ galaxies (or, indeed, between any two populations). Moreover, unlike [\ion{O}{3}]/H$\beta$, which is sensitive to changes in both the kinetic energy of the gas and $U$, R23 and O32 offer relatively independent probes of excitation and ionization, respectively. However, because both O32 and R23 require nebular extinction corrections, the indices can only be determined for $z\sim2.3$ galaxies with observations in $J$, $H$, and $K$ band.

Figure~\ref{o32r23_plots} shows $z\sim2.3$ KBSS-MOSFIRE galaxies in the O32-R23 plane, compared with galaxies from SDSS. The lefthand panel shows the observed ratios, corrected for relative slit losses but not for dust extinction, while the righthand panel shows the line ratios corrected for differential extinction due to dust as described above in \S\ref{sfrs_and_masses}. The KBSS-MOSFIRE galaxies are largely coincident with the high-O32, high-R23 tail of SDSS (similar to the S2-BPT plane, Figure~\ref{s2_bpt}), although this region of parameter space is relatively sparsely populated by $z\sim0$ galaxies overall, with the majority of the SDSS sample exhibiting much lower values of both indices; similar results were also reported for $z\sim2$ MOSDEF galaxies by \citet{shapley2015}.

This agreement reflects fundamental similarities between the most extreme SDSS galaxies and KBSS-MOSFIRE galaxies. At low gas-phase O/H, R23 increases with metallicity as the number of oxygen atoms increases; at high gas-phase O/H, R23 declines again, because the oxygen present in the gas functions as an efficient cooling pathway, reducing the gas temperature and thus the number of collisionally-excited oxygen ions. Thus, R23 reaches a maximum value at intermediate O/H, with the value of R23 at the turnaround depending sensitively on the hardness of the incident radiation field at fixed gas-phase O/H: a harder ionizing photon distribution will result in greater kinetic energy per ionized electron than a softer distribution. Because there is a maximum degree of hardness for the spectra of massive stars set by stellar evolutionary processes, there is a corresponding maximum R23 achievable for star-forming galaxies, reflected in the upper envelope at ${\rm R23}\sim1$ for both samples in the righthand panel of Figure~\ref{o32r23_plots}.

Thus, although there exist SDSS galaxies occupying the same high-O32, high-R23 region of parameter space as the majority of $z\sim2.3$ galaxies, both samples must be characterized by large ionization parameters and the combination of hard ionizing radiation fields and moderate gas-phase O/H needed to produce values of R23 near the turnaround. We test this assumption in \S\ref{twin_text} and \S\ref{model_bpt} and evaluate the utility of the combination of O32 and R23 for determining gas-phase oxygen abundance at high-redshift.

\section{Physical Characteristics of the Most Offset Galaxies in the $z\sim2.3$ N2-BPT Plane}
\label{offset_text}

\begin{figure*}
\centering
\includegraphics{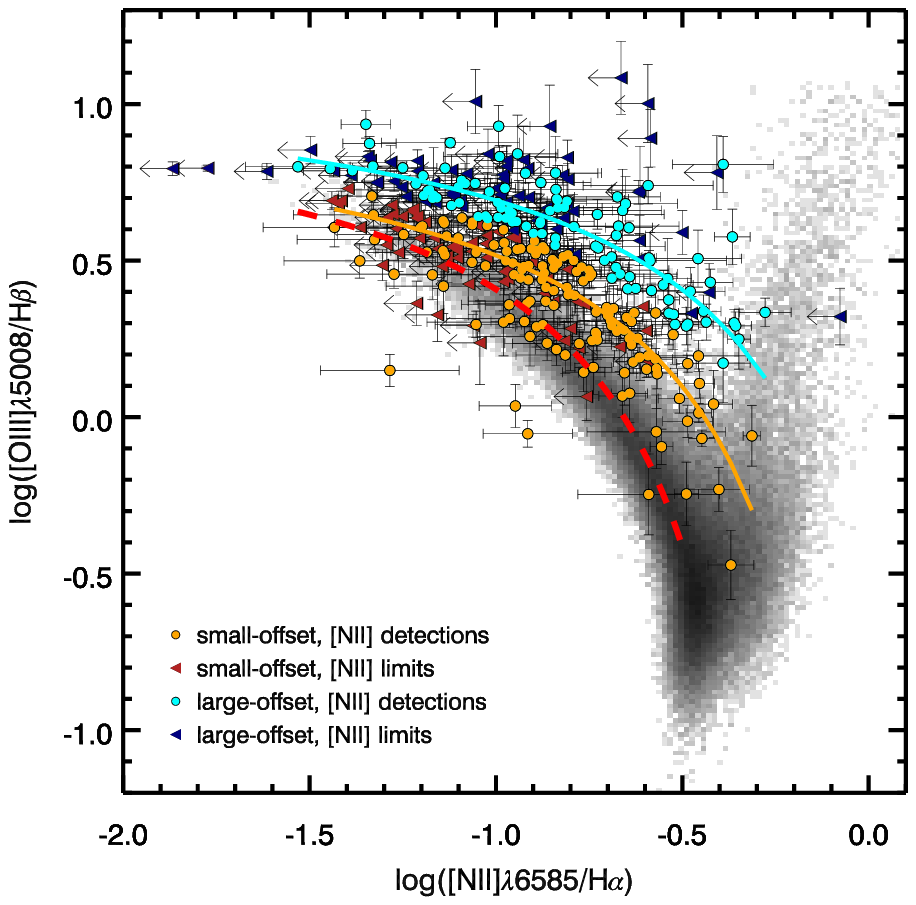}
\hspace{-0.8in}
\includegraphics{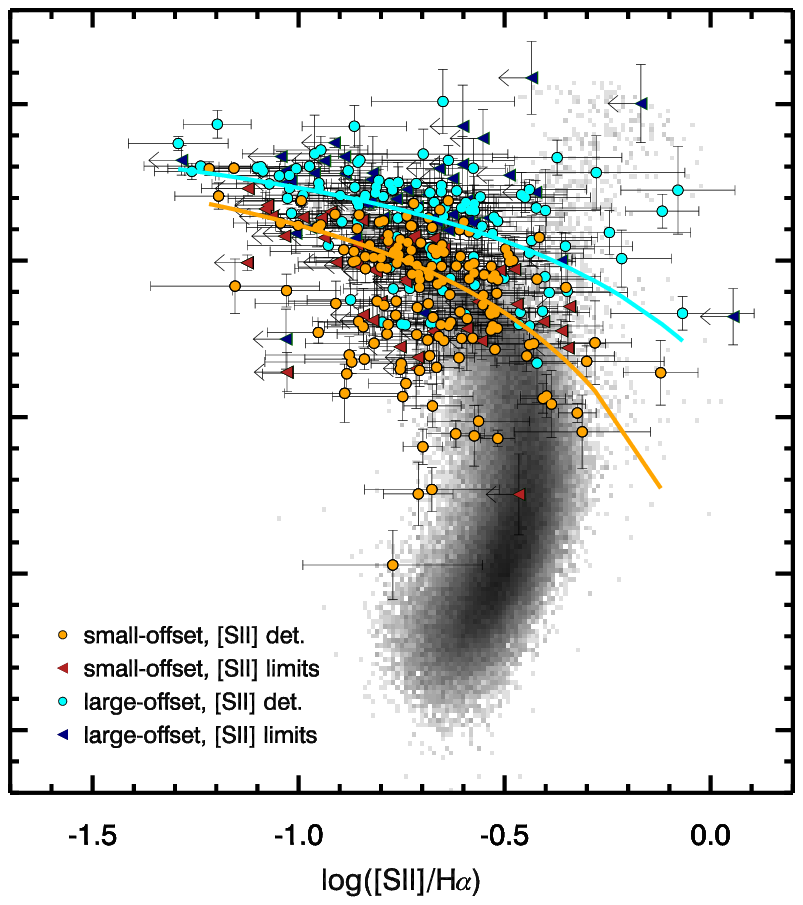}
\caption{The N2-BPT (left) and S2-BPT (right) diagrams, showing the location of $z\sim2.3$ KBSS-MOSFIRE galaxies color-coded on the basis of their location in the N2-BPT plane. Those falling above the KBSS ridge-line (Equation~\ref{kbss_n2_bpt_eq}, cyan curve in Figure~\ref{n2_bpt}) are referred to as the ``large-offset" subsample (cyan/navy points), and those falling below the ridge-line are instead referred to as ``small-offset" galaxies (orange/red points). For reference, the $z\sim0$ N2-BPT locus from Figure~\ref{n2_bpt} is shown in red, and separate fits to the loci of both the small- and large-offset subsamples are illustrated by the solid orange and cyan curves, respectively. A comparison between the $z\sim2.3$ loci reveals a significant shift even with respect to one another: up to $\Delta \log (\textrm{[\ion{O}{3}]/H$\beta$}) = 0.19$~dex or $\Delta \log (\textrm{[\ion{N}{2}]/H$\alpha$}) = 0.34$~dex. In the S2-BPT plane, a shift of $\Delta \log (\textrm{[\ion{O}{3}]/H$\beta$}) = 0.16$~dex or $\Delta \log (\textrm{[\ion{S}{2}]/H$\alpha$}) = 0.36$~dex would be required to bring the two loci into agreement. AGN in the KBSS-MOSFIRE sample have been excluded from both panels.}
\label{farselect}
\end{figure*}

In addition to comparing typical $z\sim2.3$ galaxies with those commonly seen in the local universe (such as star-forming galaxies from SDSS), comparisons between high-redshift samples with different properties are also important. The latter is necessary to understand whether the physical process(es) driving the differences observed between typical galaxies at different epochs are also responsible for the diversity of galaxy properties observed at a single epoch. To facilitate this exercise, we divide the $z\sim2.3$ KBSS-MOSFIRE galaxies into two subsamples on the basis of their position in the N2-BPT plane. High-redshift samples show the greatest deviations relative to SDSS galaxies in this parameter space and we might therefore expect to identify correlated differences in other line ratios and physical properties between subsamples identified in such a manner.

For the purposes of this comparison, galaxies with log([\ion{O}{3}]/H$\beta$) ratios that fall above the $z\sim2.3$ N2-BPT ridge-line (Equation~\ref{kbss_n2_bpt_eq}, cyan curve in Figure~\ref{n2_bpt}) are designated the ``large-offset" subsample, and those below the ridge-line form the ``small-offset" subsample; AGN are excluded from both. Figure~\ref{farselect} shows the location of these two subsamples in the N2-BPT plane (where they are selected) and the S2-BPT plane. Galaxies with low-significance ($<2\sigma$) detections of [\ion{N}{2}] or [\ion{S}{2}] are identified by darker-colored triangles at the position of their 2$\sigma$ upper limits. Note that if a more stringent selection is applied, such that galaxies must lie $>3\sigma$ away from the ridge-line in log([\ion{O}{3}]/H$\beta$) in order to be classified as large-offset or small-offset, the nature of the results discussed below remains unchanged.

\subsection{BPT Offsets}
\label{farselect_offsets}

The loci of small-offset galaxies (orange curve in Figure~\ref{farselect}) and large-offset galaxies (cyan curve) in the N2-BPT and S2-BPT planes are independently fit using the method and functional form described in \S\ref{bpt_text}. As before, objects with upper limits on [\ion{N}{2}] or [\ion{S}{2}] are not included in determining the fits. The results for both BPT diagrams are summarized in Table~\ref{bpt_table}.

Even the locus of small-offset KBSS-MOSFIRE galaxies remains noticeably different from typical SDSS galaxies in the N2-BPT plane, as shown by the offset of orange curve relative to the $z\sim0$ locus (red dashed curve) in the lefthand panel of Figure~\ref{farselect}. The additional separation between the small-offset and large-offset loci is consistent with an offset of either
\begin{eqnarray}
\Delta \log (\text{[\ion{O}{3}]/H}\beta)_{\rm N2}&=&0.19{\rm~dex~or } \nonumber \\
\Delta \log (\text{[\ion{N}{2}]/H}\alpha)_{\rm N2}&=&0.34{\rm~dex}, \nonumber
\end{eqnarray}
if the shift is restricted to one axis.

The two subsamples are also clearly differentiated in the S2-BPT diagram (righthand panel of Figure~\ref{farselect}), with large-offset galaxies exhibiting systematically higher log([\ion{O}{3}]/H$\beta$) at fixed log([\ion{S}{2}]/H$\alpha$) compared with the small-offset subsample. If only one fit parameter is allowed to vary, the separation between the two loci is equivalent to either
\begin{eqnarray}
\Delta \log (\text{[\ion{O}{3}]/H}\beta)_{\rm S2}&=&0.16{\rm~dex~or } \nonumber \\
\Delta \log (\text{[\ion{S}{2}]/H}\alpha)_{\rm S2}&=&0.36{\rm~dex}. \nonumber
\end{eqnarray}

This significant displacement between the subsample loci in the S2-BPT diagram is intriguing because of its implications for the physical origin of the offset in the N2-BPT plane. Assuming the offset between the small-offset and large-offset subsamples in the N2-BPT plane is entirely a horizontal displacement toward higher log([\ion{N}{2}]/H$\alpha$) resulting from an enhancement in N/O at fixed O/H implies $\Delta \log (\text{[\ion{O}{3}]/H}\beta)\approx0$. Provided all other abundance ratios (particularly S/O) are similar between small- and large-offset galaxies, $\Delta \log (\text{[\ion{S}{2}]/H}\alpha)$ should also be close to zero. The assumption of similar S/O abundances among the two subsamples is reasonable, as sulfur and oxygen are both alpha process elements released primarily by Type II SNe. 

Consequently, the two ridge-lines in the righthand panel of Figure~\ref{farselect} ought to lie coincident with one another if the offset between the subsamples were due primarily to differences in N/O at fixed O/H. Indeed, when \citet{sanders2016} performed the same exercise using galaxies from the MOSDEF survey, they found that high-$z$ galaxies separated on the basis of their offset from local galaxies in the N2-BPT diagram are well-mixed in the S2-BPT and O32-R23 planes. The authors interpret this result as evidence in favor of elevated N/O at fixed O/H in $z\sim2.3$ galaxies, but such an explanation is inconsistent with the results shown for the KBSS-MOSFIRE subsamples in Figure~\ref{farselect}.

While some horizontal separation between the KBSS-MOSFIRE S2-BPT loci might occur if a larger amount of [\ion{S}{2}] emission originated from diffuse ionized gas in large-offset galaxies than in small-offset galaxies, the contribution from such gas would need to be substantial to account for the 0.36~dex difference in log([\ion{S}{2}]/H$\alpha$) at fixed log([\ion{O}{3}]/H$\beta$) between the subsamples. Instead, it is likely that the clear separation between the two loci corresponds to important differences in the shape and/or normalization of the ionizing radiation in small- and large-offset galaxies, and similar differences may be important between $z\sim0$ and $z\sim2.3$ galaxies. In \S\ref{twin_text}, we test whether such differences can account for the entirety of the offset observed between KBSS-MOSFIRE and SDSS in the N2-BPT diagram.

\begin{deluxetable}{lccc}
\tablewidth{\columnwidth}
\tablecaption{Best-fit parameters for BPT ridgelines}
\tablehead{
\colhead{} &\colhead{$p_0$} &\colhead{$p_1$} &\colhead{$p_2$}}
\startdata
SDSS N2-BPT & 0.61 & 0.09 & 1.08 \\
KBSS-MOSFIRE N2-BPT & 0.61 & -0.22 & 1.12 \\
KBSS-MOSFIRE small-offset (N2) & 0.61 & -0.14 & 1.05 \\
KBSS-MOSFIRE large-offset (N2) & 0.61 & -0.31 & 1.16 \\
KBSS-MOSFIRE S2-BPT & 0.72 & -0.53 & 1.15 \\
KBSS-MOSFIRE small-offset (S2) & 0.72 & -0.41 & 1.12 \\
KBSS-MOSFIRE large-offset (S2) & 0.72 & -0.73 & 1.15 
\enddata
\label{bpt_table}
\end{deluxetable}

\subsection{Galaxy Properties}
\label{farselect_properties}

\begin{figure}
\centering
\includegraphics{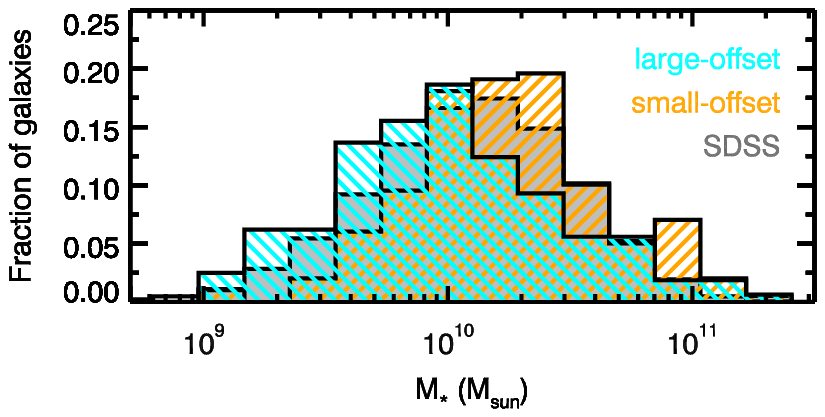}
\includegraphics{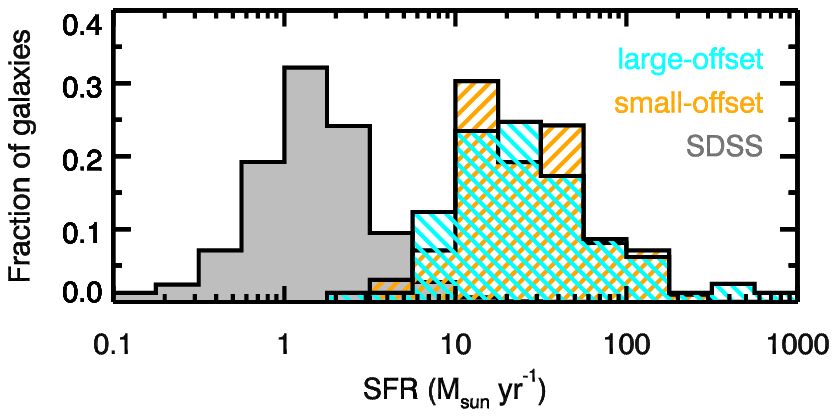}
\includegraphics{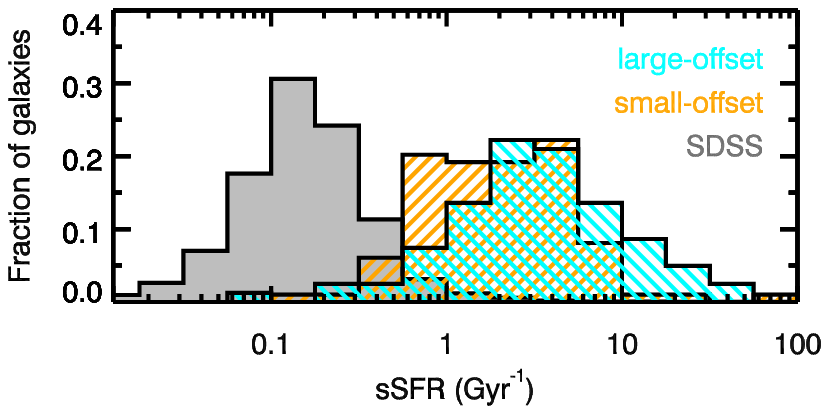}
\caption{The distributions of M$_{\ast}$, SFR, and sSFR for SDSS galaxies (in grey) and the KBSS-MOSFIRE subsamples divided by the degree of their N2-BPT offset.  Although the two $z\sim2.3$ subsamples have very similar distributions in SFR, they exhibit clear differences in M$_{\ast}$ and sSFR. All KBSS-MOSFIRE galaxies have substantially higher SFRs and sSFRs than SDSS galaxies.}
\label{farselect_hists}
\end{figure}

We can also compare the global properties of galaxies as a function of their N2-BPT offset. In general, large-offset galaxies have significantly lower M$_{\ast}$ than small-offset galaxies, despite having nearly identical distributions in SFR (see Figure~\ref{farselect_hists}). The separation as a function of M$_{\ast}$ is consistent with the results reported by \citet{shapley2015} for $z\sim2$ MOSDEF galaxies, but the resulting separation in sSFR for KBSS-MOSFIRE galaxies is especially notable. This difference highlights the fact that recent star formation has contributed a larger fraction of the integrated stellar mass of large-offset galaxies than for small-offset galaxies. For comparison, the distributions of M$_{\ast}$, SFR, and sSFR for the SDSS galaxies are shown in grey in Figure~\ref{farselect_hists}. Interestingly, the SDSS, small-offset KBSS-MOSFIRE, and large-offset KBSS-MOSFIRE samples appear to form a sequence in sSFR that mirrors their separation in N2-BPT diagram (from least to most offset).

\begin{figure}
\centering
\includegraphics{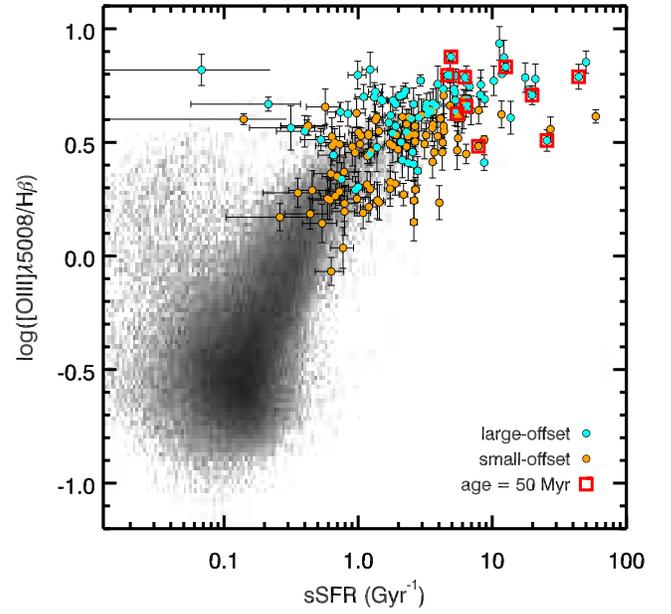}
\caption{The sSFR-excitation relation for SDSS (in greyscale) and KBSS-MOSFIRE galaxies, with the $z\sim2.3$ galaxies color-coded by the degree of their N2-BPT offset (large-offset galaxies in cyan, small-offset galaxies in orange; see lefthand panel of Figure~\ref{farselect}). KBSS-MOSFIRE galaxies with an imposed minimum age of 50~Myr from SED-fitting are identified by red squares; this sample of galaxies is discussed at greater length in Section~\ref{sed_shapes} and Figures~\ref{farselect_colors} and \ref{farselect_seds}. All of the KBSS-MOSFIRE galaxies overlap only with SDSS galaxies with the highest log([\ion{O}{3}]/H$\beta$) and sSFR. However, both $z\sim2.3$ and $z\sim0$ samples appear to form a single excitation sequence.}
\label{o3hb_ssfr_plot}
\end{figure}

The importance of recent star formation is also apparent in the behavior of [\ion{O}{3}]/H$\beta$---which we have already shown is significantly higher for KBSS-MOSFIRE galaxies compared with SDSS and serves as a measure of excitation---as a function of sSFR, which is shown in Figure~\ref{o3hb_ssfr_plot}. Unlike M$_{\ast}$-excitation space (Figure~\ref{mex_plots}), KBSS-MOSFIRE galaxies appear to follow the sSFR-excitation trend established by local galaxies in SDSS, with large-offset galaxies exhibiting the highest sSFRs and ratios of [\ion{O}{3}]/H$\beta$. \citet{dickey2016} found the same trend for a sample of $z\sim2$ star-forming galaxies from the 3D-$HST$ survey.

The interquartile range in M$_{\ast}$ for SDSS galaxies with sSFR $>2$~Gyr$^{-1}$ ($\sim1\%$ of the total sample) is log(M$_{\ast}$/M$_{\odot})=8.40-9.06$, while high sSFRs are found in significantly more massive $z\sim2.3$ galaxies. KBSS-MOSFIRE galaxies with sSFR$>2$~Gyr$^{-1}$ represent $\sim57\%$ of the $z\sim2.3$ sample and are $\sim10$ times more massive than high-sSFR $z\sim0$ SDSS galaxies on average, with an interquartile range in log(M$_{\ast}$/M$_{\odot})=9.59-10.11$. If the offset of the two $z\sim2.3$ subsamples with respect to one another (and with respect to $z\sim0$) is tied to sSFR, the implication is that the physical cause of the observed differences between samples is also correlated with the relative importance of recent star formation in galaxies at fixed M$_{\ast}$. This highlights the importance of the details of star formation history generally, as the high sSFRs seen in KBSS-MOSFIRE galaxies become increasingly difficult to sustain in older, massive galaxies and/or galaxies that have begun to deplete their gas supply.  Consequently, galaxies with the most extreme sSFRs are likely to have rising star formation histories or intense bursts of recent star formation, conditions that are seemingly commonplace at $z\simeq2-2.7$ \citep[e.g.,][]{reddy2012} but rare in the local universe.

\subsection{SED Shapes}
\label{sed_shapes}

\begin{figure}
\centering
\includegraphics{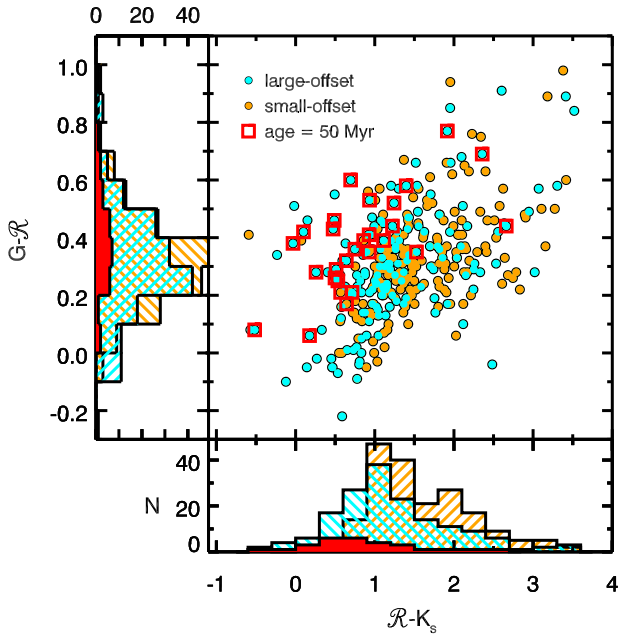}
\caption{The distribution of $\mathcal{R}-K_s$ and $G-\mathcal{R}$ color for the two subsamples of KBSS galaxies, flanked by 1D histograms to highlight the differences. The large-offset galaxies (cyan) have generally bluer colors than their small-offset counterparts (orange), with fewer small-offset galaxies exhibiting blue optical-NIR colors ($\mathcal{R}-K_s\leq1$). Galaxies with an imposed minimum age of 50~Myr (which are identified by red squares and are among the quartile of galaxies most offset from the $z\sim0$ locus) have the bluest $\mathcal{R}-K_s$ colors, which suggests that the maturity of the stellar population is inversely correlated with the degree of offset from SDSS galaxies in the N2-BPT diagram.}
\label{farselect_colors}
\end{figure}
\begin{figure}
\centering
\includegraphics{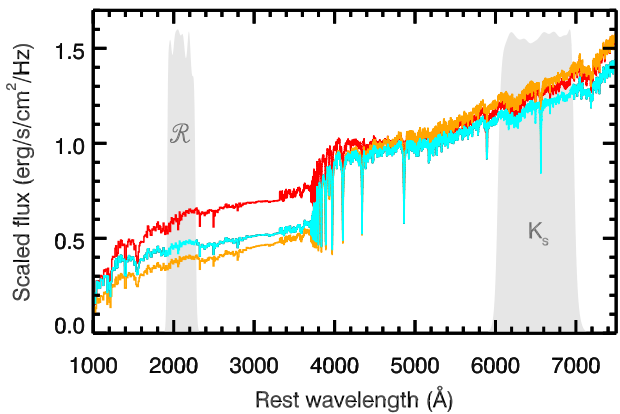}
\caption{Typical SEDs for subsamples of KBSS-MOSFIRE galaxies (small-offset galaxies in orange, large-offset galaxies in cyan, and minimum-age galaxies in red), showing how differences in global spectral shape correlate with the degree of offset in the N2-BPT plane.}
\label{farselect_seds}
\end{figure}

The KBSS-MOSFIRE subsamples also exhibit significant differences in the characteristic shape of their SEDs, with a larger fraction ($\sim40$\%) of large-offset galaxies appearing blue in the rest-optical ($(\mathcal{R}-K_s)_{\rm AB}\leq1$) than small-offset galaxies ($\sim19$\%). Figure~\ref{farselect_colors} presents the distributions of both $\mathcal{R}-K_s$ and $G-\mathcal{R}$ color for the two subsamples; despite a statistically significant difference in $\mathcal{R}-K_s$, the $G-\mathcal{R}$ distributions are nearly identical.

Because $\mathcal{R}-K_s$ color probes the depth of the Balmer and 4000\AA\, breaks, it traces both age and stellar mass \citep[e.g.,][]{shapley2005}, which is part of the motivation for the RK selection described in \S\ref{sample_text}.  The bluer colors of the large-offset galaxies are therefore consistent with their smaller overall M$_{\ast}$ (Figure~\ref{farselect_hists}). Additionally, although age estimates from SED fitting are not very robust in an absolute sense, 15\% of large-offset galaxies have SEDs that are best fit by a constant star formation history with an age of 50~Myr, compared to only 3\% of small-offset galaxies; as discussed in \S\ref{sfrs+masses}, this age is a minimum imposed to avoid unrealistically young inferred ages.  In total, 17/28 (61\%) ``minimum-age" galaxies (identified by red squares in Figure~\ref{o3hb_ssfr_plot} and Figure~\ref{farselect_colors}) fall above even the large-offset N2-BPT locus (cyan curve in Figure~\ref{farselect}) and are therefore among the quartile exhibiting the most dramatic offset with respect to SDSS galaxies. 

Figure~\ref{farselect_seds} shows the average SED model for small-offset (orange) and large-offset (cyan) galaxies alongside the SED model characteristic of minimum-age galaxies (red), illustrating both the similarities in shape in the rest-UV and the differences across the Balmer break and at rest-optical wavelengths. The SEDs have been constructed by averaging the reddened high-resolution best-fit stellar population synthesis models from \citet{bruzual2003}, after scaling each to match between $4500-4600$\AA; the continuum extinction for small- and large-offset galaxies is roughly similar, with $\langle E(B-V) \rangle_{\rm small}=0.18$ and $\langle E(B-V) \rangle_{\rm large}=0.16$.  The templates shown in Figure~\ref{farselect_seds} do not include nebular continuum emission, which will further diminish the stellar Balmer break.

The over-representation of minimum-age systems with blue $\mathcal{R}-K_s$ colors among the most offset galaxies suggests that selection techniques that rely on the presence of a significant Balmer break may not reliably identify such galaxies. Conversely, the fact that small- and large-offset KBSS-MOSFIRE galaxies show no obvious separation in their rest-UV colors---despite differences in their rest-optical colors and nebular emission line properties---suggests that the selection techniques described in \S\ref{sample_text} are not strongly biased in favor of large-offset galaxies; in fact, the RK selection described in \S\ref{sample_text} will prefer galaxies with larger Balmer breaks, like the small-offset galaxies.
 
 \subsection{Balmer Emission Line Luminosity}

\citet{juneau2014} showed that a Balmer emission line luminosity selection applied to SDSS galaxies could result in a sample that exhibits an N2-BPT offset of similar magnitude to that reported for $z\sim2$ galaxies. More recently, \citet{cowie2016} suggested that Balmer line luminosity may be an efficient way of identifying low-redshift analogs of typical $z\sim2$ galaxies. Given these results, the magnitude of the N2-BPT offset observed in KBSS-MOSFIRE could be artificially inflated if the sample were biased toward galaxies with the highest Balmer luminosities. It is interesting therefore to consider whether large-offset galaxies are overrepresented in KBSS-MOSFIRE due to such observational biases.

\begin{figure}
\centering
\includegraphics{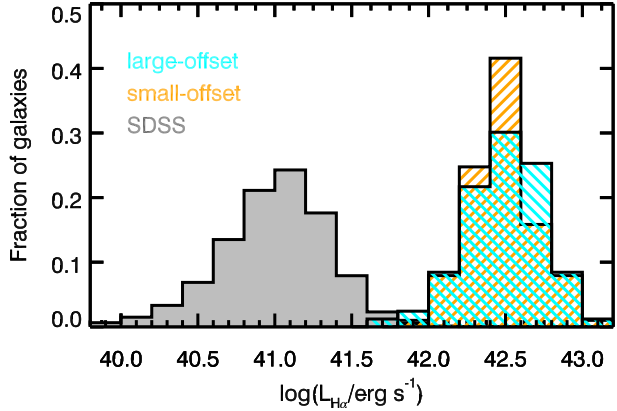}
\caption{Histograms of H$\alpha$ luminosity for SDSS galaxies and KBSS-MOSFIRE galaxies, divided on the basis of their N2-BPT offset into large- and small-offset subsamples. Although a Balmer luminosity selection in SDSS mimics the N2-BPT offset observed for $z\sim2.3$ galaxies, the same behavior is not observed for KBSS-MOSFIRE galaxies, with the two samples having statistically consistent distributions.}
\label{farselect_lum}
\end{figure}

Figure~\ref{farselect_lum} shows the distribution of observed H$\alpha$ luminosity, uncorrected for dust extinction, for the small- and large-offset KBSS-MOSFIRE subsamples, as well as that for the full SDSS comparison sample. For SDSS galaxies, $L_{\rm H\alpha}$ has been corrected for aperture effects using the ratio of the total inferred SFR to the SFR measured from the fiber. All KBSS-MOSFIRE galaxies exhibit significantly higher line luminosities than typical $z\sim0$ galaxies, but the $z\sim2.3$ subsamples are statistically consistent with being drawn from the same parent population as each other ($p=0.55$ from a two-sample Kolmogorov-Smirnoff test). Moreover, when the full KBSS-MOSFIRE sample is divided by H$\alpha$ luminosity, fits to the N2-BPT loci do not show a significant offset with respect to one another in either log([\ion{O}{3}]/H$\beta$) or log([\ion{N}{2}]/H$\alpha$). Together, these results suggest that observational biases do not significantly impact the degree of the observed N2-BPT offset and that H$\alpha$ luminosity is not strongly correlated with the location of $z\sim2.3$ galaxies in the N2-BPT plane. In fact, the observing strategy outlined in \S\ref{sample_text} mitigates this kind of incompleteness, as observations of individual galaxies are repeated until the strongest lines (including H$\alpha$) are significantly detected.

\subsection{Electron Density}
\label{density_text}

Finally, we consider differences between the large- and small-offset samples in terms of the electron density ($n_e$) in their \ion{H}{2} regions. \citet{sanders2016} offered a thorough discussion of electron density estimates for $z\sim2$ star-forming galaxies using observations from the MOSDEF survey, noting that electron densities inferred from both the [\ion{O}{2}]$\lambda\lambda3727,3729$ and [\ion{S}{2}]$\lambda\lambda6718,6732$ doublets are an order of magnitude higher for $z\sim2$ galaxies in their sample relative to a comparison sample from SDSS. At the same time, a number of studies have demonstrated that galaxies with higher electron densities ($n_e\sim10^2$~cm$^{-3}$) exhibit higher log([\ion{N}{2}]/H$\alpha)$ and log([\ion{O}{3}]/H$\beta$) than typical SDSS galaxies \citep[e.g.][]{brinchmann2008,liu2008,yuan2010}, often associated with higher inferred ionization parameters and differences in ISM pressure \citep{lehnert2009,kewley2013}. One may therefore expect large-offset KBSS-MOSFIRE galaxies to exhibit higher electron densities than small-offset galaxies.

\begin{figure}
\centering
\includegraphics{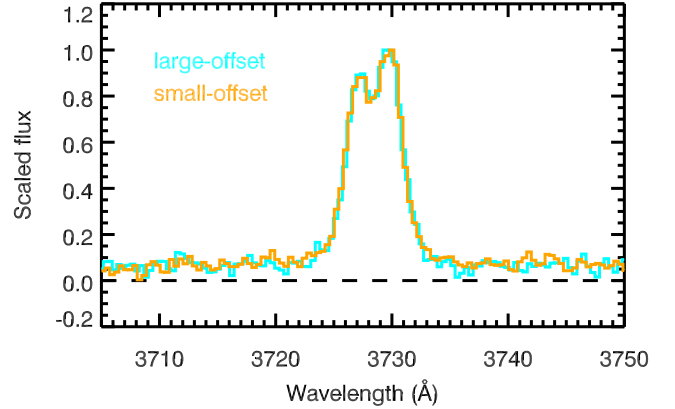}
\caption{The [\ion{O}{2}]$\lambda\lambda3727,3729$ doublet, commonly used to infer $n_e$ in the emitting gas, as observed in stacks of small- and large-offset KBSS-MOSFIRE galaxies. That the $n_e$ estimates for both composites are consistent with one another (values for each subsample are listed in the text) suggests that electron density---and, by extension, ISM pressure---do not play an important role in determining the location of $z\sim2.3$ galaxies in the N2-BPT plane.}
\label{ratios_paper_ne}
\end{figure}

To estimate $n_e$ for the KBSS-MOSFIRE subsamples, we rely on inferences from spectral stacks, in order to consistently account for individual galaxies with lower S/N observations of the density-sensitive [\ion{O}{2}] doublet. Galaxies from the small- and large-offset subsamples were corrected for slit losses then averaged separately, masking regions contaminated by OH emission lines. The spectral region near the [\ion{O}{2}] doublet is shown in Figure~\ref{ratios_paper_ne}, where the $J$ band stacks have been scaled by the peak flux in the [\ion{O}{2}]$\lambda$3729 line for ease of comparison.

For the small-offset composite, $I(3729)/I(3727)=1.13\pm0.03$, and for the large-offset composite, $I(3729)/I(3727)=1.14\pm0.04$. Using the diagnostic relation from \citet{sanders2016}, which assumes $T_e=10^4$~K, these values correspond to 
\begin{eqnarray}
n_{e,{\rm small}} &=& 281_{-39}^{+43}{\rm~cm}^{-3} \nonumber \\
n_{e,{\rm large}} &=& 267_{-43}^{+48}{\rm~cm}^{-3}. \nonumber
\end{eqnarray}
These values are consistent with one another within measurement uncertainties and also with the median value reported by \citet{sanders2016} for MOSDEF galaxies using the [\ion{O}{2}] diagnostic ($n_{e,{\rm MOSDEF}} = 225$~cm$^{-3}$).

The similarity between the electron density estimates for the two KBSS-MOSFIRE subsamples suggests that $n_e$ is not strongly correlated with the physical process(es) that determine the position of $z\sim2.3$ galaxies in the N2-BPT diagram, in contrast to observations of galaxies in the local universe.

\section{Nitrogen-to-oxygen Abundance Ratios}
\label{nitrogen_ratios}

Nitrogen is formed as part of the CNO cycle in intermediate mass and massive stars, both as a primary and secondary element. Primary nitrogen production is regulated by the C, N, and O created within stars formed out of relatively metal-poor gas, whereas secondary production is seeded by the C, N, and O initially present in stars (as the result of prior enrichment of the ISM). This double nature is reflected by the well-documented behavior of the N/O abundance ratio as a function of O/H for samples of galaxies and \ion{H}{2} regions \citep[e.g.,][]{diaz1986,vanzee1998}, where N/O remains at a constant value (log(N/O)$\approx-1.5$) for low values of 12+log(O/H) ($\lesssim8.0$), but rapidly increases with increasing oxygen abundance at higher values of O/H.

The precise relationship between N/O and O/H is a matter of some debate in the literature, depending sensitively on the choice of sample and method of measurement, making it difficult to place the N/O reported for individual sources or small samples in context. At $z\sim0$, many studies rely on measurements from the spectra of individual compact galaxies or local \ion{H}{2} regions \citep[e.g.,][]{perez-montero2009,pilyugin2012} or from composite spectra of relatively metal-poor galaxies \citep{andrews2013}, but the typical value of N/O at a given measured value of O/H can vary by up to 0.5~dex. As we discussed in \citetalias{steidel2014}, the differences between such studies may be dominated by differences in the method used to measure O/H. It has been argued separately that scatter in the N/O-O/H relation is due instead to the effects of pristine gas inflow \citep[e.g.,][]{koppen2005}, which could lower O/H while leaving N/O largely unaffected, particularly if much of the nitrogen is formed in intermediate mass stars that take longer to return their nucleosynthetic products to the ISM.

In \S\ref{farselect_offsets}, we briefly addressed the idea that the N2-BPT offset observed for $z\sim2.3$ galaxies could be due primarily to an enhancement in N/O relative to typical local galaxies with the same gas-phase oxygen abundance \citep[as proposed by, e.g.,][]{masters2014,shapley2015,cowie2016,sanders2016,masters2016}. If found to be present in high-redshift galaxies, such systematic deviations in N/O would have serious implications for calibrations linking nebular emission line ratios that include nitrogen lines (e.g., [\ion{N}{2}]$\lambda6585$) to oxygen abundance. Strong-line abundance calibrations rely on the implicit assumption that N/O varies similarly with O/H for both the calibration sample and the objects of interest, and many photoionization models likewise impose a prior on the value of N/O expected at a given O/H. 

From an astrophysical perspective, it is also a concern that there is presently no consensus regarding the physical process(es) that might result in enhanced N/O for the majority of high-redshift galaxies. Although \citet{masters2014} favored the existence of a large population of Wolf-Rayet stars as one possible explanation, the lifetimes of such stars are short, meaning that the effect would be limited to systems with extremely young ages \citep[as discussed by][]{shapley2015}. More recently, \cite{masters2016} have suggested that the existence of a redshift-invariant or slowly-evolving N/O-M$_{\ast}$ relation, when combined with the strong evolution toward lower O/H at fixed M$_{\ast}$ observed at $z>1$ \citep[e.g.,][]{erb2006metal,wuyts2014,steidel2014,sanders2015}, might more plausibly explain a systematic increase in N/O at fixed O/H for high-redshift galaxies.

The \citet{masters2016} study relied on a detailed analysis of $\sim100,000$ star-forming galaxies from SDSS-DR12 and compared the behavior of $z\sim0$ galaxies with the results from stacked spectra from the FMOS-COSMOS survey at $z\sim1.6$ reported by \citet{kashino2016}. However, to fairly evaluate the likelihood of the proposed explanations, observations of a statistical sample of individual $z\sim2$ galaxies are necessary. In this section, we present the first efforts to explicitly examine N/O for such a sample, using observations of the $z\sim2.3$ KBSS-MOSFIRE sample to investigate the behavior of N/O as a function of other measured galaxy properties.

\subsection{Strong-line Abundance Calibration using N2O2}
\label{pilyugin_text}

For ionized gas, the most reliable measurements of O/H and N/H---and thus N/O---are derived from the measurement of electron temperature ($T_e$), which is often described as the ``direct" method. Unfortunately, these measurements are extremely challenging even for galaxies in the local universe, as the auroral emission lines necessary for measuring $T_e$ of both singly- and doubly-ionized species are relatively weak compared to their nebular counterparts and become more difficult to measure for objects with increasing gas-phase abundance. The situation is even worse for distant objects, including star-forming galaxies at high redshift. 

The most practical alternative is to rely on strong-line indices known to exhibit tight correlations with intrinsic O/H and N/O. For oxygen abundance, two of the most widely-used indices are N2 and O3N2 (Table~\ref{strongline}). For N/O, N2O2 and N2S2 (also defined in Table~\ref{strongline}) are the most common choices. One of the most serious outstanding challenges to using strong-line calibrations for these indices is knowing whether or not they are reliable for sources that differ from those used to establish the original calibration between emission line ratio and metallicity. The existence of the strong N2-BPT offset cautions against the application of locally-calibrated relations for N2 and O3N2 at high-redshift, as they will inevitably return inconsistent estimates of 12+log(O/H) with respect to one another for objects that are offset from the $z\sim0$ locus. Studies of local objects have also shown that a factor of $\sim2-3$ enhancement in N/O at fixed O/H (measured using the direct method) is not recovered when both N/O and O/H are estimated using strong-line methods, because O/H is overestimated for such objects \citep{pilyugin2010}. Furthermore, as we previously discussed in \citetalias{steidel2014}, photoionization models suggest that many of the strong-line ratios are more strongly correlated with the details of the ionizing radiation field than gas-phase oxygen abundance, particularly for high-excitation nebulae. 

In contrast, the calibration for N/O using N2O2 relies primarily on the assumption that, due to the similarity in the ionization potentials for O and N, the relative ionization correction factors (ICFs) are also similar and, thus, log(N/O) $\simeq$ log(N$^+$/O$^+$). This makes N2O2 considerably less sensitive to changes in the ionization parameter and/or the shape of the ionizing radiation than O3N2 or N2. 

We employ a new calibration for N/O derived using a sample of extragalactic \ion{H}{2} regions compiled by \citet[][hereafter Pil12]{pilyugin2012}. The 414 sources described therein have direct method measurements of both nitrogen and oxygen abundance as well as measurements of [\ion{O}{2}]$\lambda\lambda 3727$,3729, H$\beta$, [\ion{O}{3}]$\lambda\lambda$4960,5008, [\ion{N}{2}]$\lambda\lambda 6549,6585$, H$\alpha$, and [\ion{S}{2}]$\lambda\lambda 6718$,6732. In most cases, the electron temperature was measured directly for only one ion (and thus only one temperature zone); since the authors adopt a two-zone model, the temperature in the second zone is determined using the relations from \citet{campbell1986} and \citet{garnett1992}.

\begin{figure}
\centering
\includegraphics{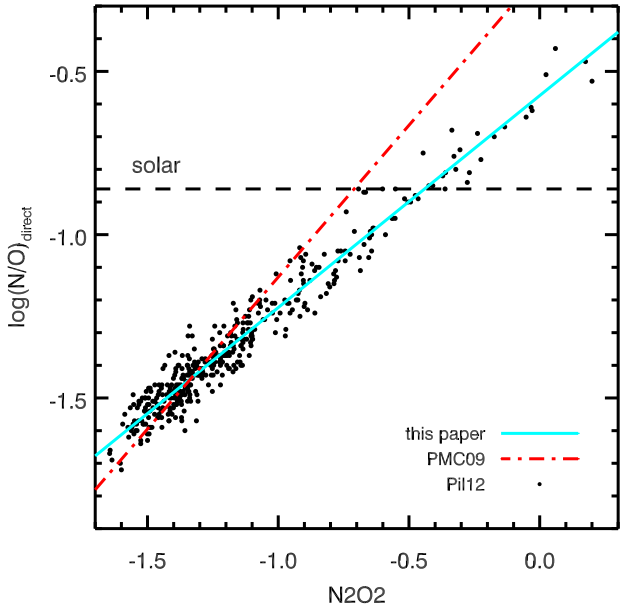}
\caption{The observed N2O2 index and $T_e$-based N/O measurements for the extragalactic \ion{H}{2} regions from \citet{pilyugin2012}. The best-fit linear relation, which is used in this paper, is shown in cyan. The relation from \citet{perez-montero2009} is shown by the dashed red line for comparison.}
\label{pilyugin_cals}
\end{figure}

Figure~\ref{pilyugin_cals} shows the N2O2 and N/O measurements for the \citetalias{pilyugin2012} sample and the corresponding calibration used in this paper (in cyan). For reference, the calibration from \citet[][PMC09]{perez-montero2009} is shown in red, highlighting the differences that can arise from the choice of calibration sample, particularly when both individual \ion{H}{2} regions and entire galaxies are considered, as is the case with \citet{perez-montero2009}. For the new calibration, we have fit to the entire range in N/O, since the spread in observed N2O2 is nearly identical for KBSS-MOSFIRE and the \citetalias{pilyugin2012} sample. The resulting relation is
\begin{equation}
\log{(\textrm{N/O})} = 0.65\times {\rm N2O2}-0.57,
\label{no_eq}
\end{equation}
with an accompanying RMS scatter about the best-fit relation of $\sigma_{\rm RMS} = 0.05~{\rm dex}$. Adopting the \citetalias{perez-montero2009} calibration instead of Equation~\ref{no_eq} would imply values of log(N/O) larger by up to $\sim0.3$~dex.

The use of N2S2 as an alternative to N2O2 presents many practical advantages, as it requires observations in only one NIR band, and thus needs fewer observations, no cross-band calibration, and is much less sensitive to relative extinction due to dust. However, the mapping of N2S2 to N/O relies on the expectation that S/O remains constant and, therefore, that sulfur may be used as a proxy for oxygen. Because the ionization potentials of O (13.62~eV) and S (10.4~eV) differ significantly, though, [\ion{S}{2}]-emitting gas may not be entirely spatially coincident with the region where the [\ion{O}{2}] emission originates and could lie well outside of the H-ionizing region. Thus, despite the advantages of requiring observations in only one NIR window and no extinction correction, we rely exclusively on the N2O2 index in this paper.

\subsection{Correlation with Galaxy Properties}
\label{no_text}

We infer N/O for 151 galaxies in the $z\sim2.3$ KBSS-MOSFIRE sample using Equation~\ref{no_eq}. Galaxies were included in the sample if they had a $\geq~3\sigma$ detection of the combined [\ion{O}{2}]$\lambda\lambda3726,3729$ doublet and H$\beta$, a $\geq5\sigma$ detection of H$\alpha$, and a robust measurement of the Balmer decrement (as described in \S\ref{sfrs+masses}); objects with $<2\sigma$ detections of [\ion{N}{2}] were assigned $2\sigma$ upper limits.

\begin{figure}
\centering
\includegraphics{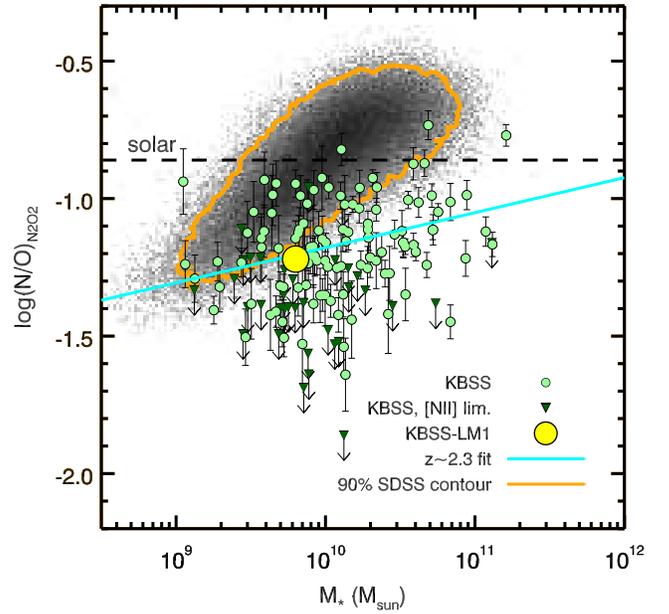}
\caption{N/O measurements for KBSS-MOSFIRE galaxies (excluding AGN) as a function of M$_{\ast}$, plotted alongside the $z\sim0$ SDSS sample (with the orange contour enclosing 90\% of the local galaxies). KBSS-MOSFIRE galaxies exhibit significantly lower N/O at fixed M$_{\ast}$ relative to SDSS galaxies, which show a strong positive correlation between both parameters. The more moderate positive correlation observed for the $z\sim2.3$ KBSS-MOSFIRE galaxies can be approximated by a linear relation (Equation~\ref{no_mstar_eq}), which is shown in cyan. The location of the \citet{steidel2016} composite spectrum in this parameter space is represented by the large yellow point, showing that it is fully consistent with the $z\sim2.3$ relation.}
\label{no_mstar}
\end{figure}

Figure~\ref{no_mstar} shows the N/O-M$_{\ast}$ relation for $z\sim2.3$ KBSS-MOSFIRE galaxies compared with $z\sim0$ SDSS galaxies. As for local galaxies, the KBSS-MOSFIRE sample exhibits a positive correlation between N/O and M$_{\ast}$ (with Spearman correlation coefficient $\rho=0.24$ and $p=0.01$). The best-fit linear relation describing the sample is
\begin{equation}
\log({\rm N/O)} = -1.18+0.13\times[\log(M_{\ast}/M_{\odot})-10],
\label{no_mstar_eq}
\end{equation}
which has $\sigma_{\rm RMS}=0.17$~dex.

The $z\sim2.3$ locus is noticeably offset relative to typical SDSS galaxies (enclosed by the orange contour) at fixed M$_{\ast}$. At low stellar masses, the difference between the two samples is small, but becomes larger with increasing M$_{\ast}$, reaching an offset of $\sim0.5$~dex toward lower log(N/O) at M$_{\ast}\sim10^{11}$M$_{\odot}$; if the $z\sim0$ locus is also approximated by a linear relation, the average separation between $10^9-10^{11}$M$_{\odot}$ is $\sim0.32$~dex in log(N/O). This is a much larger shift than that observed (albeit using N2S2) for the $z\sim1.6$ FMOS-COSMOS stacks from \citet{kashino2016}. Moreover, those authors reported the largest differences relative to SDSS at low stellar masses (M$_{\ast}\sim10^{9.5}$), the opposite of the trend seen here for KBSS-MOSFIRE galaxies.

\begin{figure}
\centering
\includegraphics{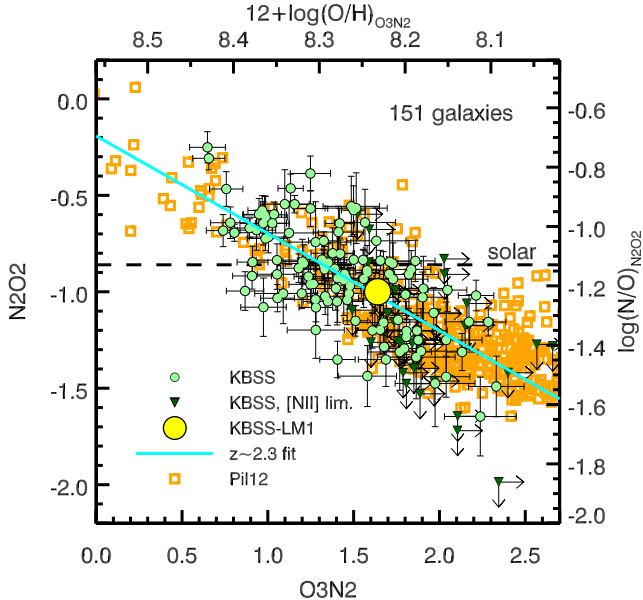}
\caption{N2O2 as a function of O3N2 for both $z\sim2.3$ KBSS-MOSFIRE galaxies (green points, with $2\sigma$ upper limits in [\ion{N}{2}] shown as dark green triangles) and extragalactic \ion{H}{2} regions from \citetalias{pilyugin2012} (orange squares). Both samples behave similarly and exhibit comparable degrees of scatter in this observed line ratio space. For comparison, the location of the composite spectrum of KBSS galaxies from \citetalias{steidel2016}, which shows no enhancement in N/O at fixed O/H relative to the \citetalias{pilyugin2012} sample, is shown by the large yellow point. The best-fit linear relation for the full KBSS-MOSFIRE sample is shown in cyan; if interpreted as a N/O-O/H relation using log(N/O) derived from N2O2 (see right axis, Equation~\ref{no_eq}) and estimates of 12+log(O/H) from O3N2 (see top axis, Equation~\ref{oh_calib}), this relation is fully consistent with the $T_e$-based N/O-O/H relation observed for the \citet{pilyugin2012} sample.}
\label{no_o3n2}
\end{figure}

Clearly, the N/O-M$_{\ast}$ relation evolves significantly between $z\sim0$ and $z\sim2.3$. However, high-$z$ galaxies would still exhibit higher values of N/O at fixed O/H relative to $z\sim0$ galaxies if the N/O-M$_{\ast}$ relation evolves more slowly than the O/H M$_{\ast}$-metallicity relation (MZR). Although it remains difficult to measure O/H independently of N for many individual high-$z$ galaxies, there is some evidence to suggest that the redshift evolution of the N/O-M$_{\ast}$ relation and the MZR are comparable: \citetalias{steidel2014} showed that oxygen abundances measured using the O3N2 index exhibit the least bias and scatter relative to the direct method for individual KBSS-MOSFIRE galaxies with $T_e$-based measurements and that when O3N2 is used to estimate O/H, $z\sim2.3$ galaxies exhibit a 0.32~dex offset toward lower 12+log(O/H) at fixed M$_{\ast}$ relative to SDSS. Furthermore, in \citetalias{steidel2016}, we reported a direct, $T_e$-based O/H abundance for a stack of KBSS galaxies and N/O independently inferred using both photoionization modeling and the strong-line calibration from Equation~\ref{no_eq} of this paper. These measurements show that the KBSS stack, which is representative of the full KBSS-MOSFIRE sample discussed here (c.f. the yellow point in Figure~\ref{no_mstar}), is consistent with the N/O-O/H relation observed for the sample of \ion{H}{2} regions from \citetalias{pilyugin2012} (see Figure~16 of \citealt{steidel2016}).

Taken together, these results suggest that the N/O-O/H relation is redshift-invariant. Confirming the universality of the N/O-O/H relation must await independent measurements of N/O and O/H for a large sample, and we discuss the possibility of obtaining such measurements for KBSS-MOSFIRE galaxies in \S\ref{oh_constraints}. For now, we consider the behavior of N2O2 as a function of O3N2 for the KBSS-MOSFIRE galaxies (green) and the \citetalias{pilyugin2012} sample (orange) in Figure~\ref{no_o3n2}. The best-fit linear relation describing the KBSS-MOSFIRE locus (cyan line) also traces the behavior of the \citetalias{pilyugin2012} sample in terms of observed line ratios. Indeed, there is virtually no separation in either N2O2 or O3N2 between the samples, except for ${\rm O3N2}>2.5$, where there are no KBSS-MOSFIRE galaxies. The reason for the lack of $z\sim2.3$ galaxies in that region of parameter space is easily understood as a selection effect in the KBSS-MOSFIRE sample; high values of O3N2 correspond to low values of 12+log(O/H), and at such low metallicities, the strength of both oxygen and nitrogen emission lines begins to decline below practical detection limits. 

We can establish a calibration mapping O/H to O3N2 for the \citetalias{pilyugin2012} sample as we did for N/O and N2O2 in \S\ref{pilyugin_text}, resulting in the following linear relation\footnote{We include only \ion{H}{2} regions with 12+log(O/H)$>8.0$ in determining the calibration.}:
\begin{equation}
12+\log{(\textrm{O/H})} = 8.56-0.20\times {\rm O3N2};~\sigma_{\rm RMS} = 0.08~{\rm dex}.
\label{oh_calib}
\end{equation}
Such a calibration will disguise a potential enhancement in N/O at fixed O/H, but if we assume that the N/O-O/H relation is indeed the same at all redshifts, the linear fit to the $z\sim2.3$ KBSS-MOSFIRE locus in Figure~\ref{no_o3n2} can then be described as
\begin{equation}
\log{(\textrm{N/O})} = -1.29+1.64\times {\rm [(12+log(O/H)_{O3N2})-8.2]}.
\label{kbss_no_oh}
\end{equation}
This relation is slightly steeper but statistically consistent with a linear relation determined using only $T_e$-based measurements of N/O and O/H for the \citetalias{pilyugin2012} sample. Because of this similarity, we use Equation~\ref{kbss_no_oh} in Section~\ref{model_bpt}, but note that the results presented there would be the same if we used a fit to the $T_e$-based measurements from \citetalias{pilyugin2012}.

\subsection{N/O at Fixed Ionization and Excitation}
\label{twin_text}

We have already discussed the likelihood that SDSS galaxies occupying the same region of parameter space in the O32-R23 plane possess similar ionization and excitation properties as neighboring KBSS-MOSFIRE galaxies. This claim was based on the nature of the O32 and R23 indices: sources with maximal R23 are expected to have similar (moderate) oxygen abundances, so that O32 independently tracks ionization parameter, with higher values of O32 corresponding to higher degrees of ionization. The value of R23 at fixed $U$ and O/H is likewise set by the shape of the ionizing spectrum, as we reasoned in \S\ref{o32_r23_text1} and show explicitly in \S\ref{models} below.

However, given the obvious discrepancy between typical values of O32 and R23 for $z\sim0$ and $z\sim2.3$ galaxies (Figure~\ref{o32r23_plots}) and the similarly dramatic separation 
in [\ion{O}{3}]/H$\beta$ (Figures~\ref{mex_plots} and \ref{o3hb_ssfr_plot}) it is necessary to test whether differences in the shape and normalization of the ionizing spectrum account for all (or most) of the differences observed between the nebular properties of $z\sim0$ and $z\sim2.3$ galaxies, including the N2-BPT offset. To make a fair comparison between KBSS-MOSFIRE and SDSS, it is imperative to select a sample of SDSS galaxies in a way that matches the high-redshift distribution of these quantities. Because high-ionization, high-excitation galaxies are common at $z\sim2.3$ yet relatively rare at $z\sim0$, imposing a simple cut in both line ratios will sample only the tail of local galaxies and lead to significantly different distributions in O32 and R23 (and, therefore, ionization and excitation) for the low- and high-redshift samples.

\begin{figure}
\centering
\includegraphics{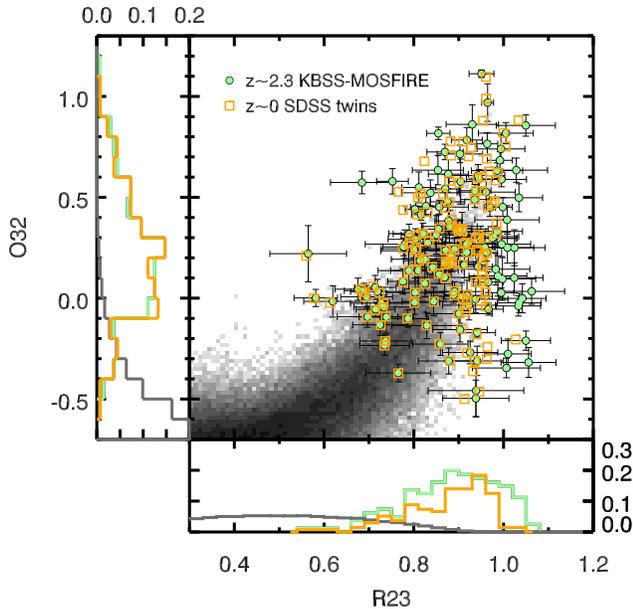}
\caption{The distribution of $z\sim2.3$ KBSS-MOSFIRE galaxies and $z\sim0$ SDSS twins in the O32-R23 plane. Each SDSS comparison galaxy (shown by the open orange squares) is chosen based on proximity to a KBSS-MOSFIRE galaxy (represented by the green points with error bars) in both parameters. This selection technique ensures that the distributions of R23 and O32 are statistically consistent for the two samples, as can be seen in the flanking histograms of both line ratios. For comparison, the distributions of O32 and R23 for the full SDSS sample are shown in grey.}
\label{twin_select_figure}
\end{figure}

\begin{figure*}
\centering
\includegraphics{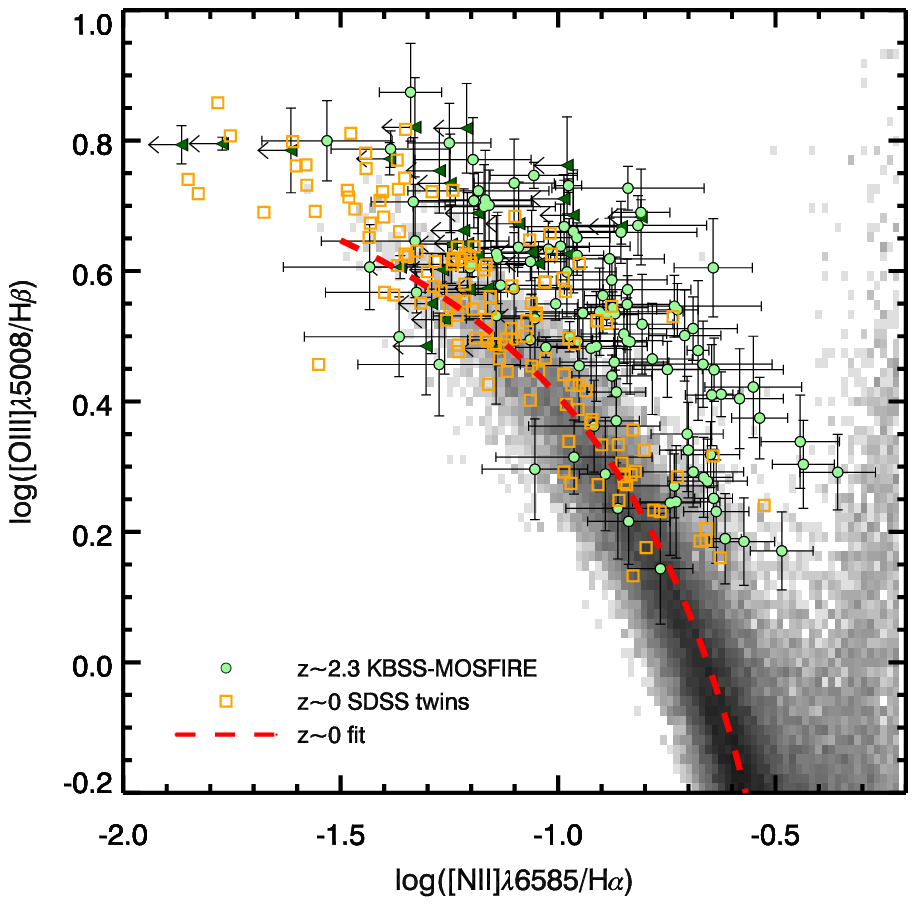}
\hspace{-0.8in}
\includegraphics{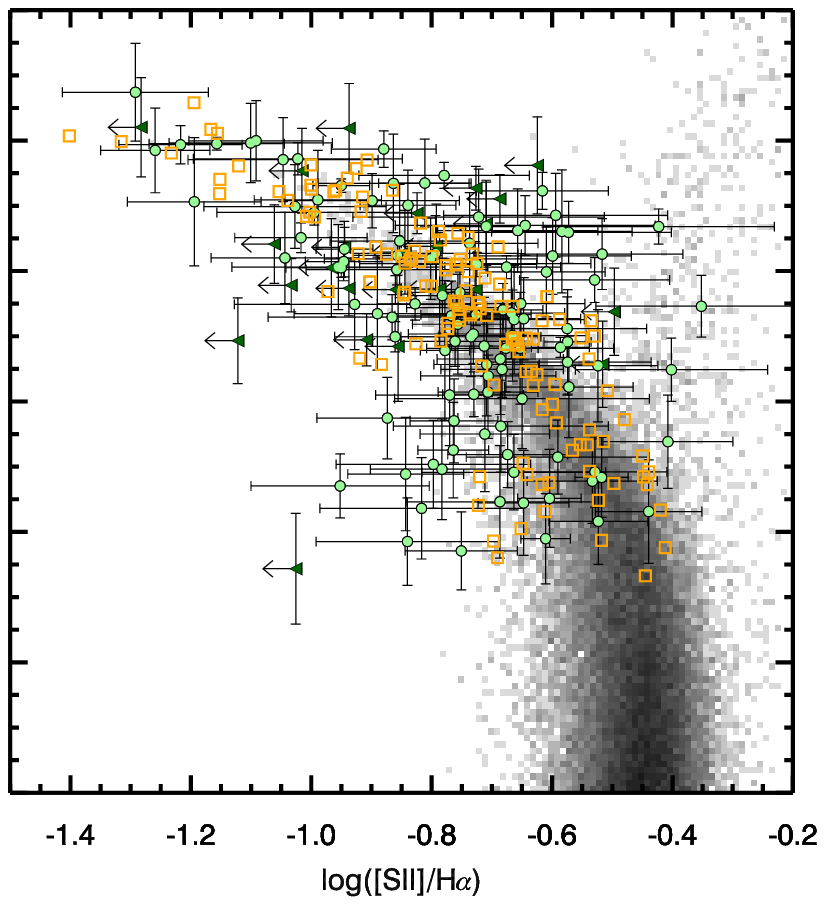} \\
\includegraphics{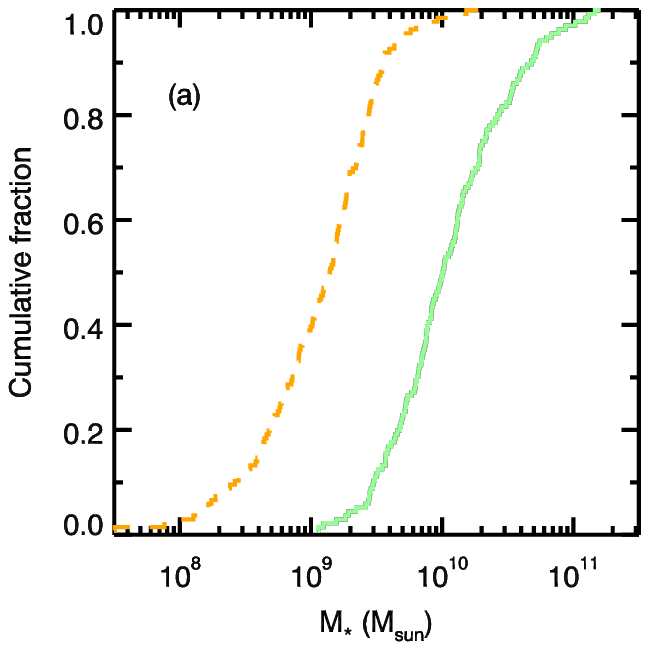}
\hspace{-0.8in}
\includegraphics{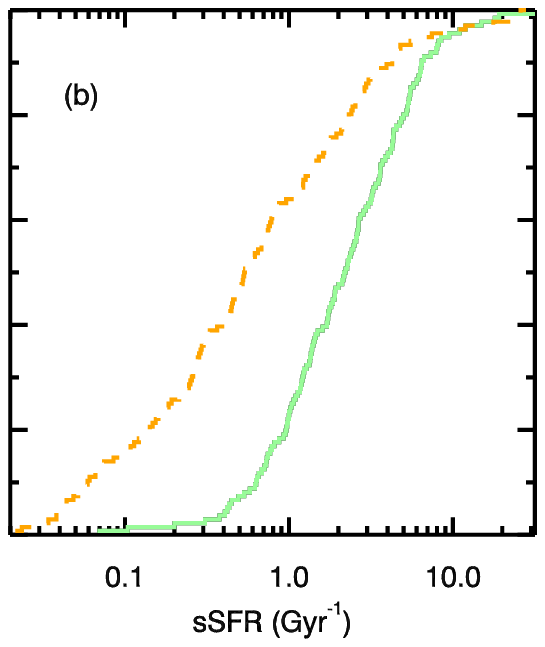}
\hspace{-0.8in}
\includegraphics{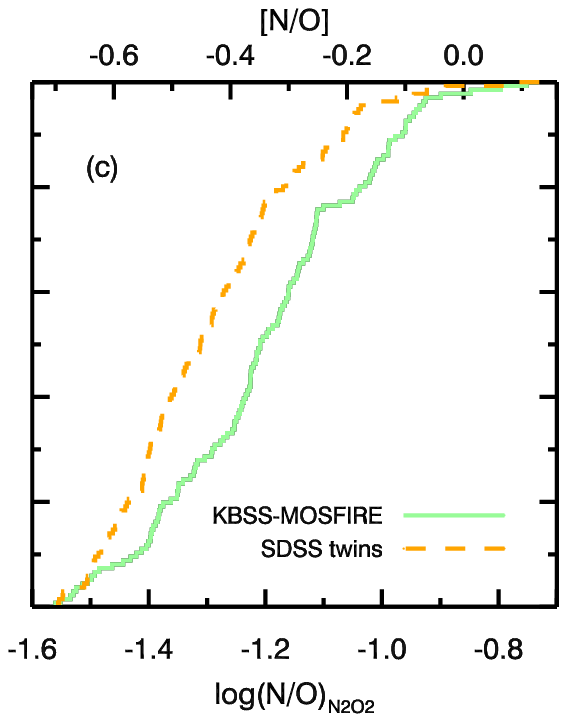}
\caption{A comparison of the nebular, physical, and chemical properties of $z\sim2.3$ KBSS-MOSFIRE galaxies and their $z\sim0$ SDSS twins matched in O32 and R23, as described in the text and shown in Figure~\ref{twin_select_figure}. The upper row illustrates the differences in the N2- and S2-BPT diagrams, which are consistent with the results described in \S\ref{bpt_text}: in the N2-BPT plane, there is a strong offset observed between KBSS-MOSFIRE galaxies (green points with error bars, 2$\sigma$ upper limits in [\ion{N}{2}] drawn as dark green triangles) and the SDSS twins (open orange squares), with no significant separation in the S2-BPT plane (here the dark green triangles represent 2$\sigma$ upper limits in [\ion{S}{2}]). The bottom panels show differences in the physical properties of KBSS-MOSFIRE galaxies (solid green curves) and SDSS twins (dashed orange curves); from left to right: M$_{\ast}$, sSFR, and N/O determined from N2O2 (Equation~\ref{no_eq}). The $\sim0.10$~dex higher values of log(N/O) observed in KBSS-MOSFIRE galaxies fall short of accounting for the horizontal displacement of $z\sim2.3$ galaxies relative to SDSS in Figure~\ref{n2_bpt} and are much lower than would be expected if the N/O-M$_{\ast}$ relation were redshift-invariant.}
\label{twin_result_figure}
\end{figure*}

Instead, we identify a SDSS ``twin" for each of the 150 $z\sim2.3$ KBSS-MOSFIRE galaxies with $>5\sigma$ measurements of H$\alpha$, $>3\sigma$ measurements of H$\beta$, [\ion{O}{3}], and [\ion{O}{2}], and $>5\sigma$ significance measurements of the Balmer decrement (see Figure~\ref{twin_select_figure}). For each KBSS-MOSFIRE galaxy (identified by green points with error bars), the closest SDSS galaxy in the O32-R23 diagram within 2$\sigma$ measurement uncertainties is selected (represented by open orange squares). To ensure a unique comparison sample, the next closest SDSS galaxy is chosen in cases where a single SDSS galaxy is the closest ``twin" for more than one KBSS-MOSFIRE galaxy. There are 17 KBSS-MOSFIRE galaxies without a SDSS galaxy within 2$\sigma$ measurement uncertainties, due largely to the absence of SDSS galaxies with very large values of R23 at fixed O32; KBSS-MOSFIRE galaxies without SDSS twins were removed from the sample described here and excluded from the analysis in this section. The resulting distributions of O32 and R23 for the KBSS-MOSFIRE sample and the SDSS twins are shown in the flanking panels of Figure~\ref{twin_select_figure} and are statistically consistent with being drawn from the same parent distributions of these quantities. For comparison, the distributions of O32 and R23 for the full SDSS sample are shown in grey.

We consider the nebular and chemical properties of KBSS-MOSFIRE galaxies and the SDSS twins in Figure~\ref{twin_result_figure}. The upper panels show the distribution of both samples in the N2-BPT and S2-BPT planes. Even for galaxies matched in O32 and R23, there is a large offset between the KBSS-MOSFIRE galaxies and their SDSS twins in the N2-BPT plane whereas there is no apparent separation in the S2-BPT plane, consistent with the behavior observed for the full samples of $z\sim0$ and $z\sim2.3$ galaxies presented in \S\ref{bpt_text}. Notably, the SDSS twins fall largely above the ridge-line of the full SDSS sample (red dashed line), such that the offset between the orange squares and green points in the upper left panel of Figure~\ref{twin_result_figure} is almost entirely horizontal. The median horizontal offset of KBSS-MOSFIRE galaxies relative to their SDSS twins is equivalent to a $0.18$~dex shift toward higher log([\ion{N}{2}]/H$\alpha$), whereas the median $\Delta \log \textrm{[\ion{O}{3}]/H$\beta$} < 0.01$~dex. This result is largely in agreement with \citet{shapley2015}, who reported an offset in the N2-BPT plane for low-mass (M$_{\ast}<10^{10.11}$M$_{\odot}$) MOSDEF galaxies relative to SDSS galaxies matched in O32 and R23. Contrary to the MOSDEF results, however, both low- and high-mass KBSS-MOSFIRE galaxies are offset with respect to their SDSS twins, with median offsets in log([\ion{N}{2}]/H$\alpha$) of 0.17~dex for M$_{\ast}<10^{10.11}$M$_{\odot}$ and 0.19~dex for M$_{\ast}\geq10^{10.11}$M$_{\odot}$.

KBSS-MOSFIRE galaxies are also $\sim10$ times more massive than SDSS galaxies matched in O32 and R23 (see panel (a) in the bottom row of Figure~\ref{twin_result_figure}). While SDSS twins have higher sSFRs than the majority of SDSS galaxies (Figure~\ref{farselect_hists}), panel (b) shows that KBSS-MOSFIRE galaxies still have $0.58$~dex higher sSFRs on average. Panel (c) in the bottom row of Figure~\ref{twin_result_figure} presents the distribution of N/O for both samples, showing that the $z\sim2.3$ KBSS-MOSFIRE galaxies have higher N/O than their SDSS twins ($p=1.1\times10^{-5}$ that they are drawn from the same distribution), with a median difference of 0.10~dex. These results are broadly consistent with the trends observed for all SDSS galaxies \citep{masters2016}; however, based on the difference in M$_{\ast}$ observed between KBSS-MOSFIRE galaxies and the SDSS twins, the $z\sim2.3$ galaxies should exhibit N/O close to solar (log(N/O)$=-0.86$, [N/O]$=0.0$) if the N/O-M$_{\ast}$ is redshift-invariant as the authors propose. Additionally, the observed difference in N/O falls markedly short of the 0.37~dex enhancement in N/O that would be required to explain the separation between the $z\sim0$ and $z\sim2.3$ N2-BPT loci in Figure~\ref{n2_bpt} if the difference were entirely in log([\ion{N}{2}]/H$\alpha$).

Although we do not report O/H measurements for individual KBSS-MOSFIRE galaxies here, we note that 0.10~dex higher N/O would correspond to only a 0.06~dex difference in O/H if $z\sim2.3$ galaxies follow the N/O-O/H relation in Equation~\ref{kbss_no_oh}. As we will show in the next section, such a difference in O/H would not be readily apparent in S2-BPT and O32-R23 diagnostic diagrams, as those nebular line ratios are relatively insensitive to changes in O/H in high-excitation nebulae, particularly at moderate gas-phase metallicities.

\section{Photoionization Models}
\label{models}

Many inferences regarding the properties of high-$z$ \ion{H}{2} regions can be made by comparing observations of their nebular spectra with carefully-chosen samples of analogous objects, usually low-redshift sources whose characteristic physical conditions are more easily established due to the availability of higher S/N data and/or large sample sizes; we employed this technique in \S\ref{nitrogen_ratios} to infer the distribution of N/O in the $z\sim2.3$ KBSS-MOSFIRE sample. Still, there remain limitations to using calibrations based on comparison samples that may differ systematically in some key parameter---including the inability to construct suitable calibrations for characteristics that depend on several correlated variables, like ionization parameter ($U$).

Photoionization models predict the nebular spectrum produced by a specific combination of input ionizing spectrum, ionization parameter, chemical abundance pattern, and physical conditions in the emitting gas (e.g., electron density). A common shortcoming of photoionization model comparisons, however, is that the large number of free parameters makes it difficult to infer physically meaningful results from the agreement between a given model grid and the data. Ideally, some of the inputs to the photoionization model can be fixed using independent measurements, leaving only a small number of parameters to vary; in this case, the agreement between a specific set of photoionization models and the data constrains the allowed range of those parameters.

One of the strengths of conducting a survey of $z\simeq2-3$ galaxies is the ability to simultaneously obtain spectra in the rest-UV and rest-optical wavebands. Together, observations of both spectral windows provide probes of the massive stars and the surrounding gas that is being ionized and heated by radiation from the same stars. In \citetalias{steidel2016}, we described the results from a campaign of deep rest-UV spectroscopic follow-up of KBSS galaxies with MOSFIRE spectroscopy, specifically focusing on demonstrating the ability to account simultaneously for the observed stellar FUV spectrum and the UV-optical nebular spectrum using a set of physically-motivated photoionization models. The models used here represent a small subset of those described in that paper, where we discussed them in detail; for completeness, we briefly summarize the salient features.

We use Cloudy \citep[v13.02,][]{ferland2013} to predict the nebular spectrum of the irradiated gas given an incident radiation field with a fixed spectral shape. The models assume a plane-parallel geometry where the intensity of the radiation field is parametrized by the ionization parameter $U$($\equiv n_\gamma/n_H$). For ionized gas, $n_H$ is roughly equivalent to the electron density ($n_e$) and is assumed to be 300 cm$^{-3}$ based on measurements of the density-sensitive [\ion{O}{2}] and [\ion{S}{2}] doublets from the rest-optical composite spectrum of 30 KBSS-MOSFIRE galaxies \citepalias{steidel2016}. This value is also consistent with the densities reported in \citetalias{steidel2014} for a early sample of KBSS-MOSFIRE galaxies and with the stacked spectra of the small-offset and large-offset subsamples presented in \S\ref{density_text}. Because $n_H$ is fixed, $U$ may be interpreted as the normalization of the ionizing radiation field.

Our earlier work \citepalias{steidel2014} relied on similar models, but eschewed a specific choice of spectral synthesis model, instead using blackbodies to parametrize the shape of the input ionizing spectrum. Here, we employ stellar population models from the most recent version of ``Binary Population and Spectral Synthesis" \citep[BPASSv2;][Eldridge et al. in prep.]{stanway2016}. The most important aspect of BPASSv2 relative to other stellar population synthesis codes is the inclusion of interacting massive binary stars, which has the net effect of boosting the overall ionizing flux and producing a significantly harder ionizing spectrum for models with continuous star formation histories, particularly at low stellar metallicity. We adopt the default IMF for BPASSv2, with an IMF slope of $-2.35$ over the range $0.5\leq{\rm M_{\ast}/M_{\odot}}\leq300$.\footnote{We discuss the effects of restricting the BPASSv2 IMF to a maximum of 100~M$_{\odot}$ in \citetalias{steidel2016}, which do not impact any of the results of this paper.}

The metallicity of the gas ($Z_{\rm neb}$) is allowed to vary independently and is not required to match the stellar metallicity ($Z_{\ast}$) of the input population synthesis model, where $Z_{\ast}$ is the fraction of metals by mass and $Z_{\odot}=0.0142$ \citep{asplund2009}. Although the metallicity of the gas near young, massive stars should closely trace the stellar metallicity, decoupling $Z_{\rm neb}$ from $Z_{\ast}$ reflects the understanding that high-redshift galaxies may not exhibit solar abundance ratios, particularly between elements synthesized by stars of different masses (as is the case for O/Fe).  We specifically highlight O/Fe because the cooling of hot, ionized gas (and thus many features in the nebular spectrum) is regulated largely by the abundance of O, whereas the shape of the ionizing spectrum is instead determined by the abundance of Fe, which accounts for much of the total opacity in stellar atmospheres and, in turn, governs mass loss rates. As we discussed in \citetalias{steidel2016}, this means that since $Z_{\ast}$ effectively traces the Fe abundance in stars, a combination of low $Z_{\ast}$ and moderate $Z_{\rm neb}$ in high-$z$ galaxies does not require the stellar O/H to differ from O/H in the gas---because O/Fe may be enhanced relative to (O/Fe)$_{\odot}$ by up to a factor of $\sim5.5$ \citep{nomoto2006}. 

In the case of the composite spectrum of 30 KBSS galaxies from \citetalias{steidel2016}, we demonstrated that the UV-optical nebular properties were consistent with $Z_{\rm neb}/Z_{\odot}=0.5$, while the shape of the rest-UV spectrum was globally best fit by population synthesis models with $Z_{\ast}/Z_{\odot}\approx0.1$. Together, these results imply O/Fe$\simeq4-5$(O/Fe)$_{\odot}$, consistent with the \citet{nomoto2006} yields for low-$Z_{\star}$ core-collapse SNe. Since we are now interested in comparing the model predictions with observations of the \textit{ensemble} of $z\sim2.3$ KBSS-MOSFIRE galaxies, we consider two separate models with $Z_{\ast}/Z_{\odot}=0.07$ (BPASSv2-z001) and $Z_{\ast}/Z_{\odot}=0.28$ (BPASSv2-z004) in order to account for a range in the Fe content of massive stars in high-$z$ galaxies. As we will show in \S\ref{o32-r23_models}, significantly higher values of $Z_{\ast}(\gtrsim0.5$~Z$_{\odot}$) can be ruled out for the majority of $z\sim2.3$ galaxies by comparison with their nebular spectra alone.

 \subsection{Ionization Parameter}
 \label{ionization_text}

When the input ionizing spectrum is fixed, the photoionization model predictions are restricted to a surface in line ratio space, with each point on the surface defined by a unique combination of $U$ and $Z_{\rm neb}$. The space occupied by the photoionization model surface can be compared with line ratio measurements from the full $z\sim2.3$ KBSS-MOSFIRE sample to understand the allowed range in both parameters and also how they are correlated with one another across the galaxy population.

Inconveniently, the photoionization model surface ``folds over" in many 2D line ratio spaces, requiring constraints from many individual line ratios to break the degeneracy between combinations of $U$ and $Z_{\rm neb}$ that result in at least some of the same observed line ratios. Fortunately, O32 and Ne3O2 (see Table~\ref{strongline}) are substantially more sensitive to changes in $U$ than changes in $Z_{\rm neb}$, and we use measurements of these line ratios to independently constrain the likely values of ionization parameter.

\begin{figure}
\centering
\includegraphics{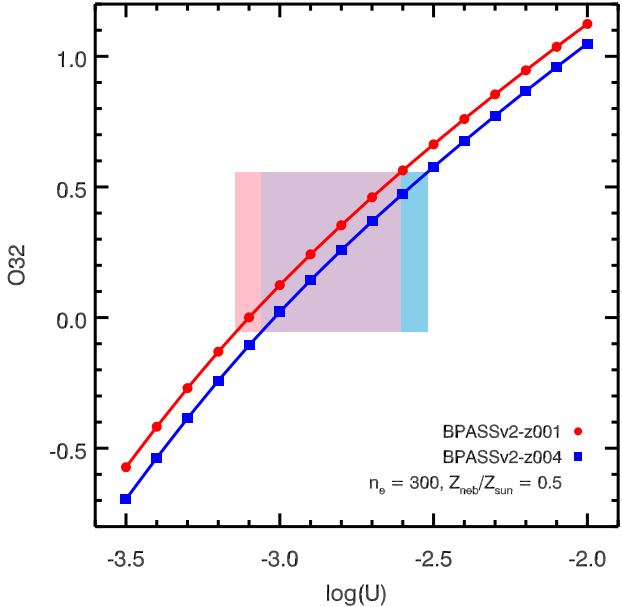}
\caption{O32 as a function of log($U$) for BPASSv2 models with $Z_{\ast}/Z_{\odot}=0.07$ (red points) and $Z_{\ast}/Z_{\odot}=0.28$ (blue squares). The metallicity of the gas is assumed to be $Z_{\rm neb}/Z_{\odot}=0.5$ in both cases. The vertical extent of the shaded region denotes the 68\% HDI in O32 for the KBSS-MOSFIRE sample; the width of the shaded region shows the corresponding intervals in log($U$) for the models (values provided in the text).}
\label{o32_probe}
\end{figure}

Figure~\ref{o32_probe} illustrates how the predicted value of the O32 index varies as a function of log($U$), assuming $Z_{\ast}/Z_{\odot}=0.07$ (red points) or $Z_{\ast}/Z_{\odot}=0.28$ (blue squares); the nebular oxygen abundance is assumed to be $Z_{\rm neb}/Z_{\odot}=0.5$ in both cases. The 68\% highest density interval (HDI, the narrowest range that includes 68\% of the sample) for O32 in the KBSS-MOSFIRE sample is indicated by the vertical extent of the shaded region, with a median ${\rm O32}=0.24$. Using the predictions from the models, the corresponding ranges in log($U$) needed to reproduce the observed range in O32 are
\begin{eqnarray}
\log(U)_{\rm z001,68\%}&=&[-3.14,-2.61] \nonumber \\
\log(U)_{\rm z004,68\%}&=&[-3.06,-2.52]. \nonumber
\end{eqnarray}
Note that the inferences made assuming $Z_{\ast}/Z_{\odot}=0.28$ are shifted toward higher log($U$) by $\sim0.08$~dex. This trend extends to higher $Z_{\ast}$ as well, such that higher values of $U$ are required to produce the same value of O32 as $Z_{\ast}$ increases. This trend---highlighting the trade-off between ionization parameter and hardness of the ionizing radiation field---is discussed more generally by \citet{sanders2016}.

The Ne3O2 index serves as a powerful cross-check on ionization parameters determined using O32, as it is significantly less affected by differential extinction due to dust, does not require any cross-band calibration, and provides an additional constraint on the shape of incident ionizing radiation. As with O32, however, the translation from Ne3O2 to $U$ depends sensitively on the details of the photoionization model, particularly on the choice of ionizing spectrum. In brief, because the ionization potential of Ne$^+$ (40.96~eV) is larger than that of O$^+$ (35.12~eV), which is the probe of high-ionization emission in the O32 index, the Ne3O2 index responds to changes in the shape of the EUV stellar spectrum at higher energies than O32. Thus, Ne3O2 is predicted to increase relative to O32 with increasing hardness of the ionizing spectrum.

\begin{figure}
\centering
\includegraphics{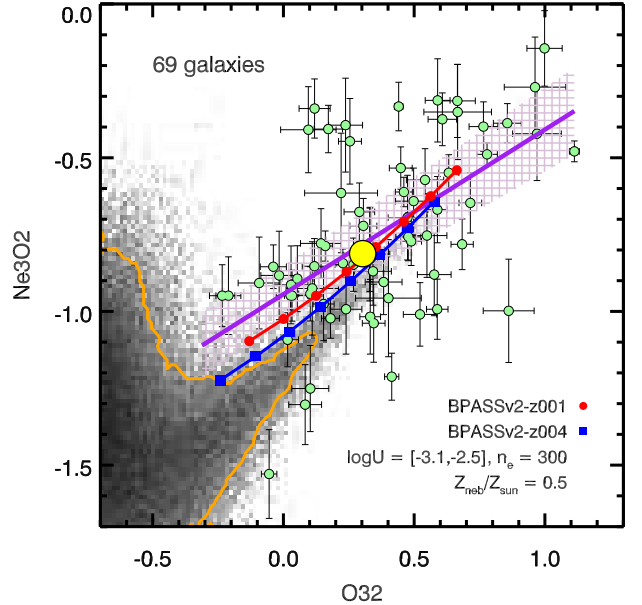}
\caption{The distribution of Ne3O2 vs. O32 for KBSS-MOSFIRE (green points with error bars) and SDSS (in greyscale) galaxies, with the orange contour enclosing 90\% of the $z\sim0$ sample; the location of the \citetalias{steidel2016} stack is identified by the large yellow point. The solid purple line shows the fit to the KBSS-MOSFIRE sample, with the 1$\sigma$ errors in the fit parameters represented by the hatched purple region. The nebular line ratio predictions for BPASSv2 models with $Z_{\ast}/Z_{\odot}=0.07$ (red points) and $Z_{\ast}/Z_{\odot}=0.28$ (blue squares) are shown for  $Z_{\rm neb}/Z_{\odot}=0.5$ and log($U)=[-3.1,-2.5]$, with the lowest values of log($U$) corresponding to low O32 and Ne3O2.}
\label{neon_plot}
\end{figure}

Figure~\ref{neon_plot} shows both indices, Ne3O2 and O32, for $z\sim0$ SDSS galaxies (in greyscale, 90\% contour in orange) and the $z\sim2.3$ KBSS-MOSFIRE sample selected as outlined in Table~\ref{samples} (green points with error bars). The linear fit to the data, accounting for errors in both line indices, is shown by the solid purple line, with the 1$\sigma$ uncertainty in the fit parameters represented by the hatched purple region.

For comparison, the nebular line ratio predictions from the $Z_{\ast}/Z_{\odot}=0.07$ (red points) and $Z_{\ast}/Z_{\odot}=0.28$ (blue squares) models shown in Figure~\ref{o32_probe} are also included in Figure~\ref{neon_plot}, with log($U$) limited to the range identified using O32 measurements alone. The $Z_{\ast}/Z_{\odot}=0.07$ model predicts slightly higher values of Ne3O2 at fixed O32 than the $Z_{\ast}/Z_{\odot}=0.28$ model, but both agree well with the ridgeline of $z\sim2.3$ galaxies.

\begin{figure}
\centering
\includegraphics{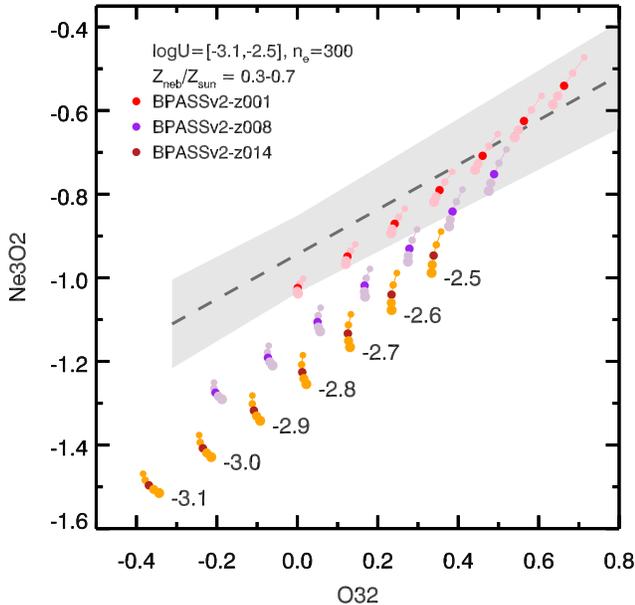}
\caption{Photoionization model predictions for Ne3O2 and O32 for a range of $Z_{\ast}$ and $Z_{\rm neb}$. Each color reflects a different stellar metallicity: $Z_{\ast}/Z_{\odot} = 0.07$ (red), $Z_{\ast}/Z_{\odot} = 0.56$ (purple), and $Z_{\ast}/Z_{\odot} = 1.0$ (orange). The gas-phase metallicity sequence at each log($U$) is represented by the connected points, with the increasing size of the symbol reflecting an increase in $Z_{\rm neb}/Z_{\odot}$ from $0.3-0.7$; the grid points with $Z_{\rm neb}/Z_{\odot}=0.5$ are identified by darker-colored symbols. For comparison, the locus of $z\sim2.3$ galaxies is represented by the grey line and shaded region, which is identical to the hatched region in Figure~\ref{neon_plot}.}
\label{logu_probe_comp}
\end{figure}

Although we have thus far assumed a single gas-phase metallicity, the inferred range of log($U$) is not particularly sensitive to $Z_{\rm neb}$. Figure~\ref{logu_probe_comp} shows the effect of changing $Z_{\rm neb}$ on the predicted line indices for three separate stellar metallicities: $Z_{\ast}/Z_{\odot} = 0.07$ (red), $Z_{\ast}/Z_{\odot} = 0.56$ (purple), or $Z_{\ast}/Z_{\odot} = 1.0$ (orange). For each value of log($U$), the sequence in $Z_{\rm neb}$ is illustrated by a series of connected points, with increasing symbol size reflecting larger values of $Z_{\rm neb}$; grid points with $Z_{\rm neb}/Z_{\odot}=0.5$ are identified by darker-colored symbols. Regardless of $Z_{\ast}$, the differences in the predicted line indices are substantially larger for changes in $U$ than for changes in $Z_{\rm neb}$, particularly at low ionization parameters. At high $U$, the effect of decreasing $Z_{\rm neb}$ is more pronounced, but degenerate with changes in $U$.

However, the choice of $Z_{\ast}$ does have a noticeable impact on the model predictions for Ne3O2 and O32. Figure~\ref{logu_probe_comp} shows that increasing $Z_{\ast}/Z_{\odot}$ from 0.07 to 0.56 decreases the predicted value of Ne3O2 by $\sim0.4$~dex at every value of log($U$); the corresponding decrease in O32 is $\sim0.3$~dex. Although it is possible to reproduce separate values of Ne3O2 or O32 with other combinations of $Z_{\ast}$, $Z_{\rm neb}$, and $U$, the model assuming $Z_{\ast}/Z_{\odot}=0.07$ (red points) is best able to match the \textit{combinations} of Ne3O2 and O32 observed in KBSS-MOSFIRE galaxies, represented by grey shaded region in Figure~\ref{logu_probe_comp}.

\subsection{Comparison with Other Nebular Diagnostics}
\label{model_bpt}

\subsubsection{The O32-R23 Diagram}
\label{o32-r23_models}

\begin{figure}
\centering
\includegraphics{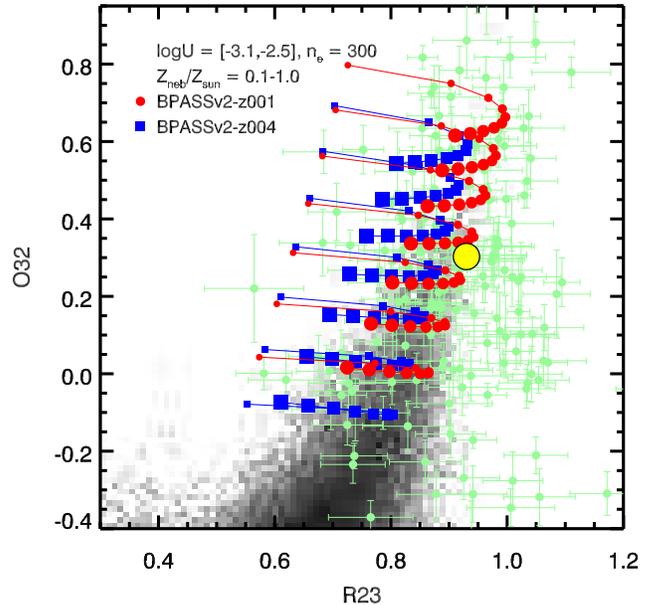}
\caption{A comparison between the photoionization model predictions for O32 and R23 and the location of $z\sim2.3$ KBSS-MOSFIRE and $z\sim0$ SDSS galaxies, with the location of the \citetalias{steidel2016} stack identified by the large yellow point. A range of ionization parameters and gas-phase metallicities are considered, with the $Z_{\rm neb}$ sequence at each value of $U$ connected by a line; the increasing size of the points reflects increasing $Z_{\rm neb}/Z_{\odot}$ from $0.1-1.0$. At fixed O32 and $Z_{\rm neb}$, the only way to increase the maximum value of R23 is to change the shape of the ionizing radiation, here parametrized by $Z_{\ast}$.}
\label{models2}
\end{figure}

Figure~\ref{models2} compares predictions for O32 and R23 from the photoionization models with observations of the KBSS-MOSFIRE and SDSS galaxies; the sample of KBSS-MOSFIRE galaxies is identical to the sample introduced in \S\ref{o32_r23_text1} and is represented by the green points with error bars. The allowed range in ionization parameter has been set by the results from the previous section and matches the values shown in Figure~\ref{neon_plot}, but we now consider a range in gas-phase metallicity: $Z_{\rm neb}/Z_{\odot}=0.1-1.0$.

The metallicity sequences for both models are shown by the series of connected points  in Figure~\ref{models2} and are nearly horizontal, with changes in $Z_{\rm neb}$ (reflected by the symbol size) resulting in motion to the left or right (in R23). Independently, changes in $U$ move points vertically (in O32). Thus, extending models with fixed $Z_{\ast}$ to higher values of $U$ and/or lower $Z_{\rm neb}$ cannot produce the high values of R23 at high O32 that are characteristic of both the $z\sim2.3$ KBSS-MOSFIRE sample and the extreme tail of SDSS. Models with the lowest gas-phase metallicities (the smallest red points and blue squares) are among the most discrepant with respect to the data; only grid points with $Z_{\rm neb}/Z_{\odot}\approx0.3-0.9$ can match the line indices observed for the majority of galaxies, but the location of individual objects will be equally (or more) sensitive to changes in $U$ and $Z_{\ast}$.

These results present severe challenges to using the combination of O32 and R23 as a metallicity determination method for objects in this region of parameter space. Because of the narrow high-ionization sequence at both $z\sim0$ and $z\sim2.3$ and the double-valued nature of the R23 index, galaxies with different $Z_{\rm neb}$ but similar values of $U$ may exhibit similar values of R23 at fixed O32 and thus be spatially coincident in Figure~\ref{models2}. As discussed in \S\ref{twin_text}, the difference in N/O between SDSS twins (identified in the O32-R23 diagram) and KBSS-MOSFIRE galaxies imply that $z\sim2.3$ galaxies have 0.06~dex higher 12+log(O/H) than $z\sim0$ galaxies with the same ionizing spectral shape. Thus, if a KBSS-MOSFIRE galaxy has $Z_{\rm neb}/Z_{\odot}=0.5$, its $z\sim0$ twin should have $Z_{\rm neb}/Z_{\odot}=0.44$, and the insensitivity of O32 and R23 to $Z_{\rm neb}$ (particularly near the turnaround in R23) will render them virtually indistinguishable in Figure~\ref{models2}. An apparent metallicity sequence in the O32-R23 plane could arise due to an anti-correlation between ionization parameter and O/H, as proposed by \citet{sanders2016}, but the relationship between those parameters may not be redshift-invariant and is sensitive to the shape of the incident ionizing radiation. Because the shape of the EUV radiation is regulated largely by the Fe content of massive stars ($Z_{\ast}$), it can differ greatly for galaxies with the same O/H ($Z_{\rm neb}$) if they have different values of O/Fe.

\begin{figure}
\centering
\includegraphics{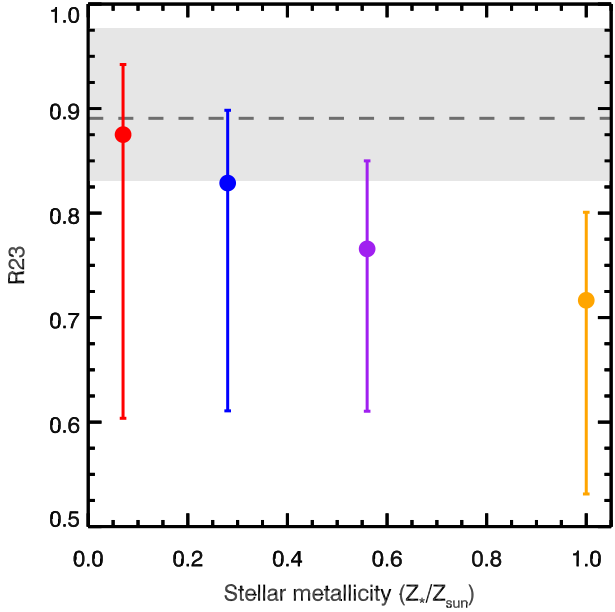}
\caption{The range of predicted R23 indices from BPASSv2 models with a range of $Z_{\ast}$ compared to the interquartile range observed in $z\sim2.3$ KBSS-MOSFIRE galaxies with O32 ratios within $2\sigma$ of the median, O32$=0.10-0.40$ (horizontal grey band, with the median R23 shown by the dashed line). The colored points represent the median R23 value predicted for all combinations of $Z_{\rm neb}$ and $U$ at fixed $Z_{\ast}$ for grid points with equivalent O32 ratios; the error bars show the minimum and maximum R23 predicted for the same grid points. Notably, models assuming $Z_{\ast}/Z_{\odot}\gtrsim0.5$ predict peak values of R23 that are too low to match the majority of $z\sim2.3$ galaxies, suggesting high $Z_{\ast}$ is not typical in this sample.}
\label{r23_at_o32}
\end{figure}

The O32-R23 diagram serves as a much more sensitive diagnostic of the shape and normalization of the ionizing radiation fields in \ion{H}{2} regions and galaxies, in part because it is so insensitive to $Z_{\rm neb}$ for objects with high values of both indices. Figure~\ref{r23_at_o32} compares the interquartile range of R23 for $z\sim2.3$ KBSS-MOSFIRE galaxies with O32 ratios within $2\sigma$ of the median, ${\rm O32}=0.10-0.40$ (represented by the horizontal grey band) with R23 predictions from photoionization models spanning $Z_{\ast}/Z_{\odot}=0.07-1.0$. The colored points represent the median R23 value for all model grid points ($Z_{\rm neb}/Z_{\odot}=0.1-1.0$ and log($U$)=[-3.5,-2.0]) with ${\rm O32}=0.10-0.40$ at each $Z_{\ast}$; the error bars show the minimum and maximum R23 values. For $Z_{\ast}/Z_{\odot}\gtrsim0.5$, even the maximum value of R23 in this range of O32 is too low to match the majority of $z\sim2.3$ galaxies, which suggests that the ionizing spectra in most KBSS-MOSFIRE galaxies are consistent with low-$Z_{\ast}$ BPASSv2 models. It is important to note that the actual constraint imposed by the comparison of the data with the model predictions is on the \textit{shape} of the stellar population EUV SED in the 1-4 Ryd range, and the implementation of low-metallicity binaries in BPASSv2 is simply one realistic way of achieving that spectral shape. 

\subsubsection{The BPT Diagrams}
\label{bpt_models}

\begin{figure*}
\centering
\includegraphics{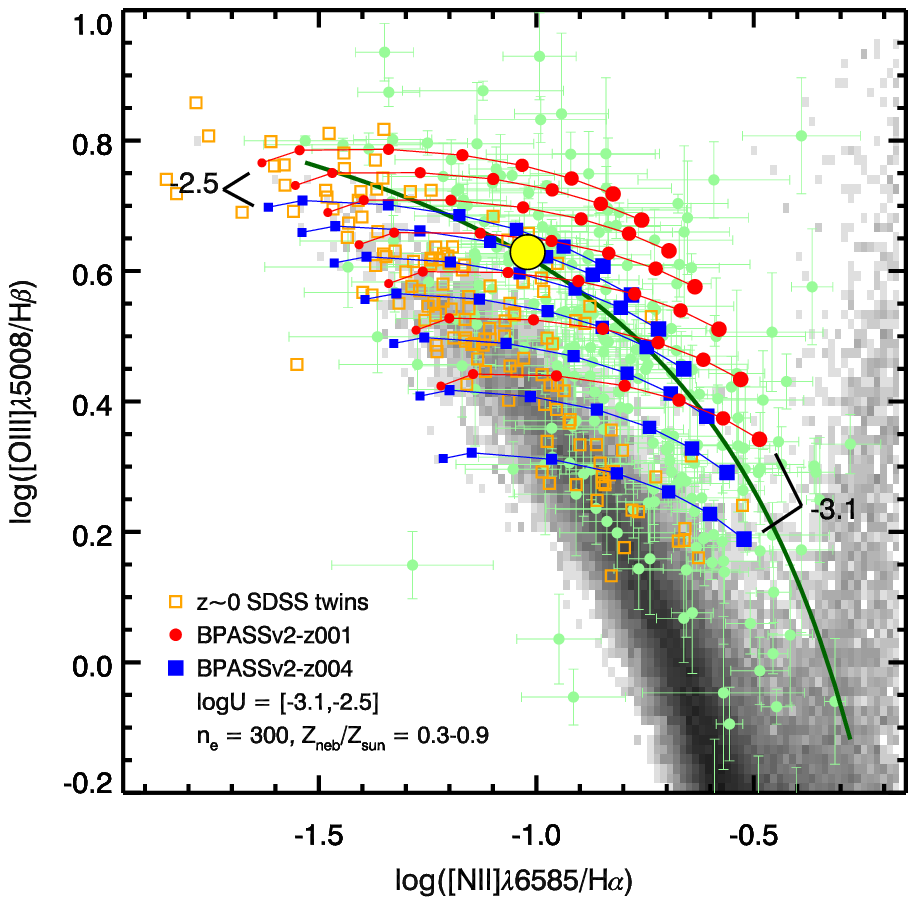}
\hspace{-0.8in}
\includegraphics{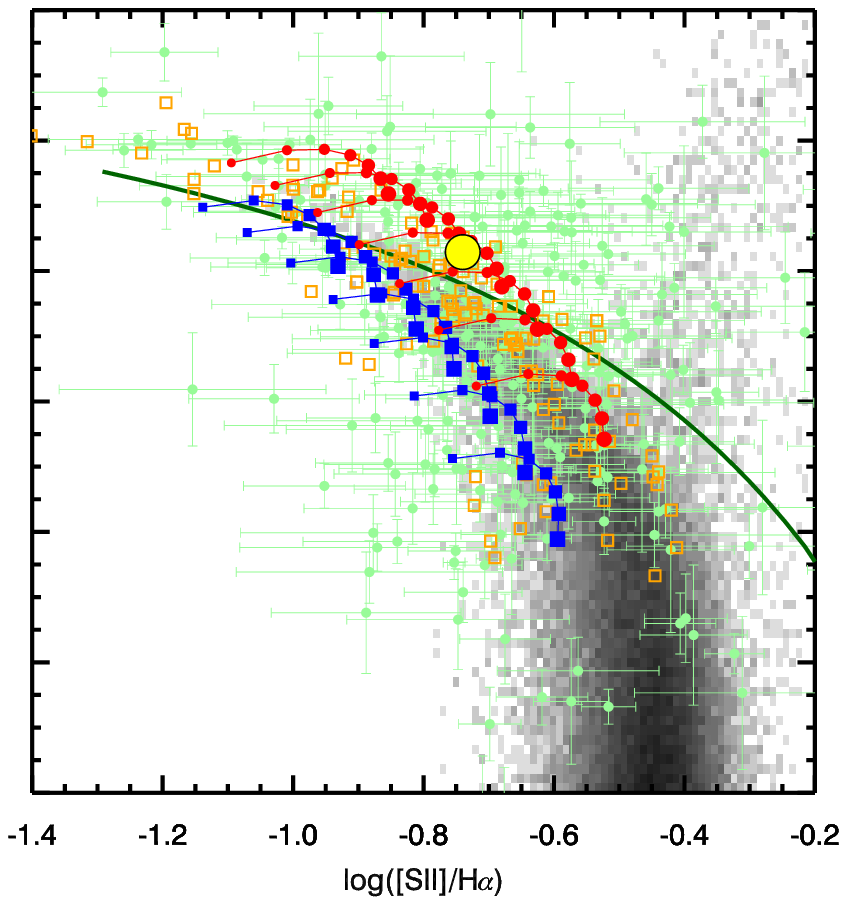}
\caption{A comparison between the range of nebular emission-line ratios predicted by the photoionization models described in the text and those observed in $z\sim2.3$ KBSS-MOSFIRE galaxies (represented by the green points with error bars, with the ridge-line shown by the solid green curve and the location of the \citetalias{steidel2016} stack identified by the large yellow point). The model predictions are presented in the same manner as Figure~\ref{models2}, with the $Z_{\rm neb}$ sequence at each value of $U$ connected by a line; the size of the point reflects the level of gas-phase enrichment, which has been restricted to $Z_{\rm neb}/Z_{\odot}=0.3-0.9$. The model predictions agree well with the full range of observations in both panels and support the results from \S\ref{twin_text}: $z\sim2.3$ galaxies and SDSS twins (matched in O32 and R23, Figure~\ref{twin_select_figure}) are consistent with having the same relatively hard ionizing spectra, but KBSS-MOSFIRE galaxies require higher $Z_{\rm neb}$ to explain their nebular line ratios. The most offset $z\sim2.3$ galaxies in the N2-BPT are consistent with values of $Z_{\rm neb}$ that make them indistinguishable from the SDSS locus in the S2-BPT plane, meaning that an offset in [\ion{N}{2}]/H$\alpha$ with no corresponding offset in [\ion{S}{2}]/H$\alpha$ is a prediction of the models.}
\label{models1}
\end{figure*}

We now return to the BPT diagrams, using constraints from the rest-UV spectral analysis presented in \citetalias{steidel2016} combined with the analysis presented earlier in this paper. To summarize, we adopt
\begin{eqnarray}
n_e &=& 300{\rm~cm}^{-3} \nonumber \\
\log(U)&=&[-3.1,-2.5] \nonumber \\
Z_{\ast}/Z_{\odot} &=& 0.07-0.28 \nonumber \\
Z_{\rm neb}/Z_{\odot} &=& 0.3-0.9 \nonumber
\end{eqnarray}
as the optimal combination of parameters capable of reproducing the nebular properties of $z\sim2.3$ KBSS-MOSFIRE galaxies.

The relation between N/O and O/H in the photoionization models is motivated by the agreement between the \citetalias{pilyugin2012} sample and KBSS-MOSFIRE galaxies shown in \S\ref{no_text}. As shown in Figure~\ref{no_o3n2}, however, the apparent N/O-O/H relation for the KBSS-MOSFIRE sample would imply super-solar values of N/O at solar O/H, a feature common to many empirical relations for N/O-O/H.  Evidence suggests that this discrepancy arises from the tendency of direct $T_e$-based abundances to underestimate the true nebular O/H, in part because the highest temperature regions of nebulae will contribute more to the total flux in collisionally-excited lines than cooler regions. We discussed this topic in detail in Section 8.1.2 of \citetalias{steidel2016}, but we refer to one specific example from the literature here: \citet{esteban2014} found that direct-method oxygen abundances based on measurements of collisionally-excited lines from the \ion{H}{2} regions of nearby galaxies were $0.24\pm0.02$~dex lower than the oxygen abundances determined from measurements of nebular recombination lines. Recombination line methods are less temperature-sensitive and often agree better with stellar abundances when gas-phase and stellar abundances are measured for the same systems. Motivated by this significant difference, we introduce a $+0.23$~dex shift in 12+log(O/H) to Equation~\ref{kbss_no_oh}, which forces the relation to produce (N/O)$_{\odot}$ at (O/H)$_{\odot}$. The final relation used in the photoionization models is then
\begin{eqnarray}
\log({\rm N/O)} = 1.68\times\log(Z_{\rm neb}/Z_{\odot})-0.86; Z_{\rm neb}/Z_{\odot}>0.4 \nonumber \\
\log({\rm N/O)} = -1.5; Z_{\rm neb}/Z_{\odot}\leq0.4 \nonumber \\
\end{eqnarray}

Figure~\ref{models1} shows the range of line ratios observed in KBSS-MOSFIRE galaxies in the N2-BPT and S2-BPT planes compared with predictions from the photoionization models. The best-fit loci for the KBSS-MOSFIRE sample are represented by the dark green curves, with the KBSS-MOSFIRE sample plotted as light green points with error bars. The models are illustrated in a similar manner to Figure~\ref{models2}, where the metallicity sequence at each value of log($U$) is connected by a line and the size of the point reflects the value of $Z_{\rm neb}/Z_{\odot}$.

Foremost, we note that the model predictions agree well with the range of KBSS-MOSFIRE observations in both diagrams, as they have for the other combinations of line indices presented in this section. The ability of the BPASSv2 models to self-consistently reproduce many of the commonly-used strong rest-optical emission line ratios across the $z\sim2.3$ sample is encouraging. It also confirms that the radiation fields in the majority of high-redshift galaxies must at least resemble the spectra produced by stellar populations with $Z_{\ast}/Z_{\odot}=0.07-0.28$ that include massive binaries. 

In contrast to the locus of $z\sim2.3$ galaxies in the O32-R23 diagram, which provides only weak constraints on the likely range of $Z_{\rm neb}$, the left panel of Figure~\ref{models1} shows that moderately-high $Z_{\rm neb}$ is required to reproduce the high values of log([\ion{N}{2}]/H$\alpha$) observed at fixed log([\ion{O}{3}]/H$\beta$). However, despite exhibiting an offset relative to the locus of SDSS galaxies in the N2-BPT diagram, the photoionization model predictions for $Z_{\rm neb}/Z_{\odot}\approx0.6-0.9$ overlay the high-[\ion{O}{3}]/H$\beta$ tail of SDSS in the S2-BPT diagram---as observed for the KBSS-MOSFIRE galaxies. The strong offset of $z\sim2.3$ galaxies relative to SDSS in the N2-BPT diagram with little to no corresponding offset in the S2-BPT plane is actually a prediction of photoionization models that include hard ionizing radiation from relatively low-$Z_{\ast}$ stars and relatively high-$Z_{\rm neb}$ gas.

We may also use observations of SDSS galaxies in the N2- and S2-BPT diagrams to constrain the combination of photoionization model parameters that would successfully describe $z\sim0$ galaxies with the highest ratios of [\ion{O}{3}]/H$\beta$, such as the SDSS twins introduced in \S\ref{twin_text}. The location of the SDSS twins in the O32-R23 diagram (Figure~\ref{twin_select_figure}) provides a compelling argument that the shape and normalization of their ionizing radiation fields (and, thus, $Z_{\ast}$ and $U$) are similar to those observed in KBSS-MOSFIRE galaxies and that predictions from the same BPASSv2 models are appropriate. In the left panel of Figure~\ref{models1}, the SDSS twins are identified by open orange squares and most closely matched by models with $Z_{\rm neb}/Z_{\odot}\approx0.3-0.5$; the same model parameters also provide good agreement with observations of the SDSS twins in the S2-BPT diagram. That SDSS galaxies matched in O32 and R23 appear to have lower $Z_{\rm neb}$ than KBSS-MOSFIRE galaxies (which require $Z_{\rm neb}/Z_{\odot}$ up to $\approx0.9$) is compatible with the significantly different distributions of N/O for the SDSS twins and KBSS-MOSFIRE galaxies (panel (c) in Figure~\ref{twin_result_figure}) and the N/O-O/H relation from Equation~\ref{kbss_no_oh}.

\subsection{O/H Constraints from Photoionization Models}
\label{oh_constraints}

The discussion in this section explicitly highlights the separate impact that gas chemistry (including both N/O and O/H) and the shape and normalization of the ionizing radiation (parametrized by $Z_{\ast}$ and $U$) have on the nebular spectra of \ion{H}{2} regions. Although the combination of O32 and R23 is relatively insensitive to gas-phase oxygen abundance in high-excitation nebulae (like those in $z\sim2.3$ galaxies and in the SDSS twins), the O32-R23 diagram is especially powerful for constraining the shape of the EUV radiation field because the line indices trace ionization parameter and excitation almost independently.

Conversely, the location of galaxies in the N2-BPT diagram depends sensitively on almost every parameter of interest---gas-phase oxygen abundance, N/O, ionization parameter, and the shape of the ionizing radiation field---making it difficult to disentangle the contributions of each. Fortunately, surveys like KBSS-MOSFIRE allow us to leverage observations of $z\sim2.3$ galaxies in a multidimensional nebular line ratio space to constrain these quantities \textit{for individual galaxies} through comparisons with photoionization models, without requiring a recalibration of the relationship between O/H and the strong-line indices. The utility of this method has already been demonstrated for composite spectra by \citetalias{steidel2016}, where the oxygen abundance could be measured directly using $T_e$, and independent constraints on the most likely input ionizing radiation field were available from comparisons between stellar population models and the rest-UV spectrum. The agreement between the BPASSv2 models and the extent of the full KBSS-MOSFIRE sample in multiple parameter spaces suggests that similar families of models may be applied directly to observations of individual galaxies, which is the focus of future work. Self-consistent measurements of O/H, N/O, and $U$ for a large sample of $z\sim2.3$ galaxies will allow us to directly investigate the N/O-O/H relation and the correlation between metallicity and $U$.

\section{What Causes the $z\sim2.3$ BPT Offset?}
\label{whyoffset}

Returning to the question that has been the subject of much scientific debate over the last few years---``What causes the BPT offset?"---we are now better equipped to determine which mechanism or combination of mechanisms is likely to be primarily responsible, both for explaining the differences between samples of galaxies at $z\sim0$ and $z\sim2.3$ and between galaxies at the same redshift. We have already shown that the offset of $z\sim2.3$ galaxies with respect to SDSS in the N2 BPT plane cannot be attributed only to selection and observational biases (Figures~\ref{farselect_colors} and \ref{farselect_lum}). In this section, we use the combined analysis of the rest-optical spectra of KBSS-MOSFIRE galaxies to evaluate the relative importance of some of the most frequently proposed physical explanations. 

\textbf{Ubiquitous AGN activity.} Like $z\sim2$ galaxies from other surveys, the KBSS-MOSFIRE sample occupies a region in the N2-BPT diagram that is populated almost exclusively by AGN or composite galaxies at $z\sim0$ (Figure~\ref{n2_bpt}). Additionally, the values of log([\ion{O}{3}]/H$\beta$) observed in the $z\sim2.3$ galaxies are $\sim0.8$~dex higher than local star-forming galaxies at fixed M$_{\ast}$ and, thus, similar to local AGN (Figure~\ref{mex_plots}). However, there is no evidence of AGN activity in the rest-UV spectra of most KBSS-MOSFIRE galaxies with complementary LRIS observations (see \citetalias{steidel2014}); very high values of log([\ion{N}{2}]/H$\alpha$) or large rest-optical line widths are also rare. In total, only 7/377 KBSS-MOSFIRE galaxies with measurements that place them on the N2-BPT diagram are identified as AGN or QSOs (\S\ref{agn_text}). Although the supermassive black holes in these galaxies are almost certainly accreting gas, the resulting effect on the total nebular spectrum is sub-dominant with respect to ongoing star formation.

\textbf{Higher electron density.} In the local universe, $n_e$ is correlated with the location of galaxies in the N2-BPT plane \citep{brinchmann2008,liu2008,yuan2010}. Because the characteristic $n_e$ observed in high-redshift galaxies is 10 times higher than galaxies at $z\sim0$, with typical values of $\approx200-300$~cm$^{-3}$ at $z\sim2.3$ \citep[e.g.,][]{steidel2014,sanders2016}, it is important to quantitatively assess whether this difference is sufficient to explain the observed N2-BPT offset. As shown in Figure~\ref{ratios_paper_ne}, however, measurements of the density-sensitive [\ion{O}{2}] doublets for small- and large-offset KBSS-MOSFIRE galaxies (Figure~\ref{farselect}) are nearly identical. Given that the inferred electron densities for both subsamples are consistent within errors, we conclude that a difference in $n_e$ cannot account for the still significant offset between the small- and large-offset N2-BPT loci and, thus, is unlikely to be the primary driver of the overall $z\sim2.3$ offset.

\textbf{Enhanced N/O at fixed O/H.} Several authors have interpreted the strong offset of high-$z$ galaxies in the N2-BPT plane (and lack of corresponding offset in the S2-BPT and O32-R23 diagrams) as evidence for larger N/O at fixed O/H \citep[including][]{masters2014,shapley2015,cowie2016,sanders2016,masters2016}, an explanation that naturally results in larger values of log([\ion{N}{2}]/H$\alpha$) while leaving line ratios not including N unaffected. This enhancement, if it exists, should also be reflected in the correlation between N/O and other galaxy properties (such as M$_{\ast}$), particularly given the strong, but evolving correlation between O/H and M$_{\ast}$ observed at all redshifts.

We showed in Figure~\ref{no_mstar} that, at fixed M$_{\ast}$, $z\sim2.3$ KBSS-MOSFIRE galaxies exhibit values of log(N/O) that are lower by an average of $\sim0.32$~dex relative to SDSS galaxies. Galaxies at $z\sim2.3$ also have lower 12+log(O/H) than SDSS galaxies by a comparable amount \citep{steidel2014,wuyts2014,sanders2015}, suggesting that there is a relatively constant N/O-O/H relation for all galaxies. Additional evidence in support of this interpretation comes from the comparison of the composite UV-optical nebular spectrum from \citetalias{steidel2016} with a sample of $z\sim0$ \ion{H}{2} regions from \citet{pilyugin2012}, with the $z\sim2.3$ stack exhibiting N/O at fixed O/H consistent with the local relation.

KBSS-MOSFIRE galaxies do show 0.10~dex larger values of log(N/O) with respect to SDSS galaxies matched in O32 and R23 (as described in \S\ref{twin_text}), consistent with the results reported by \citet{shapley2015} for $z\sim2$ MOSDEF galaxies, but we argue that samples matched in O32 and R23 are more likely to share ionizing radiation fields than gas-phase oxygen abundance. Assuming the N/O-O/H relation from Equation~\ref{kbss_no_oh}, this difference in N/O would accompany a 0.06~dex difference in 12+log(O/H), which would not be readily apparent from the locations of galaxies in the O32-R23 diagram (Figure~\ref{models2}). We note that, even if the higher N/O measurements represent an enhancement at fixed O/H, it can account for $<55\%$ of the 0.37~dex horizontal displacement of the $z\sim2.3$ locus with respect to the $z\sim0$ locus in Figure~\ref{n2_bpt}. Thus, even if high-$z$ galaxies were moderately N/O-enhanced, an additional mechanism (such as harder ionizing radiation fields) would still be required to explain the differences between $z\sim2.3$ and $z\sim0$.

\textbf{Hard ionizing spectra of Fe-poor massive binaries.} The analysis in \S\ref{models} uses predictions from a subset of the photoionization models we considered in \citetalias{steidel2016} to constrain the likely combination of $U$ and $Z_{\rm neb}$ in the ensemble of KBSS-MOSFIRE observations. Adopting two BPASSv2 models with $Z_{\ast}/Z_{\odot}=0.07$ and $Z_{\ast}/Z_{\odot}=0.28$, we find that the line ratios observed in $z\sim2.3$ galaxies are consistent with log($U)=[-3.1,-2.5]$ and $Z_{\rm neb}/Z_{\odot}\approx0.3-0.9$. We are not able to place direct constraints on O/Fe in individual galaxies in the manner of \citetalias{steidel2016}, but we note that the upper range in $Z_{\rm neb}/Z_{\ast}$ obtained from comparisons with the full KBSS-MOSFIRE sample broadly agrees with the O/Fe reported for the composite spectra ($\simeq4-5$(O/Fe)$_{\odot}$).
\\

We argue, as we did in \citetalias{steidel2014} and \citetalias{steidel2016}, that the primary cause of the differences observed between typical galaxies at high-redshift and galaxies in the local universe is the degree of nebular excitation, with $z\sim2.3$ galaxies exhibiting significantly higher values of [\ion{O}{3}]/H$\beta$ and R23 relative to the majority of $z\sim0$ galaxies. These line ratios---especially the high values of R23 observed at high O32 in KBSS-MOSFIRE galaxies---can be explained by hard ionizing radiation fields, such as those produced by Fe-poor stellar populations that include massive binaries. Yet, when compared with SDSS galaxies with similar nebular excitation (like the SDSS twins from \S\ref{twin_text}), KBSS-MOSFIRE galaxies are still substantially offset to higher log([\ion{N}{2}]/H$\alpha$) at fixed log([\ion{O}{3}]/H$\beta$) due to larger values of N/O (and, likely, O/H). Thus, $z\sim2.3$ galaxies must contain harder ionizing radiation fields \textit{at fixed gas-phase N/O and O/H} to explain the N2-BPT offset relative to the locus of typical $z\sim0$ galaxies.

\subsection{Implications for Local Analogs}

These results are a natural consequence of systematic differences in star formation history between high-redshift star-forming galaxies and galaxies occupying the low-redshift locus. As we reasoned in \citetalias{steidel2016}, the ISM in galaxies with roughly constant SFR and stellar population ages $\lesssim1$~Gyr will reflect enrichment primarily from core-collapse SNe, with the resulting super-solar O/Fe abundances responsible for harder ionizing spectra (from Fe-poor stars) at a given O/H compared to galaxies with lower O/Fe. If the characteristic star formation histories of galaxies were SFR increasing with time \citep[as might be more common at high-redshift;][]{reddy2012,steidel2016}, high values of O/Fe could easily be maintained for longer than 1~Gyr. This is possible because, to first order, the enrichment rate of O is proportional to the current SFR, whereas the rate of enrichment of Fe (which comes primarily from Type Ia SNe) is proportional to the SFR 1~Gyr earlier. Thus, given typical inferred ages of a few hundred Myr \citep{reddy2012}, most $z\sim2.3$ galaxies should be forming stars that are relatively $\alpha$-enhanced (and O-enhanced), consistent with core-collapse SNe yields.

Specific star formation rate serves as a crude, but easily-measured probe of the overall star formation histories of galaxies. By definition, young stars make up a larger fraction of the stellar populations in galaxies with high sSFRs, and 1/sSFR is roughly equivalent to the age of the stellar population (assuming a constant SFR). Thus, galaxies with ${\rm sSFR}>1$~Gyr$^{-1}$ will have a substantial fraction of stars that formed from gas enriched almost entirely by Type II SNe. The median sSFR for the KBSS-MOSFIRE sample is 2.4~Gyr$^{-1}$, corresponding to a timescale of $\sim400$~Myr, substantially less than the 1~Gyr timescale for the onset of Type Ia SNe; moreover, as we showed in \S\ref{farselect_properties}, nearly $60\%$ of KBSS-MOSFIRE galaxies have ${\rm sSFR}>2$~Gyr$^{-1}$, corresponding to timescales $<500$~Myr, with only $\sim20\%$ having ${\rm sSFR}<1$~Gyr$^{-1}$. In contrast, only $\sim1\%$ of SDSS galaxies have ${\rm sSFR}>2$~Gyr$^{-1}$, and such galaxies also have $\sim10$ times smaller M$_{\ast}$ compared to high-sSFR KBSS-MOSFIRE galaxies. For the majority of SDSS galaxies, ${\rm sSFR}\sim0.1$~Gyr$^{-1}$; at these low sSFRs, the youngest stars represent a small contribution to the overall M$_{\ast}$, and their O/Fe abundances will closely reflect the end state of the galaxy's chemical evolution.

It is clear that sSFR can be used to identify galaxies whose nebular properties are similar in some ways to typical galaxies at $z\sim2.3$. However, the KBSS-MOSFIRE galaxies that are \textit{most} offset from the SDSS star-forming sequence in the N2-BPT diagram represent an interesting population without exact analogs in the local universe: galaxies at $z\sim0$ with similar ionizing spectra are $\sim10$ times less massive than typical $z\sim2.3$ galaxies and have significantly lower N/O and O/H, but $z\sim0$ galaxies with similar M$_{\ast}$ or gas-phase metallicity will have substantially softer ionizing radiation fields. The absence of a suitable comparison sample---with both high levels of nebular excitation (high [\ion{O}{3}]/H$\beta$ and R23) \textit{and} high gas-phase metallicity---limits the utility of $z\sim0$ analogs for studying the conditions in $z\sim2.3$ systems significantly offset in the N2-BPT plane. Specifically, it raises concerns about using extreme $z\sim0$ galaxies (like the SDSS twins) to construct an empirical strong-line metallicity calibration intended to be used at $z\gtrsim2$. Any study using calibrations based on nebular line ratios that are sensitive to both gas-phase chemistry and the shape and normalization of the ionizing radiation field must take care to understand the impact of differences between the calibration and test samples.

\section{Summary}
\label{summary}

We have presented a detailed analysis of the rest-optical ($3600-7000$\AA) spectra of $\sim380$ $z\simeq2-3$ star-forming galaxies drawn from the MOSFIRE component of the Keck Baryonic Structure Survey. Combining multiwavelength photometric observations and SED fitting with robust measurements of many of the commonly-used strong nebular emission lines, we have shown that:
\begin{itemize}
\item KBSS-MOSFIRE galaxies at $z\sim2.3$ exhibit a clear offset in the N2-BPT diagram relative to typical galaxies in the local universe, represented by SDSS (Figure~\ref{n2_bpt}). If one assumes the difference is along only one axis, the magnitude of this offset is $\Delta \log(\textrm{[\ion{N}{2}]/H$\alpha$})= 0.37$~dex or $\Delta \log(\textrm{[\ion{O}{3}]/H$\beta$}) = 0.26$~dex.
\item The same high-$z$ galaxies have nebular properties consistent with the high-excitation tail of the $z\sim0$ galaxy distribution in the S2-BPT (Figure~\ref{s2_bpt}) and O32-R23 (Figure~\ref{o32r23_plots}) diagnostic diagrams.
\item The offset of $z\sim2.3$ galaxies relative to the $z\sim0$ N2-BPT locus is inversely correlated with M$_{\ast}$ and positively correlated with sSFR. Likewise, sSFR is also strongly correlated with the degree of excitation (as probed by [\ion{O}{3}]/H$\beta$ and R23) for all galaxies (Figure~\ref{o3hb_ssfr_plot}).
\item Many of the most offset $z\sim2.3$ KBSS-MOSFIRE galaxies have blue optical-NIR colors ($\mathcal{R}-K_s\leq1$), high sSFRs ($\sim7$~Gyr$^{-1}$), and are best-fit by young ($50$~Myr) stellar population ages, assuming constant or increasing star formation histories (Figures~\ref{o3hb_ssfr_plot}, \ref{farselect_colors}, and \ref{farselect_seds}).
\item KBSS-MOSFIRE galaxies at $z\sim2.3$ have 0.10~dex higher N/O than SDSS galaxies matched in O32 and R23 (panel (c) in Figure~\ref{twin_result_figure}). If galaxies at all redshifts exhibit the same N/O-O/H relation (as we argue in \S\ref{no_text}), a 0.10~dex increase in log(N/O) corresponds to a 0.06~dex increase in 12+log(O/H). Such a small change in O/H would have only a minor effect on the observed locations of galaxies in the O32-R23 and S2-BPT diagrams.
\item Observations of most $z\sim2.3$ KBSS-MOSFIRE galaxies are consistent with photoionization models that use a binary population synthesis model (BPASSv2) with stellar metallicity $Z_{\ast}/Z_{\odot}=0.07-0.28$ (Figure~\ref{r23_at_o32}), gas-phase metallicity $Z_{\rm neb}/Z_{\odot}\approx0.3-0.9$, and ionization parameter log($U)=[-3.1,-2.5]$ (Figure~\ref{models1}). These models reproduce the large observed shift with respect to local star-forming galaxies in the N2-BPT diagram, but result in no appreciable offset between the predicted line ratios for $z\sim2.3$ galaxies and the local sequence in the S2-BPT plane; thus, this apparent discrepancy can be explained without needing to invoke elevated N/O at fixed O/H.

\end{itemize}
From these results, we conclude that the principal cause of the ``BPT offset" is an increase in the hardness of the ionizing radiation at fixed N/O and O/H, consistent with the EUV spectra produced by Fe-poor stellar populations that include massive binaries.

\acknowledgements
We are grateful to the referee for their thoughtful and constructive feedback during the review process. We also thank the organizers and attendees of the Carnegie Symposium in honor of Leonard Searle, ``Understanding Nebular Emission in High-Redshift Galaxies", held at the Carnegie Observatories in Pasadena, for many interesting conversations which contributed to the work presented here. This work has been supported in part by a US National Science Foundation (NSF) Graduate Research Fellowship (ALS), by the NSF through grants AST-0908805 and AST-1313472 (CCS, ALS, RFT, GCR), and by an Alfred P. Sloan Research
Fellowship (NAR). Finally, the authors wish to recognize and acknowledge the significant cultural role and reverence that the summit of Mauna Kea has within the indigenous Hawaiian community.  We are privileged to have the opportunity to conduct observations from this mountain.

\bibliography{all_refs}

\end{document}